# Hydrogen-bond relaxation dynamics: Resolving mysteries of water ice

Yongli Huang[1,*], Xi Zhang[2,3,4,*], Chang Q Sun[2,#]

**Abstract**
We present recent progress in understanding the anomalous behavior of water ice under mechanical compression, thermal excitation, and molecular undercoordination (with fewer than four neighbors in the bulk) from the perspective of hydrogen (O:H-O) bond cooperative relaxation. Extending the Ice Rule suggests a tetrahedral block that contains two $H_2O$ molecules and four O:H-O bonds. This block unifies the length scale, geometric configuration, and mass density of molecular packing in water ice. This extension also clarifies the flexible and polarizable O:H-O bond that performs like a pair of asymmetric, coupled, H-bridged oscillators with short-range interactions and memory. Coulomb repulsion between electron pairs on adjacent oxygen atoms and the disparity between the O:H and the H-O segmental interactions relax the O:H-O bond length and energy cooperatively under stimulation. A Lagrangian solution has enabled mapping of the potential paths for the O:H-O bond at relaxation. The H-O bond relaxation shifts the melting point, O 1s binding energy, and high-frequency phonon whereas the O:H relaxation dominates polarization, viscoelasticity, and the O:H dissociation energy. The developed strategies have enable clarification of origins of the following observations: (i) pressure-induced proton centralization and phase transition–temperature depression; (ii) thermally-induced four-region oscillation of the mass density and the phonon frequency over the full temperature range; and (iii) molecular-undercoordination-induced supersolidity that is elastic, hydrophobic, thermally stable, with ultra-low density. The supersolid skin is responsible for the slipperiness of ice, the hydrophobicity and toughness of water skin, and the bi-phase structure of nanodroplets and nanobubbles. Molecular undercoordination mediates the O:H and H-O bond Debye temperatures and disperses the liquid-solid transition phase boundary, resulting in supercooling at freezing and superheating at melting. O:H-O bond memory and water-skin supersolidity ensures a solution to the Mpemba paradox – hot water freezes faster than its cold. These understandings will pave the way towards unveiling anomalous behavior of $H_2O$ interacting with other species such as salts, acids and proteins, and excitation of $H_2O$ by other stimuli such as electrical and magnetic fields.

**Keywords**: Water structure; ice rule; H-bond potentials; phonon relaxation; pressure; temperature; molecular cluster; Raman; FTIR; XPS; phase transition; viscoelasticity; polarization; specific heat; slipperiness of ice; water surface tension; Coulomb coupling; multiple fields; correlation and fluctuation; Fourier fluid thermodynamics; water-protein interaction; hydrophobicity; polarization; negative thermal expansion; electro- and magneto-freezing; supersolidity; superheating and supercooling; Mpemba paradox; Hofmeister series; Leidenfrost effect.

Contents

# 1 Introduction

- *Water ice has attracted much attention because of its importance in astrophysics, biology, climate, environment, galaxy, geology, and our daily lives.*
- *Clarification, correlation, formulation, and quantification of hydrogen-bond relaxation dynamics and its associated anomalies of water ice remain largely unexplored.*
- *Cooperative relaxation of the O:H-O bond in length and stiffness and the associated electronic energetics mediate the behavior of water ice.*
- *Focusing on the statistical mean of all the correlated parameters simultaneously is more reliably revealing than on the instantaneous accuracy of a parameter at a given time for the strongly correlated and fluctuating water ice system.*

## 1.1 Scope

This report starts with a brief overview in section 1 on the challenges, significance and status in understanding the structure order, local potentials, vibration attributes, molecular images, and physical anomalies of water ice. Section 2 describes the notation of hydrogen bond (O:H-O) cooperativity. An extension of the Ice Rule produces a tetrahedral block containing two $H_2O$ molecules and four O:H-O bonds. This building block not only unifies the size, separation, structure order and mass density of packed molecules in water and ice, but also the flexible and polarizable O:H-O bond that performs like a pair of asymmetric, coupled, H-bridged oscillators with short-range interactions and memory [1]. The O:H-O bond short-range interactions and the O–O Coulomb repulsion discriminate water and ice from other materials in functionalities such as the critical temperature ($T_C$) for phase transition, O 1s electron binding energy and the O:H and the H-O stretching phonon frequency shift, as basic concerns. Section 3 describes the analysis and formulation strategies that correlate the performance of water ice to the O:H-O bond-electron-phonon attribute, in addition to computational and experimental procedures and skills. Section 4 deals with anomalies of compressed ice in proton centralization, $T_C$ and compressibility depression, and band gap expansion [2]. Section 5 is focused on the anomalies demonstrated by water molecules with fewer than four neighbors presented in clusters, hydration shells, skins, and ultrathin films [3]. Particular attention is given on verifying skin supersolidity [4]. The supersolid skin that is elastic, hydrophobic, thermally stable and less dense lubricates ice and toughens water skin. Section 6 is focused on the oscillation dynamics of mass density and phonon stiffness of water ice over the full temperature range, which clarifies why ice floats [5]. Section 7 deals with the coupling effect of compression, undercoordination, and thermal excitation on the phonon stiffness and O 1s binding energy. The size dispersion of the extreme-density temperatures of nanodroplets and nanobubbles are of particular concern, which clarifies why superheating and supercooling occur in the liquid-solid transition phase. Section 8 describes the Lagrangian-Laplace transformation that maps the potential paths of the O:H-O bond at relaxation [6]. With the known segmental lengths and their vibration frequencies as input, one can obtain the respective force constants and binding energies of these two segments at each quasi-equilibrium state. Section 9 verifies correlations between the size, separation, coordination order, and mass density of packed molecules in water ice and water structure solution uniqueness [1]. Water prefers the fluctuating tetrahedral structure with a supersolid skin except at the nanometer scale that shows bi-phase dominance in a core-shell configuration. Section 10 presents a solution to the Mpemba paradox [7]. Reproduction of observational evidence that the O:H-O bond has memory to emit energy at a rate depending on its initial storage and that water-skin supersolidity raises the thermal diffusivity, favoring outward heat flow from the liquid. Section 11 addresses prospects on open questions including Hofmeister series, dielectric relaxation, electro-, magnetico-, mechanico-stimulated freezing, water-protein and water-cancer tissue interactions, and negative thermal expansion of other species. Section 12 summarizes the quantification of the concerned properties and the consistent understanding of the anomalous behavior of water ice, which is evidence of the reality of O:H-O bond cooperativity. Further efforts to reveal the anomalies of water ice and their functionality in other areas would be even more fascinating, promising and rewarding.

1.2     Overview
1.2.1   Significance of water and ice

Water is the source and central part of all lives. Life can never evolve without liquid water. As the key component of water and biomolecules, the O:H-O bond is ubiquitously important. The O:H-O bond gives water unique properties, accelerates or slows reactions, and holds together the three-dimensional configurations of deoxyribonucleic acid (DNA), proteins, and other supramolecular structures [8-10]. Given its importance in nature [11-22] and in geochemical sciences [23,24], and its role in DNA and protein folding [8,25-27], gene delivering [28-30], cell culturing [31], drug target binding [32], ion channel activating and deactivating [33], etc., $H_2O$ and the O:H-O bond have been studied since the dawn of scientific thought. The perspectives of classical thermodynamics and quantum mechanics have considerably advanced this field.

Currently active areas include: i) crystal structure optimization [34], phase formation and transition [35,36]; ii) reaction dynamics with other ingredients [37,38]; iii) O:H-O bond weak interactions [39,40]; iv) binding energy determination [41-47]; and v) phonon relaxation dynamics under various conditions [15,38,48-50].

Authoritative reviews on the advancement in this field have focused on the following issues from various perspectives: from water structures [51-53], molecular clusters [54-57], ice nucleation and growth [22,58], ice melting [59], slipperiness and friction of ice [60], to the behavior of water ice subjected to positive [61,62] and negative pressure [54,63]. Reviews have also covered topics on water surface charge and polarization [64,65], surface photoelectron emission [66,67], phonon relaxation [24,68], water adsorption onto inorganic surfaces [69-73], and ion effects on water properties and structures [74], among other topics.

State-of-the-art techniques, such as sum frequency generation (SFG) [65,75], microjet photoelectron emission [76], glancing-angle Raman spectroscopy [77,78] and so on, have propelled advancement in studies of surfaces and interfaces of water [79,80] with acids and salt solutions [81].

However, water and ice are too strange, too anomalous, and too challenging [11,82,83]. Property variations rarely follow the same rules as 'normal' materials. Difficulties remain in accurately determining the three-dimensional coordination order and thermodynamic behavior [84]. Insight into the anomalies of water ice from the perspective of O:H-O bond relaxation is still in its infancy; mysteries pertaining to the structure order, local potentials, physical anomalies, and their correlation, as summarized in the respective sections, remain puzzling.

1.2.2   Typical structural and potential models

The success of existing models is that they have contributed significantly to structural optimization and entropy calculation, so far leading to more than 17 phase structures. Among various models for the structure of water ice, the rigid and non-polarizable TIPnP (n varies from 1 to 5) series [53,85] and the polarizable models [86] are widely used. In the TIPnQ models, for instance, the V-shaped $H_2O$ geometry with a bond length of $r_{OH} = 0.9572$ Å and a bond angle of $\theta_{HOH} = 104.52°$ describes a water molecule in the gaseous phase.

Figure 1a illustrates the TIP4Q/2005 structure model [87] that simplifies the H₂O molecule as a dipole ($O^+$ – $M^-$) with a fixed $H^+$ point charge. Lennard-Jones (L-J) potential represents the inter-dipole interaction.

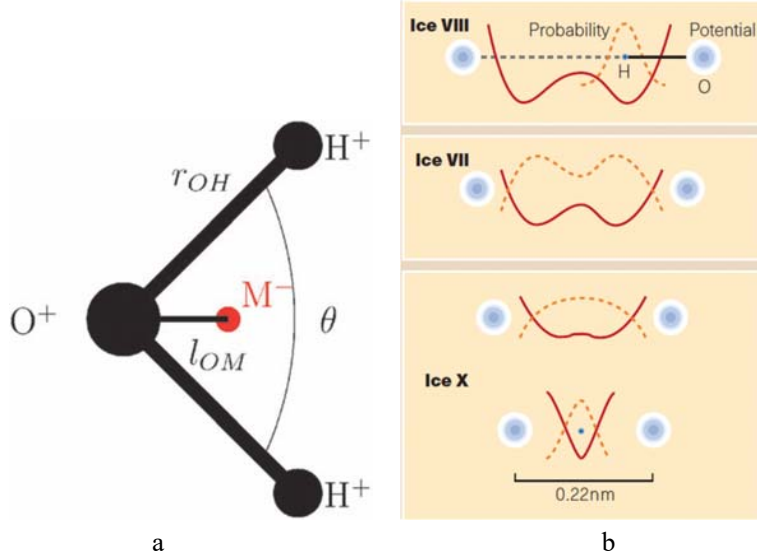

Figure 1. (a) The rigid non-polarizable TIP4Q/2005 model has a positive point charge $q_O$ on the oxygen atom, a positive point charge $q_H$ on each hydrogen atom and a negative point charge $q_M$ at site M located at a distance $l_{OM}$ from the oxygen atom along the bisector of angle ∠HOH. The molecule is electrically neutral, thus $q_M = -(2q_H + q_O)$, and $q_H$, $q_O$, $l_{OM}$, $\sigma_{OO}$, and $\varepsilon_{OO}$ are independent parameters to be determined, where $\sigma_{OO}$ and $\varepsilon_{OO}$ are the L-J potential parameters for the O—O interaction (reprinted with permission from [87]). (b) The 'double symmetrical potential well' model indicates that the $H^+$ proton is frustrated in the two identical sites between adjacent oxygen atoms. The symmetrical double wells reduce to a single well when subjected to compression, which forces the fluctuating H proton into site midway between the two oxygen ions approaching the ice-X phase [88]. (Reprinted with permission from [89].)

Figure 1b illustrates the evolution of the O:H-O bond symmetrical double-well potentials subjecting to compression [88]. Wernet et al. [90] proposed an alternative to replace the symmetrical O:H-O bond potential, which Soper [91] investigated further by assuming different charges on the $H^+$ ions. Wikfeldt et al. [92] and Leetmaa et al. [93] have also attempted . However, this assumption remains yet undemonstrated, as diffraction and phonon spectroscopy can barely probe the inter- and intramolecular local potentials.

1.2.3   Phonon frequency identities

Figure 2 shows typical phonon spectra for ambient water probed using spectroscopies of Fourier transform infrared (FTIR) absorption, Raman reflection, and neutron diffraction [94]. The characteristic frequency (or energy) features correspond to stretching and bending vibration modes. Features centered at $\omega_H \approx 3450$ and 3200 cm$^{-1}$ correspond to the H-O stretching phonons in the skin and in the bulk, respectively. The peak at $\omega_{B1}$ at 500–700 cm$^{-1}$ is the ∠O:H-O bending mode. The peak of $\omega_{B2}$ at 1600–1750 cm$^{-1}$ is the librational mode of ∠H-O-H bending. Peaks at $\omega_L < 300$ cm$^{-1}$ arise from O:H stretching. (Water molecules in the gaseous phase exhibit a peak in the vicinity of $\omega_H \approx 3650$ cm$^{-1}$, which is absent from Figure 2.)

The libration mode is insensitive to experimental conditions. The intensity of the peaks below 500 cm$^{-1}$ is rather weak, but it is very sensitive to a stimulus; any perturbation, even sunlight irradiation, changes the spectra of liquid water [95] because of the high sensitivity of the O:H nonbond. Monitoring the cooperative relaxation of high-frequency $\omega_H$ and low-frequency $\omega_L$ phonons would suffice to examine the cooperativity of the O:H-O bond under excitation. The advantage of phonon spectroscopy is the Fourier transformation between the real and the energy spaces. Each spectral feature represents all bonds with the same vibrational attributes, regardless of their number or location in real space.

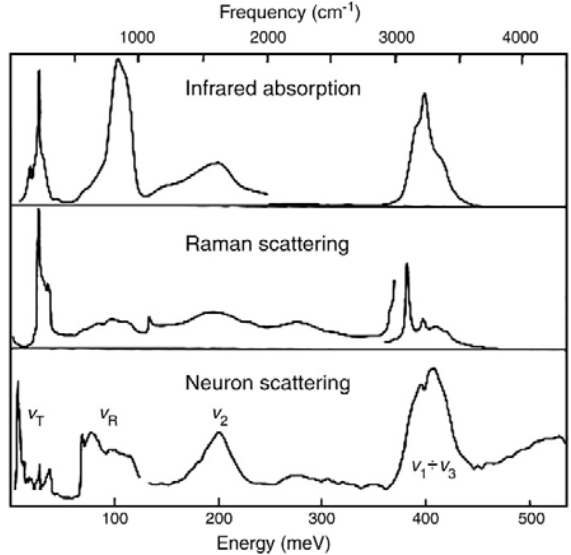

Figure 2. Typical vibrational spectra measured from bulk water at ambient temperature. Peaks at $\omega_L < 300$ cm$^{-1}$ correspond to O:H stretching; the peak at $\omega_{B1} \approx 500$–$700$ cm$^{-1}$ is from $\angle$O:H-O bending; features $\omega_{B2} \approx 1600$–$1750$ cm$^{-1}$ are the $\angle$H-O-H bending libration mode; features centered at $\omega_H \approx 3200$ and $3450$ cm$^{-1}$ are the H-O stretching mode in the bulk and in the skin of water. (Reprinted with permission from [94].)

1.2.4    Molecular images and orbital energies

Figure 3 shows the orbital images and the dI/dV spectra of a H$_2$O monomer and a (H$_2$O)$_4$ tetramer deposited on a NaCl(001) surface probed using scanning tunneling microscopy/spectroscopy (STM/S) at 5 K [96]. The highest occupied molecular orbital (HOMO) below the $E_F$ orbit of the monomer appears as a double-lobe structure with a nodal plane in between, while the lowest unoccupied molecular orbital (LUMO) above $E_F$ appears as an ovate lobe developing between the two HOMO lobes. STS spectra at different heights discriminate the tetramer from the monomer in the density of states (DOS) crossing $E_F$.

These observations [96] confirmed the occurrence of $sp^3$-orbit hybridization of oxygen in H$_2$O monomer at 5 K temperature and the intermolecular interaction involved in (H$_2$O)$_4$. According to the bond-band-barrier correlation notation [97,98], the HOMO located below $E_F$ corresponds to the energy states occupied by electron lone pairs of oxygen, and the LUMO to states yet be occupied by electrons of antibonding dipoles. The image of the monomer showing the directional lone pairs suggests that the lone pairs point into the open end of the surface. As the H$^+$ ion can only share its unpaired electron with oxygen, the Cl$^-$ ion interacts with the H$^+$ ion electrostatistically. The spectral difference between the tetramer and the monomer may indicate intermolecular action.

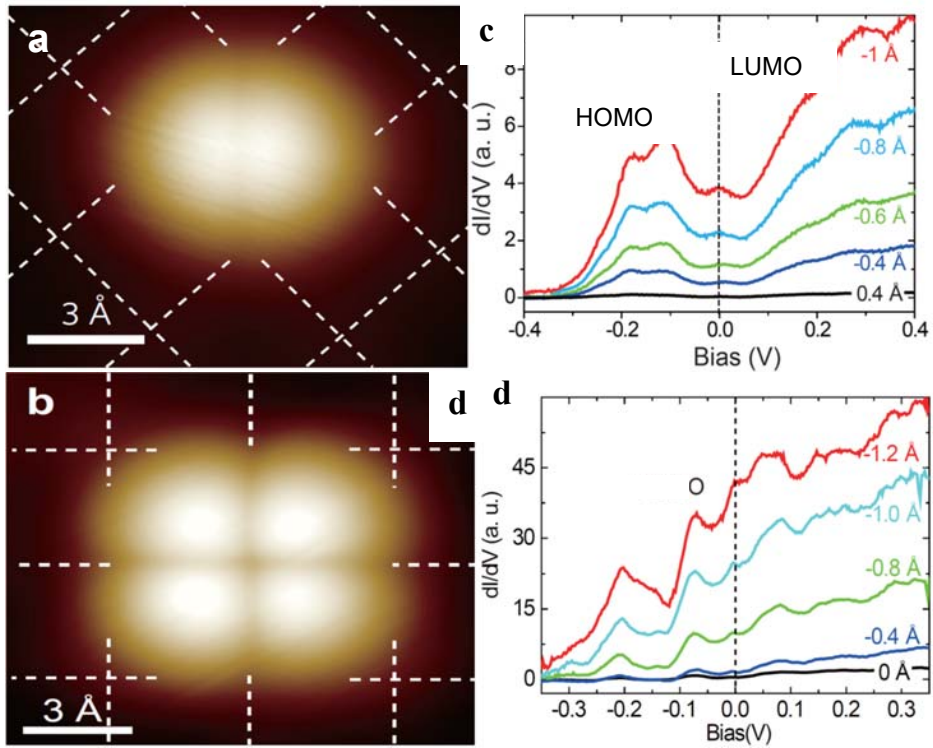

Figure 3. STM images of (a) a $H_2O$ monomer; (b) a $(H_2O)_4$ tetramer; and (c, d) the respective dI/dV spectra obtained under conditions of V = 100 mV, I = 100 pA, and dI/dV collected at 50 pA of different heights at 5 K temperature. Grid in images denotes the $Cl^-$ lattice of the NaCl(001) substrate (Reprinted with permission from [96].) LUMO (> $E_F$) and the HOMO (< $E_F$) indicated in (b) denote the orbital energy states.

1.2.5   Thermodynamic attributes

Figure 4 shows the temperature (θ) dependence of the mass density ρ, specific heat under constant pressure $C_p$, thermal conductivity κ, sound velocity v, and surface tension γ of water. Newton-Laplace equation, v = $(Y/\rho)^{1/2}$, correlates the elastic modulus Y to v and to ρ of a substance. These quantities determine the thermal diffusivity α = κ/[$C_p$ ρ] in fluid thermal transport dynamics [99].

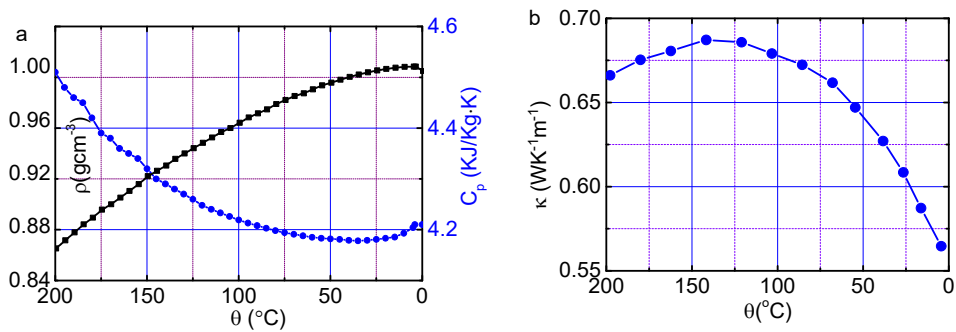

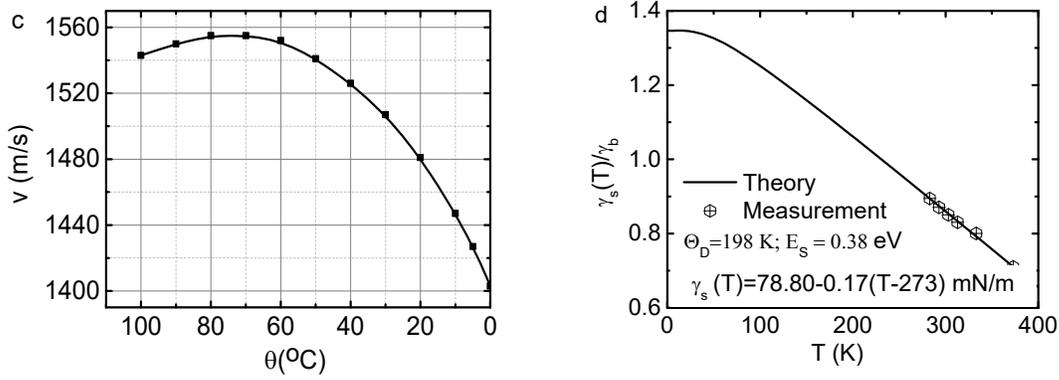

Figure 4. Temperature dependence of (a) the mass density ρ and specific heat under constant pressure $C_p$; (b) thermal conductivity κ; (c) sound velocity v [82,99]; and (d) surface tension $\gamma_s$ [100] of bulk water. Reproduction of (d) the measured temperature-dependent $\gamma_s$ results in an estimation of the Debye temperature $\Theta_{DL}$ = 198 K and molecular cohesive energy $E_S(0)$ = 0.38 eV in the skin, which gives rise to $E_L = E_b(0)/4$ = 0.095 eV/nonbond for the skin O:H nonbond dissociation. (Reprinted with permission from [101].)

1.2.6  Debye temperature and O:H dissociation energy

The following equation formulates the temperature-dependence of the elastic modulus Y and the surface tension, both of which are local energy-density dependent [101],

$$\gamma(T) \propto \frac{E_s(T)}{d^3(T)} = \frac{E_s(0) - \int_0^T \eta(t)dt}{d^3\left(1 + \int_0^T \alpha(t)dt\right)^3},$$

where η is the specific heat in Debye approximation; $\alpha$ is the coefficient of thermal expansion. The temperature-dependent $\gamma_s$ relationship is given by:

$$\frac{\gamma_s(T)}{\gamma_s(0)} \cong \left(1 + \int_0^T \alpha_s(t)dt\right)^{-3} \times \begin{cases} 1 - \dfrac{\int_0^T \eta_s(t/\theta_D)dt}{E_s(0)}, & (T \leq \theta_{DL}) \\ 1 - \dfrac{\eta_s T}{E_s(0)}, & (T > \theta_{DL}) \end{cases}.$$

(1)

Reproduction of the measured temperature-dependence of surface energy [99], see Figure 4d, results in the molecular cohesive energy $E_S(0) = 4E_L$ = 0.38 eV/molecule and the Debye temperature $\Theta_{DL}$ = 198 K [101] with the known $\alpha$ = 0.162 mJ/m²K for bulk water as input for approximation. This value of $\Theta_{DL}$ is compatible with the values of 185 ± 10 K derived from helium scattering from ice in the temperature range 150–191 K [102].

1.3  Challenges and objectives

Unlocking the mysteries of water ice regarding its structure order, O:H-O bond local potentials, and its physical anomalies and their interdependence under various stimuli has long been a challenge. For instance, bond relaxation and the associated bonding charge entrapment and nonbonding electron polarization are beyond the scope of the rigid and non-polarizable models [87,103,104]. The length, energy and charge distribution of a substance must respond, without exception, to applied stimuli such as mechanical compression, molecular undercoordination, thermal, electric, and magnetic excitation, for instance, all of which change the structure and property of a substance [98]. Modeling hypotheses and expectations, numerical calculations and experimental measurements should be consistent and correlated in addressing the property change of the highly correlated and fluctuating water system. Examination of the statistical mean of all the correlated quantities with certain rules and multiple means is more realistic and meaningful than focusing on the instantaneous accuracy of a certain quantity at a point of time for a strongly correlated and fluctuating system. Alternative ways of thinking and new strategies are necessary to deal with these difficult issues effectively.

This work deals with the correlation between the anomalous behavior of water ice and the relaxation dynamics of the O:H-O bond under stimuli of mechanical compression, molecular undercoordination, and thermal excitation. An extension of the Ice Rule towards the primary building block to unify the length scale, structural geometry, and the mass density of water ice and towards the flexible and polarizable O:H-O bond with cooperative, short-range interactions to reconcile physical anomalies is necessary. Lagrangian solution to the O:H-O bond oscillation dynamics ensures mapping of the potential paths for the O:H-O bond when it relaxes. Solving the Fourier thermal fluid transportation equation with adequate initial-and-boundary conditions would clarify the historical mystery of heating "emission-conduction-dissipation" in the Mpemba paradox. Interplay of density-functional theory (DFT) and molecular dynamics (MD) calculations, Raman and IR spectroscopy, XPS measurements has enabled clarification, correlation, formulation, and quantification of multiple puzzles demonstrated by water ice and thus verified our hypothesis and expectations consistently.

2     Principle: O:H-O bond cooperativity

- *An extended tetrahedron unifies the length scale, geometric configuration, and mass density of molecular packing in water ice.*
- *The flexible and polarizable O:H-O bond performs as an asymmetrical, H-bridged oscillator pair coupled by inter-electron pair Coulomb repulsion.*
- *The strength and specific heat disparity of the O:H-O bond and the Coulomb repulsion discriminate water ice from other 'normal' substance in physical properties.*
- *When stimulated, O:H and H-O relax in the same direction but by different amounts; the one becomes shorter will be stiffer, and vice versa.*

2.1     O:H-O bond: Asymmetrically coupled oscillator pair
2.1.1   Extension of the Ice Rule: Frustration less

An oxygen atoms prefers $sp^3$-orbital hybridization when it reacts with atoms of relatively lower electronegativity, irrespective of the structural phase [97,105]. As shown in Figure 5a, an oxygen atom ($2s^2 2p^4$) catches one electron from each of its two neighbors such as hydrogen (H) or metal atoms and then hybridizes its $sp$ orbits, creating four directional orbits [97]. In the case of $H_2O$, one O forms two intra-molecular H-O polar-covalent bonds with shared electron pairs of 4.0–5.1 eV binding energy [3,106]. Two electron lone pairs (':') is produced upon O-H bond being formed with $sp^3$-orbital hybridization. The $O^{2-}$ then fills up the remaining two orbits with its nonbonding lone pairs to form the intermolecular O:H nonbond. The O:H nonbond energy is normally lower than 0.1 eV [101]. The anisotropy of the distribution of charge and energy surrounding the central oxygen atom permits a $H_2O$ molecule only $C_{v2}$ group symmetry besides its rapid rotation and vibration in the network.

Therefore, an oxygen atom always tends to find four neighbors to form a stable tetrahedron; but the nonequivalent bond angles ($\angle$H-O-H $\leq$ 104.5° and $\angle$H:O:H $\geq$ 109.5°) and the repulsion between the electron pairs on adjacent oxygen atoms [2,3] prevent the tetrahedron from being stable in the liquid phase. This anisotropy explains why water remains liquid at temperatures above the critical temperature of other liquids such as nitrogen at 77 K and below. The strong fluctuation is analogous to the motion of a complex pendulum surrounded by four nonbonding interactions. The O:H nonbond switches on and off ceaselessly with a sub-picosecond period [90,107-109] or with a frequency of THz ($\nu = \omega_L/2\pi \sim 30$ s$^{-1}$). Therefore, it is more realistic and useful to consider the statistical mean of the structure order, length scale and mass density in a phase of question over a long time span rather than attempt to capture a snapshot of a quantity with instantaneous accuracy [107].

The packing order of H$_2$O molecules follows the Ice Rule [88,110] in all phases except, for instance, for water subjected to ultra-high temperature and high pressure [44]. Despite thermal fluctuation in the O:H length and in the $\angle$O:H-O angle, the average molecular separation and molecular size changes when the H$_2$O transits from the strongly ordered solid phase to the weakly ordered liquid and the disordered amorphous or vapor state, as the Ice Rule implicitly maintains.

Encouragingly, an extension of the Ice Rule simplifies the investigation substantially. This extension results in an ideal tetrahedron with higher C$_{3v}$ group symmetry and a flexible polarizable O:H-O bond, as shown in Figure 5b. Containing two equivalent H$_2$O molecules and four identical O:H-O bonds at different orientations, this symmetrical tetrahedron unifies the length scale, geometrical configuration and mass density in building up the bulk water and ice. Relaxation of the segmented O:H-O bond and the associated electron entrapment and polarization govern the properties and performance of water and ice.

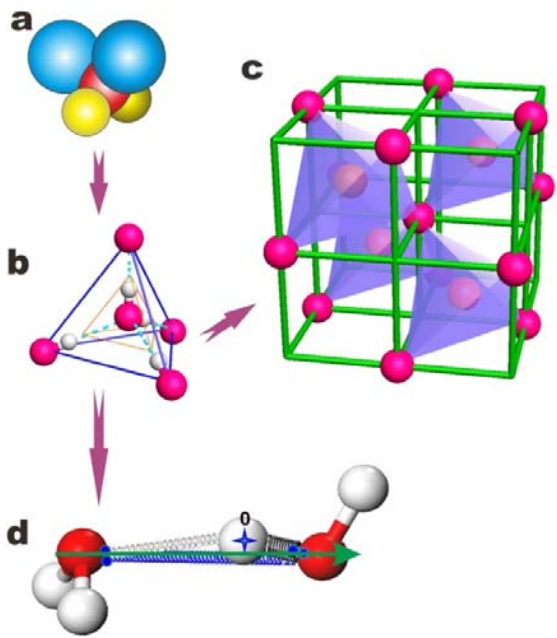

Figure 5. (a) An oxygen atom forms a quasi-tetrahedron with its neighboring H atoms through two bonding pairs (yellow) and two nonbonding lone pairs (blue) upon *sp$^3$*-orbit hybridization [97] which occurs in the temperature range from 5 K to the gaseous phase [96]. An extension of this quasi-tetrahedron yields (b) an ideal $C_{3v}$-symmetrical tetrahedron containing two $H_2O$ molecules and four O:H-O bonds. Assembling the tetrahedral blocks around the central oxygen atom in (c) yields a diamond structure that correlates the size, separation and mass density of molecular packing in water and in ice [1]. (d) The flexible and polarizable O:H-O bond forms a pair of asymmetrical, coupled, H-bridged oscillators whose relaxation in length and energy, and the associated electron entrapment and polarization, mysterizes water ice. (Reprinted with permission from [2].)

Figure 5 shows the sampling procedure in extending the Ice Rule into the basic building block and the O:H-O bond [2]. The central tetrahedron in Figure 5b represents the Ice Rule [88,110]. In hexagonal or cubic ice, $O^{2\delta-}$ ions form each a tetrahedron with an O-O bond 2.76 Å long and an H-O bond 0.96 Å long in ice-VIII phase. Every $O^{2\delta-}$ ion is surrounded by four $H^{\delta+}$ ions ($\delta \leq 1$, but to simplify discussion we consider only $\delta = 1$ from now on). One $H^+$ binds to one $O^{2-}$ through a polar-covalent bond and connects to the other $O^{2-}$ through a nonbonding lone pair ':'.

According to Pauling [88], the minimum energy position of a proton in the O:H-O bond is not half-way between two adjacent oxygen ions. There are two equivalent positions that an $H^+$ proton may occupy on the O-O bond with the same probability, a far- and a near position. Thus, the rule leads to the 'frustration' of the proton positions for the ground state configuration: for each oxygen ion, two of the neighboring protons must reside in the far positions and two of them in the near, the so-called 'two-in/two-out' frustration. The open tetrahedral structure of ice affords many equivalent states, including spin glasses that satisfy the Ice Rule. However, as it is presently understood, the $H^+$ proton undergoes only rotational and vibrational fluctuations because of the stronger H-O bond and the weaker, fluctuating O:H interaction.

2.1.2 O:H-O bond segmentation: Correlation and fluctuation

The building block in Figure 5b results in two entities. One is the geometrical structure in Figure 5c that water and ice prefer; the other is the coupled O:H-O oscillators in Figure 5d with asymmetrical and ultra-short-range interactions, analogous to springs. The $H^+$ proton at the coordinate origin donates its electron to the O shown on the right, to form the intramolecular H-O polar-covalent bond, whereas the electron lone pair ':' of the O shown on the left (blue pairing dots) polarizes the shared electron pair '–' and attracts the $H^+$ proton to form an intermolecular O:H nonbond without sharing any charge. Relaxation of either the H-O or the O:H component plays a role in mediating separately the detectable properties of water and ice. Table 1 specifies the O:H-O bond identities compared to the C-C bond in a diamond.

Table 1. Identities of the segmented O:H-O bond compared to the C-C bond in a diamond [5]. The H-O cohesive energy $E_H$ determines H-O atomic dissociation, O 1s energy shift, H-O phonon frequency $\omega_H$, and the $T_C$ for phase transition, except for evaporation. O:H cohesive energy $E_L$ determines molecular dissociation, dipole moment P, elastic modulus Y, O:H phonon frequency $\omega_L$ etc.

|        | Length, $d_x$ (Å) | Energy, $E_x$ (eV) | Phonons, $\omega_x$ (cm$^{-1}$) | $\Theta_D$ (K) | $T_m$ (K) | Interaction type | Properties |
|--------|---|---|---|---|---|---|---|
| H-O(H) | ≈ 1.00 | ≈ 3.97–5.10 | > 3000 | >3000 | ~5000 | Exchange | $T_C$, $O_{1s}$, $\omega_H$, $E_H$, etc. |
| O:H(L) | ≈ 1.70 | ≈ 0.05–0.10 | < 300 | 198 | 273 | vdW-like | P, Y, $\omega_L$, $E_L$, etc. |

| | | | | | | Coulomb | ρ |
|---|---|---|---|---|---|---|---|
| O-O | – | – | – | – | | Coulomb | ρ |
| C-C | 1.54 | 1.84 | 1331 | 2230 | 3800 | Exchange | |

### 2.1.3 Electron localization and dual polarization

Figure 6(a) shows the density functional theory (DFT)-derived trajectory of the strongly localized bonding and nonbonding electron pairs (in red) of a unit cell with gridded reference. As expected, both bonding and nonbonding electron pairs are strongly localized at sites close to oxygen ions (in blue). The localization of the electron pairs lays the foundation for Coulomb repulsion between adjacent oxygen ions introduced in this present way. Such often-overlooked repulsion dictates the unusual performance of water ice.

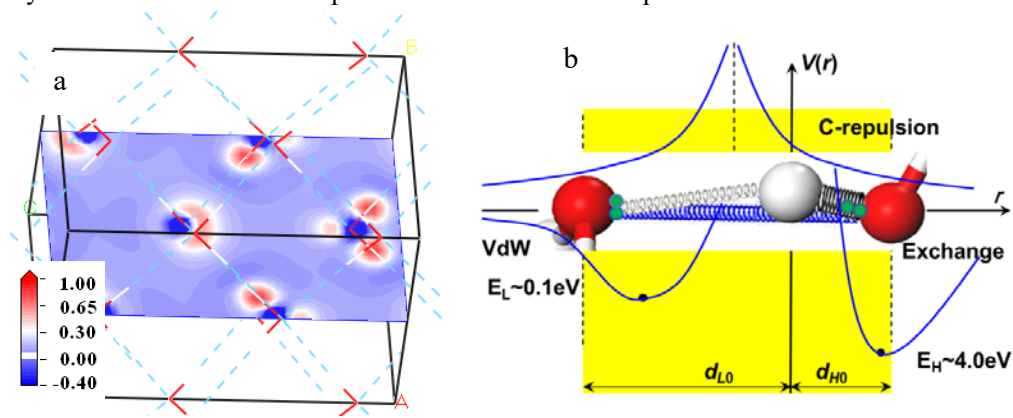

Figure 6. (a) DFT-derived residual charge density of an ice-VIII unit cell. The residual charge density is the difference between the charge of an $H_2O$ molecule and that of an isolated O atom. The positive regions (red) correspond to the gain of electrons. The negative regions (blue) correspond to the charge loss. (b) Asymmetrical, short-range interactions for the segmented O:H-O bond include the O:H van der Waals (vdW)-like nonbond interaction (left-hand side), the H-O bond exchange (right-hand side) interaction, and the Coulomb repulsion between electron pairs on adjacent $O^{2-}$ ions. One switches off a particular potential and the other on at the boundary, or at the atomic site, when it moves across the boundary. (Reprinted with permission from [1].)

The nonbonding lone pairs are associated with the process of tetrahedral-coordinated bond formation of O, N or F atoms in reactions [97,111]. Replacing the $O^{2-}$ with $N^{3-}$ or $F^-$ and replacing the $H^{+/p}$ with any element $M^{+/p}$ of lower electronegativity, an O:H-O -like bond forms. The strength of the nonbonding lone pair also varies with the local environment [112]. The O:H nonbond breaks at evaporation point of water [113]. For comparison, an O:Cu nonbond breaks at a temperature around 700 K by *sp*-orbit dehybridization, as detected by very-low-energy electron diffraction (VLEED) [114].

Weak O:H nonbonds are, to quote Hoffmann [115], "… ubiquitous, and their prevalence gives them a power that belies their modest nature. In water, they influence the global geology and climate of the Earth. In biomolecules, they regulate the folding, signaling and messaging of proteins, and hold together the DNA's double helix". Because of its relatively weak interaction, O:H binding energy $E_L$ contributes insignificantly to the Hamiltonian and associated properties [116]; however, these electrons add impurity states in the vicinity of Fermi energy ($E_F$), and neither follow usually observed dispersion relations nor occupy the allowed states of the valence band and below, and are located right within the energy scope of STM/S surrounding $E_F$ [116].

In addition to the weak interactions with energies around 50 meV, as detected using Raman and electron energy loss spectroscopy (EELS) [97], these lone pairs polarize their neighboring atoms into dipoles. The lone pair and dipole interactions not only act as the most important function groups in biological and organic molecules, but they also play important roles such as high-$T_C$ superconductivity and topological insulating in inorganic compounds. An ultraviolet light irradiation or thermal excitation can dehybridize the $sp^3$-orbit, annihilating the lone pairs and dipoles and altering their functionalities.

The relaxation of the O:H nonbond contributes to volume change, polarization, molecular dissociation, low-load elasticity, friction reduction and associated $\omega_L$ phonon relaxation dynamics. To get rid of one molecule from the bulk, it is necessary to break four intermolecular O:H nonbonds.

There are dual processes of nonbonding electron polarization pertaining to undercoordinated water molecules. As will be discussed shortly, molecular underccordination shortens and stiffens the covalent bond, which densifies and entraps the core and bonding electrons. These densely entrapped electrons of an oxygen atom polarize the lone pairs of its own. On the other hand, the polarised lone pairs on the adjacent oxygen atoms polarize and repel one another, resulting in a second round of polarization. This explains why the surface of water ice is so strongly polarized, and why it is elastic, hydrophobic, viscostic and slippery, in the case of ice.

2.1.4    Asymmetric short-range potentials

The O:H-O bond contains both the O:H nonbond and the H-O bond rather than either of them alone. Segmentation of the O:H-O bond is necessary into a shorter and stiffer H-O covalent bond with a stronger exchange interaction and a longer and softer O:H nonbond with a weaker nonbond (vdW-like) interaction, as ilusrated in Figure 6b [2,3]. The vdW-like interaction contains electrostatic interaction between the lone-pair and the H$^+$ proton, so the nonbond interaction is slightly stronger than the ideal vdW bond that denotes purely dipole–dipole interaction. The H$^+$ proton always remains closer to the O (right-hand side of Figure 6b) without any frustration and keeps away from the other O atom because of the much stronger H-O exchange interaction than the weaker O:H nonbond. The O:H-O bond links the O-O in both the solid and liquid H$_2$O phase, regardless of phase structures [1].

The following functions respectively describe the short-range interactions in an O:H-O bond [6,117,118]:

$$\begin{cases} V_L(r_L) = V_{L0}\left[\left(\dfrac{d_{L0}}{r_L}\right)^{12} - 2\left(\dfrac{d_{L0}}{r_L}\right)^6\right] & \text{(O:H vdW-like L-J potential } (V_{L0}, d_{L0})) \\[2pt] V_H(r_H) = V_{H0}\left[e^{-2\alpha(r_H - d_{H0})} - 2e^{-\alpha(r_H - d_{H0})}\right] & \text{(H-O Morse interaction } (\alpha, V_{H0}, d_{H0})) \\[2pt] V_C(r_C) = \dfrac{q_O^2}{4\pi\varepsilon_r\varepsilon_0 r_C} & \text{(O---O Coulomb repulsion } (q_O, \varepsilon_r)) \end{cases}$$

(2)

where $V_{L0}$ ($E_{L0}$) is the potential well depth for the O:H nonbond; $V_{H0}$ ($E_{H0}$) is the H-O bond energy; $r_x$ (x = L, H) and $r_C = r_L + r_H$ are the interatomic distance (the lengths of springs); and $d_{x0}$ is the length at equilibrium. The parameter $\alpha$ determines the width of the Morse potential; $q_O$ denotes the net charge on an O$^{2-}$; $\varepsilon_r$ = 3.2 is the relative dielectric constant of ice, which is subject to change with external excitation, including coordination condition; and $\varepsilon_0$ = 8.85×10$^{-12}$ F/m is the dielectric constant of the vacuum.

Having the smallest number of adjustable parameters, the Morse potential suffices for the exchange interaction. The Lennard-Jones (L-J) potential approximates the O:H nonbond interaction since no charge is shared within the O:H segment. It is difficult to say whether one potential is better than the other when considering O:H and H-O interactions; however, the equilibrium coordinates of the potentials as to the bond length and bond energy are of greatest concern, disregarding the shape of a potential curve.

Because of the short-range nature of the interactions, the solid lines in Figure 6b are valid only within the shaded range for the basic O:H–O unit. A particular potential must be switched off and the other switched on immediately when moving to the boundary of the region, or to any atomic site. No spatial decay of any potential exists, irrespective regime.

2.1.5 Forces driving O:H-O bond relaxation

Averaging the background long-range interactions due to other $H_2O$ molecules or protons [119] and omitting the nucleus quantum effect on fluctuations [120], the forces acting on the electron pairs of oxygen ions are illustrated in Figure 7:

1) Coulomb repulsion between electron pairs on adjacent $O^{2-}$ ions is the first-order differentiation of the Coulomb potential, $f_q = -\partial V_C(r)/\partial r$, out of equlibrium. Replacing one $O^{2-}$ ion with an ion of acid, salt, sugar, protein, or a biomolecule or a cell, mediates the Coulomb repulsion $f_q$ by varying the ionic size and charge quantity. This replacement may mediate the O:H dissociation energy and the functionalities of the O:H-O bond and the solubility of water solution [121].
2) Pointing either towards or away from the coordinate origin, the force dislocates oxygen atoms $f_{dx}$. An applied stimulus (e.g. mechanical compression, molecular undercoordination, thermal excitation, chemcal reaction, etc.) provides this driving force.
3) The force of deformation recovery, $f_{rx} = -\partial V_x(r)/\partial r$, approximates to the first-order differentiation of the respective $V_x(r)$ at equilibrium. The $f_{rx}$ always points opposite to the direction of deformation.

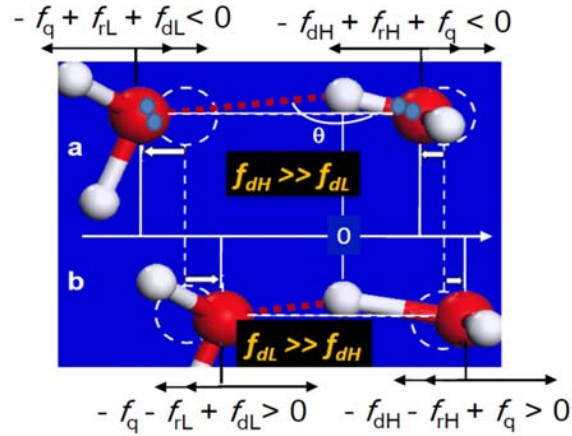

Figure 7. Forces and relaxation dynamics of the segmented O:H-O bond with $H^+$ as the coordinate origin. Forces include the Coulomb repulsion $f_q$, deformation recovery $f_{rx}$, and the external force driving relaxation $f_{dx}$. Coulomb repulsion and external stimulus dislocate O atoms in the same direction but by different amounts. The O:H always relaxes more than H-O. (a) Relaxation corresponds to situations of molecular undercoordination [3] and cooling in the liquid–solid transition phase [5]. (b) Relaxation corresponds to compression [2] and cooling in the liquid and solid phases [5]. Relaxation in the specified direction takes place when the $f_{dX}$ competition meets the condition $f_{dH} \gg f_{dL}$ or $f_{dH} \ll f_{dL}$.

The following relationships define the 'master' segment dominating the O:H-O relaxation and the O-O length gain or loss under applied stimulus (see Figure 7):

a) (freezing, molecular undercoordination)  b) (compression, liquid and solid cooling)

$$\begin{cases} -f_q + f_{rL} + f_{dL} < 0 & (1) \\ -f_{dH} + f_{rH} + f_q < 0 & (2) \end{cases} \qquad \begin{cases} -f_q - f_{rL} + f_{dL} > 0 & (3) \\ -f_{dH} - f_{rH} + f_q > 0 & (4) \end{cases}$$

or (1)+(2),                                or (3)+(4),

$$f_{rL} + f_{dL} - f_{dH} + f_{rH} < 0 \qquad\qquad -f_{rL} + f_{dL} - f_{dH} - f_{rH} > 0$$

or,                                        or,

$$f_{dH} > (f_{rL} + f_{dL} + f_{rH}) \qquad\qquad f_{dL} > (f_{rL} + f_{dL} + f_{rH})$$

$$\left.\begin{array}{c} f_{dH} > (f_{dL} + f_{rL} + f_{rH}) \\ f_{dL} > (f_{dH} + f_{rL} + f_{rH}) \\ f_{dH} = f_{dL} \end{array}\right\} \Rightarrow \Delta d_{O-O} \begin{cases} > \\ < \\ = \end{cases} 0$$

(3)

Segments of the O:H-O bond relax cooperatively because of the joint effect of Coulomb coupling and applied stimulus. One segment dominates the relaxation and the other follows under a certain stimulus. The segment that drives relaxation is assigned to be the 'master', and the other as the 'slave'. When stimulated, the master segment relaxes and pushes or pulls the electron pair of the slave $O^{2-}$ through the repulsion and the deformation recovery. Meanwhile, the repulsion widens the ∠O:H-O angle θ and polarizes the electron pairs during relaxation.

The H-O bond serves as the master to drive the relaxation, if the driving forces meet the criterion $f_{dH} \gg f_{dL}$. In this situation, the master H-O relaxes less than the slave O:H, resulting in a net O-O length gain or loss and an accompanying volume variation. If $f_{dL} \gg f_{dH}$, the master and the slave change roles and the situation reverses. At $f_{dH} = f_{dL}$, there is a transition between O-O expansion and contraction, corresponding to density extremes.

Inter-oxygen repulsion has an influence in not only O:H-O relaxation but also in the bonding dynamics of oxygen chemisorption. An STM/VLEED study of oxygen chemisorption onto Cu(001) surface revealed that the $O^{2-}$–$Cu^+$ bond and the $O^{2-}$:$Cu^p$ nonbond relax cooperatively and oppositely in lengths. The $O^{2-}$–$Cu^+$ bond contracts to 0.163 nm while the $O^{2-}$:$Cu^p$ nonbond expands to 0.195 nm in the $Cu^p$:$O^{2-}$–$Cu^+$ configuration, with the creation of the $Cu^p$ dipoles and the missing Cu atoms [97].

Three variables describe the O:H-O bond relaxation dynamics. They are the ∠O:H-O angle θ and the segmental lengths $d_x$ ($x$ = L and H). Relaxation of the angle θ contributes little to the physical properties, except for mass density. Response of $d_x$ and its bond energy $E_x$ to a stimulus changes physical properties such as phonon relaxation, density variation, $T_C$ change, O 1s energy shift, viscoelasticity, chemical reactivity, hydrophobicity, etc.

2.2    O:H-O bond segmental disparity
2.2.1  Mechanical strength disparity

Letting the compression force be $f_{dx} \approx P/s_x$, where $s_x$ is the cross-sectional area of the $x$ segment and $P$ the pressure, the mechanical disparity of the O:H-O bond may be derived. At quasi-equilibrium (compression shortens the O:H length so $f_{dL} - f_{dH} > 0$, Eq. (3b)):

$$f_{dL} - f_{dH} = P(1/s_L - 1/s_H) = (f_{rL} + f_{rH}) > 0$$
$$\text{or}$$
$$s_H - s_L > s_L s_H > 0$$

(4)

This relationship indicates that the effective cross-sectional area of the H-O bond $s_H$ is greater than that of the O:H, which explains why the O:H 'masters' the relaxation dynamics of water ice under compression and why the O:H nonbond always relaxes more than the H-O bond [2]. Hence, compression shortens and stiffens the O:H nonbond and spontaneously lengthens and softens the H-O bond; negative compression (tension) will effect reversely [2].

### 2.2.2 Undercoordination-discriminated O:H-O relaxation

According to the bond order-length-strength correlation and nonbonding electron polarization (BOLS-NEP) notation, see Eq. (5) [122], atomic undercoordination shortens and stiffens the remaining bonds between undercoordinated atoms spontaneously with an association of local densification and quantum entrapment of the bonding and core electrons. This process occurs regardless of the nature of the bond or the structure phase. Furthermore, the locally and densely entrapped bonding charge in turn polarizes the nonbonding electrons pertained to lone pairs, dangling bonds, and conduction electrons at the upper edge of the conduction band [116]. BOLS-NEP is responsible not only for the size dependence of the known bulk properties but also for the emergence of anomalies of materials at the nanometer scale. Size emergence includes catalytic enhancement, toxicity and dilute magnetism of the noble metals [123] and ZnO [124]. The polarization creates Dirac-Fermi polarons at graphene zigzag edges and graphite point defects [125], serving as carriers for topological insulators.

The following formulates the BOLS notation in terms of bond length ($d$) contraction with a coefficient $C(z)$, bond energy $E_z$ gain, and atomic cohesive energy change $E_{B,z}$, where subscript $z$ denotes an atom with $z$-coordination number (CN), and subscript $b$ denoting fully coordinated in the bulk ($z = 12$ for the fcc structure as a standards). The bond nature index $m$ correlates the bond length and energy, which remains constant for a specific substance:

$$\begin{cases} C(z) = d_z/d_0 = 2/\{1+\exp[(12-z)/(8z)]\} & (BOLS-coefficient) \\ E_z = C_z^{-m} E_b & (Single-bond-energy) \\ E_{B,z} = zE_z & (Atomic-coherency) \end{cases}$$

(5)

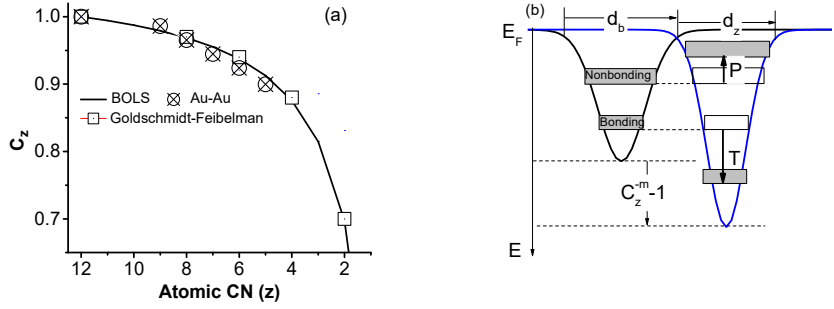

Figure 8. Schematic illustration of the BOLS-NEP notation [116]. (a) Atomic undercoordination shortens and stiffens the local bond ($d_z/d_0 = z < 1$; $E_z/E_b = C_z^{-m} > 1$). Bond contraction raises the local density of the bonding charge and binding energy. Bond stiffening deepens the local potential well and (b) entraps (T) the core and the bonding electrons accordingly. The densely entrapped bonding and core charge in turn polarizes (P) the nonbonding electrons shifting up in energy. The T and P evolution dynamics modulates the Hamiltonian by crystal potential screening and splitting and charge distribution in all bands. Scattered symbols in (a) represent observations by Goldschmidt-Feibelman [126,127] and Huang et al. [128] from gold clusters.

Water molecules with fewer than the ideal four neighbours in the bulk should follow the BOLS-NEP notation. However, the involvement of the lone-pair interaction prevents the O:H and the H-O from following the BOLS-NEP notation simultaneously because four ':' lone pairs screen an $H_2O$ molecule. The binding energy disparity means that the stronger H-O bond serves as the 'master' to contract by a different amount from what the BOLS notation predicts. The contraction of the H-O bond lengthens and softens the 'slave' O:H nonbond by Coulomb repulsion, with the dual process of polarization. The stiffness of the segment is characterized by the respective phonon frequency $\omega_x$ ($x$ = L for the O:H nonbond; $x$ = H for the H-O bond).

It is universally true that one segment of the O:H-O bond will be stiffer if it becomes shorter; it will be softer if it becomes longer [2]. Therefore, the phonon frequency shift $\Delta\omega_x$ tells directly the variation in length, strength and stiffness of the particular segment subjected to an applied stimulus. Because of the Coulomb repulsion, $\omega_L$ and $\omega_H$ shift such that if one becomes stiffer, the other will become softer.

2.2.3 Specific-heat disparity and extreme-density dispersivity

Generally, the specific heat of a substance is regarded as a macroscopic quantity integrated over all bonds of the specimen, which is also the amount of energy required to raise the temperature of the substance by 1 K. However, in dealing with the representative for all bonds of the entire specimen, it is necessary to consider the specific heat per bond that is obtained by dividing the bulk specific heat by the total number of bonds [129].

For a specimen of other usual materials, one bond represents all on average; therefore the thermal response is the same for all the bonds, without any differences in cooling contraction or thermal expansion [130]. For water ice, however, the representative O:H-O bond is composed of two segments with strong disparity in the specific heat of the Debye approximation, $\eta_x$. These two segments response to a thermal excitation differently.

The specific heat is characterized by two parameters. One is the Debye temperature $\Theta_{Dx}$, and the other is the thermal integration of $\eta_x$. It should be noted that $\Theta_{Dx}$ determines the rate at which the specific-heat curve reaches saturation. The specific-heat curve of a segment with a relatively low $\Theta_{Dx}$ value reaches saturation more rapidly than the other segment, since $\Theta_{Dx}$, which is lower than $T_{mx}$, is proportional to the characteristic vibration frequency $\omega_x$ of the respective segment.

Conversely, the integral of the specific-heat curve from 0 K to the melting point $T_{mx}$ determines the cohesive energy per bond $E_x$ [129]. The $T_{mx}$ is the temperature at which the vibration amplitude of an atom or a molecule expands abruptly to more than 3% of its diameter irrespective of the environment or the size of a molecular cluster [131,132].

Thus we have (see Table 1):

$$\begin{cases} \Theta_{DL}/\Theta_{DH} \approx 198/\Theta_{DH} \approx \omega_L/\omega_H \approx 200/3200 \sim 1/16 \\ \left(\int_0^{T_{mH}} \eta_H dt\right) / \left(\int_0^{T_{mL}} \eta_L dt\right) \approx E_H/E_L \approx 4.0/0.1 \sim 40 \end{cases}$$

(6)

Analysis of the temperature-dependence of water surface tension [101] yielded $\Theta_{DL} = 198$ K $< 273$ K ($T_m$) and $E_L = 0.095$ eV. Hence, $\Theta_{DH} \approx 16 \times \Theta_{DL} \approx 3200$ K. The O:H specific heat $\eta_L$ ends at 273 K and the H-O specific heat $\eta_H$ ends at T $\geq$ 3200 K. Reproduction of the compression-induced $T_C$ change for the ice VII–VIII phase transition results in an $E_H$ value of 3.97 eV [3]. That is, the area covered by the $\eta_H$ curve is 40 times greater that covered by the $\eta_L$ curve.

The superposition of these two ($\eta_x$) curves implies that the heat capacity of water ice differs from that of other, 'normal', materials. Such a specific-heat disparity yields four temperature regions with different $\eta_L/\eta_H$ ratios over the full temperature range; see Figure 9. These regions correspond to phases of liquid (I), liquid–solid transition (II), and solid (III, IV). At extremely low temperatures (IV) $\eta_L \approx \eta_H \approx 0$, and little relaxation occurs in the O:H-O bond length except for the angle. The intersecting temperatures correspond to extreme densities at boundaries of liquid-solid transition phase.

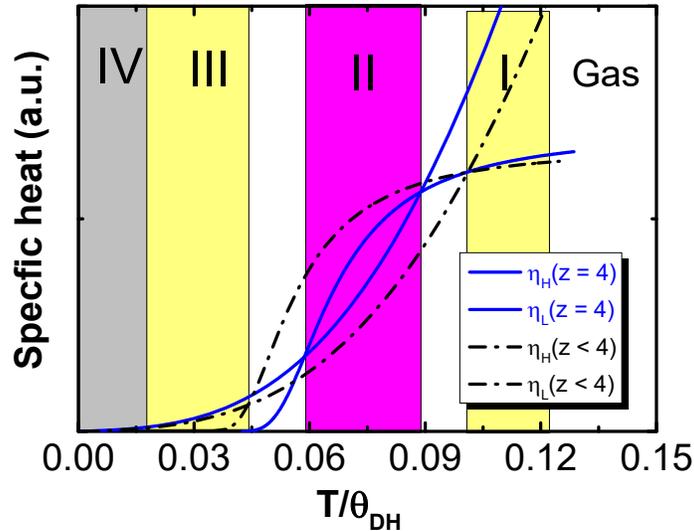

Figure 9. O:H-O bond specific-heat disparity defines four temperature regions of different $\eta_L/\eta_H$ ratios, which correspond to phases of liquid (I: $\eta_L/\eta_H < 1$), liquid–solid transition (II: $\eta_L/\eta_H > 1$), solid (III: $\eta_L/\eta_H < 1$ and IV: $\eta_L \approx \eta_H \approx 0$). Molecular undercoordination ($z < 4$), or ice under negative compression (tension), stiffens

the $\omega_H$ and softens the $\omega_L$ phonon, which modulate the respective $\varTheta_{Dx}$ to disperse the extreme-density (intersecting) temperatures. Therefore, nanodroplet and nanobubble with high-fraction of undercoordinated molecules and ice under negative compression exhibit superheating at melting and supercooling at freezing. However, ice under compression demonstrates inversely. The $\eta_L$ in the solid phase differs from the $\eta_L$ in liquid, which does not influence the validity of the hypothesis. (Reprinted with permission from [5].)

The number of specific-heat regions coincides with that demonstrated by the $\rho(T)$ profiles of water ice [133-135] over the full temperature range. This consistency suggests that the segment of lower specific heat serves as the master while the other part serves as the slave. The master segment is more active than the other when the O:H-O bond is thermally invoked. The thermal expansion or cooling contraction of the master segment drives the asymmetrical relaxation of the entire O:H-O bond. According to the specification, the O:H nonbond serves as the master to contract in the liquid (I) and in the solid (III) phases, and the slave H-O bond expands slightly, leading to the 'normal' process of cooling densification; however, an unusual mechanism governs this event. In the liquid–solid transition phase II, the master-slave roles interchange and the water volume expands, resulting in floating of ice. The intersection points correspond to a maximum density at 4°C and minimum density below the freezing point for bulk water ice [133,136]. At extremely low temperatures, both $\eta_L$ and $\eta_H$ approach zero, which means that there is no cooling contraction in either segment in this region unless the bond angle is relaxed.

The thermodynamic disparity of the O:H-O bond indicates that the H-O bond, rather than the O:H nonbond, dominates the extremely high heat capacity of water [137]. According to the current notation, cooling drives $d_{O-O}$ and the density to oscillate in these four regions:

$$\left.\begin{array}{llll} \text{II} & (\eta_H < \eta_L): & f_{dH} > (f_{dL} + f_{rL} + f_{rH}) \\ \text{I, III} & (\eta_L < \eta_H): & f_{dL} > (f_{dH} + f_{rL} + f_{rH}) \\ \text{IV} & (\eta_L \cong \eta_H \cong 0): & \Delta\theta > 0; \quad \Delta d_x = 0 \\ \text{II boundary} & (\eta_L = \eta_H): & f_{dH} = f_{dL} \end{array}\right\} \Rightarrow \Delta d_{O-O} \begin{Bmatrix} > \\ < \\ = \\ = \end{Bmatrix} 0.$$

(7)

Strikingly, molecular undercoordination or ice under negative compression (tension) stiffens the $\omega_H$ phonon and softens the $\omega_L$ phonon, which modulates the respective $\varTheta_{Dx}$ and disperse the extreme-density temperatures in Figure 9, resulting in supercooling at freezing and superheating at melting. Nanodroplets and nanobubbles of water, which have a higher proportion of undercoordinated skin molecules, should follow this trend. It is true that the least-density temperature drops from 258 to 205 K when the bulk water is a droplet of 1.4 nm size [5,133-135], and that the melting point increases from 273 K at the bulk centre to 310 K at the skin of water [4]. The $T_m$ for a monolayer water film is even higher [138,139]. Compression effects oppositely to molecular undercoordination on the phonon frequencies $\omega_x$ and the $\varTheta_{Dx}$. Facts indicate that ice melts at +6.5°C under -95 MPa negative compression and melts at -22°C under 210 MPa compression [34].

2.2.4   Verification of O:H-O bond cooperative relaxation

The equilibrium condition, $f_{dH} = f_{dL}$, rules the trend of segmental relaxation. Letting $k_x$ be the force constant and $\delta d_x$ the extent of relaxation for the respective segment, the $f_{rx}, k_x,$ and $\delta d_x$ follow the relationship:

$$f_{rH} + f_{rL} = k_H \delta d_H + k_L \delta d_L = 0,$$

which yields,

$$k_L/k_H = -\frac{\delta d_H/\delta t}{\delta d_L/\delta t} = -\frac{\delta^2 d_H/\delta t^2}{\delta^2 d_L/\delta t^2}.$$

(8)

This relationship indicates that the slopes and curvatures of the O:H and O:H relaxation curves are inversely negatively proportional to each other. The variable $t$ represents any stimulus of $T$, $P$ or $N$ for $(H_2O)_N$ clusters, or beyond.

Figure 10 confirms the universality of the O:H-O bond cooperative relaxation under various stimuli obtained from MD calculations using the force field code of Sun [140]. The slopes and curvatures of the $d_x$–$t$ curves are indeed asymmetrical and cooperative, as Eq. (8) predicts. If one segment is shortened, the other in the same panel is lengthened; the two curves in one panel relax in a manner either 'face to face' (in (a)) or 'back to back' (in (b), (c), (d)), due to the curvature correlation. Results indicate that O:H serves as the master and H-O as the slave of water in (a) under mechanical compression and in (c) cooling liquid in the region of T > $T_m$; the H-O bond serves as the master in (b) the liquid–solid transition phase (T < $T_m$) and in (d) upon molecular undercoordination. The arrows next to the 'master' segment point in the direction of density gain ((a), (c)) or loss ((b), (d)).

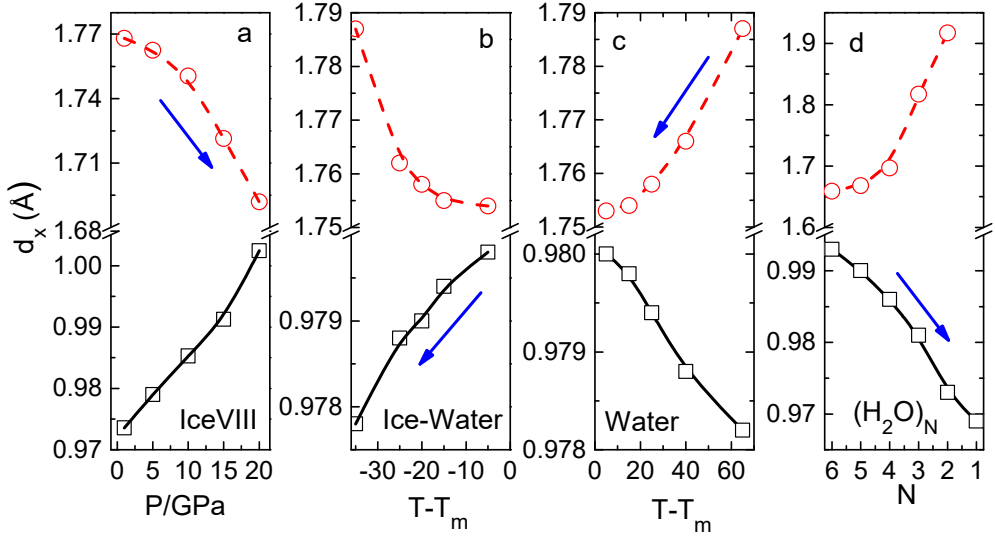

Figure 10. Universality of the O:H-O bond cooperative relaxation. Changes of length of O:H (broken red lines) and H-O (solid black lines) are shown for stimulation by: (a) mechanical compression; (b) cooling excitation in the liquid–solid transition phase; (c) cooling in the liquid phase; and (d) cluster size reduction (molecular undercoordination). The arrows indicate the master segments and their relaxation directions under excitation. If either of the O:H or the H-O shrinks, the other one expands, regardless of the applied stimulus or structural phases, because of the inter-oxygen Coulomb coupling. The arrows in (a) and (c) show the direction of density gain as $d_{O-O}$ shortens; the arrows in (b) and (d) show the direction of density loss. All processes are reversible.

The O-O distance dominates the mass density of water ice in the manner of $\rho \propto (d_{O-O})^{-3} \propto (d_H + d_L)^{-3}$. The $d_x$ is the projection along O-O without contrbution from the ∠O:H-O angle relaxation, which remains > 160° in all phases. The angle difference between 160° and 180° deviates by only 3% or less to the length scale [5]. When the structures are different, there are other possible volume changes. For example, ice VII has a smaller volume and longer intermolecular distance than ice Ic because the former has double the network of the latter. Ice VII and VIII have similar network connectivities but different crystal symmetries [141]. The transition between these two phases is of the first order [142,143]. Volume change by such structure variation is assumed to be insignificant in the present discussion, as we are focused on O:H-O bond relaxation that dictates the anomalous behavior of water ice.

The O-O distance is often measured by neutron or X-ray diffraction. Accurate measurement of the H-O or O:H length relaxation dynamics is not frequently done. It will be shown later that, by knowing the density change, these quantities are readily derived from the tetrahedrally coordinated water structure.

3    Analysis strategies, properties versus bonding identities

- *Correlation, detection, formulation and quantification of bond-phonon-electron-property relaxation dynamics are necessary.*
- *Interplay of quantum computation, phonon and electron spectroscopy, Lagrangian transformation and Fourier thermal fluid transport dynamics provides comprehensive information.*
- *Macroscopic properties are interdependent; segmental relaxation in length and stiffness and the associated electron entrapment and polarization mediate the properties of water ice.*

3.1   Quantum computations

Density functional theory and molecular dynamics (DFT-MD) computations are the main means aiding verification of our hypotheses and expectations. Relaxation of the following quantities under stimulation is the major concern in exercises:

1) Geometrical optimization of molecular clusters
2) Cooperative $d_x$ relaxation and mass density change
3) The power spectra for $\omega_x$ for phonon frequency relaxation
4) Electron and phonon DOS distribution and skin charge accumulation
5) Electron binding energy $E_{Bind}$ and segmental bond energy $E_x$, etc.

Having taken the flexibility, polarizability, and nuclear quantum effects of water into consideration, the *ab initio* optimized molecular dynamics (MD) force field, the Condensed-phase Optimized Molecular Potentials for Atomistic Simulation Studies (COMPASS) 27, [140] is an elegant approach. This package derives the phonon (power) spectra in terms of Fourier transformation of molecular velocity autocorrelation function, Cor(*t*), $I(\omega) = 2\int_0^\infty \text{Cor}(v(t))\cos \omega t \, dt$ [144]. A 360-molecule supercell of proton-disordered ice Ih was simulated (in which the proton was sufficiently optimised to avoid a net dipole moment or to minimise a net quadruple moment [145]). The unit cell was relaxed in the isothermal–isobaric ensemble (NPT) at atmospheric pressure for different temperatures. Andersen's thermostat and barostat approach maintained the temperature and pressure of the closed system [146]. The relaxation time is extended to 120 ps, in order to ensure the stability of the single-phase system in terms of temperature, density and energy. A 15 nm vacuum slab was inserted into the supercell, shown in Figure 11, to represent the skin effect, which was relaxed in the canonical ensemble (NVT) at 200 K for 100 ps to obtain equilibrium in a 0.5 fs time interval. The Nosé-Hoover thermostat algorithm with a Q ratio of 0.01 was adopted to control the temperature. The ice structure was relaxed in the NPT ensemble for 30 ps in 0.5 fs steps to converge for T, P, and energy.

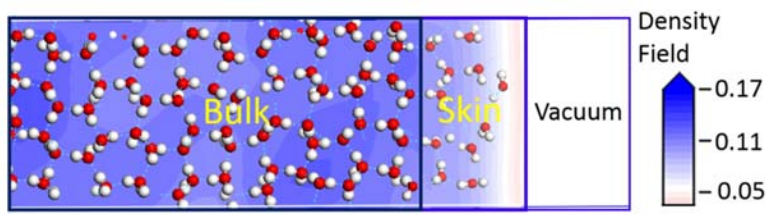

Figure 11. Schematic illustration of the water supercell with an insertion of a vacuum slab representing the supersolid skin of ice at 200 K. This comprised three regions, l. to r.: the bulk, the skin, and the vacuum. The skin contains undercoordinated molecules and free H-O radicals. The colors along the horizontal axis indicate the MD-derived mass density field in the unit cell. This unit cell also applies to the shell of a nanobubble (reprinted with permission from [4]).

The structural optimizations and the bond angle-length-stiffness relaxation of $(H_2O)_N$ clusters and ice skin were conducted using the DFT Dmol$^3$ code of Perdew and Wang (PW) [147] and the dispersion-corrected PW code of Ortmann, Bechstedt and Schmidt (OBS) [148]. The latter includes the vdW hydrogen bond interactions. The all-electron method was used to approximate the wave functions with a double numeric and polarization basis sets. The self-consistency threshold of total energy was set at $10^{-6}$ Hartree. In the structural optimization, the tolerance limits for the energy, force and displacement were set at $10^{-5}$ Hartree, 0.002 Hartree/Å and 0.005 Å, respectively. Harmonic vibrational frequencies were computed by diagonalizing the mass-weighted Hessian matrix [149].

CASTEP code [150] was used within the Perdew-Burke-Ernzerhof (PBE) [151] functional parameterization to calculate the compressed ice-VIII. The norm-conserving pseudopotential (NCP) scheme was adopted, in which the $1s^1$ and the $2s^22p^4$ are treated as valence electrons for H and O atoms, respectively. The use of the plane-wave kinetic energy cutoff of 500 eV converged in total energies. Ice-VIII consisting of two interpenetrating cubic ice lattices with eight molecules in each unit cell, was examined. The MD calculations were performed to examine the evolution of the O-H and O:H distances in a $2 \times 2$ supercell of ice-VIII unit containing 32 molecules, subjected to a pressure changing from 1 to 20 GPa. The structure was dynamically relaxed in the isoenthalpic–isobaric ensemble for 30 ps and showed sufficiently stable convergence [152]. The average O-H and O:H lengths were taken over the final 10 ps (20,000 steps).

3.2 Phonon and electron spectroscopy

The literature provides the basic database of electron and phonon spectroscopy under sorted conditions. Further Raman and FTIR measurements probed phonon spectra over the full frequency range of water during cooling, compressing, heating, and salting. A combination of these two database sets was sufficient to verify predictions of the O:H-O bond cooperative relaxation and associated phonon attributes and electron entrapment and polarization under various stimuli.

Particularly, a spectral differential strategy [153] enabled the distillation of the phonon spectral features due to the conditioning of skin and defects, heating, salting, etc., by differencing the spectra collected before and after conditioning, or collected at different reflection angles from the same specimen upon the spectral peak areas being normalized and background-corrected. This strategy also applies to calculations to discriminate the characteristic $d_x$ or $\omega_x$ peaks obtained under different conditions. For instance, it can distil the bond length and vibration frequency features for molecules with lowest CN from that of the bulk molecules with highest CN [78].

## 3.3 Skin viscosity and surface tension

Calculations using the following method quantify the tension and the viscosity of water skin. The difference between the stress components in directions parallel and perpendicular to the interface defines the surface tension $\gamma$ [154,155]:

$$\gamma = \frac{1}{2}\left(\frac{\sigma_{xx}+\sigma_{yy}}{2}-\sigma_{zz}\right)\cdot L_z ,$$

(9)

where $\sigma_{xx}$, $\sigma_{yy}$, and $\sigma_{zz}$ are the stress tensor elements and $L_z$ is the length of the supercell. The surface shear viscosity $\eta_s$ is correlated to the bulk stress $\sigma$ as [156,157]:

$$\eta_s = \frac{V}{kT}\int_0^\infty \langle \sigma_{\alpha\beta}(0)\sigma_{\alpha\beta}(t)\rangle dt ,$$

(10)

where $\sigma_{\alpha\beta}$ denotes the three equivalent off-diagonal elements of the stress tensors. The bulk viscosity $\eta_v$ depends on the decay of fluctuations in the diagonal elements of the stress tensor:

$$\eta_v = \frac{V}{kT}\int_0^\infty \langle \delta\sigma(0)\delta\sigma(t)\rangle dt$$
$$\delta\sigma = \sigma - \langle\sigma\rangle .$$

(11)

Based on these notations, $\gamma$ was first calculated using the MD method to derive the stress tensors. The auto-correlation functions of the stress tensors were also used to calculate $\eta_s$ and $\eta_v$ in accordance with Eqs. (10) and (11).

## 3.4 Potential mapping and thermal transport dynamics

Lagrangian mechanics is an ideal approach for resolving the coupled O:H-O oscillators with short-range interactions. Because ∠O:H–O in water ice is greater than 160° [5], a linear approximation of the O:H-O bond is convenient and acceptable. By averaging the surrounding background long-range interactions of the $H_2O$ molecules and protons, and the nuclear quantum effect on fluctuations [119,120], the short-range interactions dominate the O:H–O bond relaxation [3].

The reduced mass of the $H_2O$:$H_2O$ oscillator is $m_L = 18 \times 18/(18+18)m_0 = 9m_0$. For the H-O oscillator, it is $m_H = 1 \times 16/(1+16)m_0 = 16/17 m_0$, where $m_0$ is the unit mass of $1.66 \times 10^{-27}$ kg. The motion of the coupled O:H-O oscillators follows the Lagrangian equation [158]:

$$\frac{d}{dt}\left(\frac{\partial L}{\partial(dq_x/dt)}\right) - \frac{\partial L}{\partial q_x} = Q_x ,$$

(12)

where $L = T - V$, in which $T$ is the total kinetic energy and $V$ is the total potential energy. $Q_x$ is the generalized non-conservative forces, here being the compressive force $f_P$. The time-dependent $q_x(t)$ here is $u_L$ and $u_H$, representing the generalized variables of displacements of O atoms from their equilibrium positions in the $L$ and H springs. The kinetic energy T consists of two terms:

$$T = \frac{1}{2}\left[m_L\left(\frac{du_L}{dt}\right)^2 + m_H\left(\frac{du_H}{dt}\right)^2\right]$$

(13)

The potential energy V contains three terms (Eq. (2)) that approximate the coupled harmonic oscillators:

$$2V = k_L u_L^2 + k_H u_H^2 + k_C(u_L - u_C)^2 + 0(u_L^3 + ...),$$

(14)

where $k_x$ is the force constant; $u_x$ is the amplitude of vibration of the O:H nonbond ($x = L$) and the H-O bond ($x = H$); and $k_C$ is the second derivative of the Coulomb potential.

The Fourier thermal–fluid transport equation [159] with appropriate initial- and boundary conditions best describes the process of thermal–fluid transportation in the liquid water with the skin. The time-dependence of the temperature change at a site (x) in the container being cooled follows a step-function representing the bulk and the skin and proper initial-and-boundary conditions:

$$\frac{\partial \theta(x)}{\partial t} = \nabla \cdot (\alpha(\theta(x),x)\nabla \theta(x)) - v \cdot \nabla \theta(x),$$

(15)

where $\alpha$ is the thermal diffusion coefficient and $v$ is the convection velocity in the fluid. The interface between the bulk and the skin and the interface between the skin and the drain must satisfy certain conditions. The respective section gives more detail on the boundary conditions and calculation procedures. In order to examine all possible factors contributing to the Mpemba effect, this problem was solved numerically using the finite-element method [7].

3.5    Detectable properties versus bonding identities
3.5.1    $E_H$–$d_H$–$\Delta E_{1s}$ correlation

According to the BOLS notation, see Eq. (5), the cohesive energy of a specific bond denoted by subscript $x$ is proportional to the inverse $m$ power of its length under $z$ stimulus [122]:

$$\frac{E_x(z)}{E_x(\infty)} = \left(\frac{d_x(z)}{d_{x0}(\infty)}\right)^{-m} = \left[C_x(z)\right]^{-m},$$

(16)

where ∞ denotes the bulk standard, and $C_x(z)$ is the bond relaxation coefficient.

The tight-binding theory [131] formulates the shift of the $\nu$-th energy level of an isolated atom in the Hamiltonian:

$$H = \left[-\frac{\hbar^2 \nabla^2}{2m} + V_{atom}(r)\right] + V_{cry}(r).$$

The intra-atomic potential $V_{atom}(r)$ determines the $\nu$-th core level energy of an isolated atom, $E_\nu(0)$, and the crystal potential $V_{cry}(r)$ determines the core level shift $\Delta E_\nu(\infty)$. They follow the relations [131,160]:

$$E_v(0) = \langle v,i|V_{atom}(r)|v,i\rangle$$

$$\Delta E_v(\infty) = \langle v,i|V_{cry}(r)|v,i\rangle \left[1 + \frac{z\langle v,i|V_{cry}(r)|v,j\rangle}{\langle v,i|V_{cry}(r)|v,i\rangle}\right],$$

$$\cong E_b\left(1 + \left(\frac{\text{overlap integral}}{\text{exchange integral}} < 3\%\right)\right) = E_b$$

(17)

where $|v,i\rangle$ is the eigen wave function at the $i$-th atomic site, which satisfies the relationship $\langle v,i|v,j\rangle = \delta_{ij}$ because of the strong localization of the core electrons; and $E_b$ is the bond energy in ideal bulk. The $V_{cry}(r)$ sums the pairing potentials over all neighbors. A Tylor series approximates the pairing potential u(r):

$$u(r) = \left.\frac{\partial^n u(r)}{n!\partial r^n}\right|_{r=d} x^n = E_b + 0 + \left.\frac{\partial^2 u(r)}{2\partial r^2}\right|_{r=d} x^2 + \left.\frac{\partial^3 u(r)}{6\partial r^3}\right|_{r=d} x^3 + 0(x^{n\geq 4})$$

(18)

The zeroth approximation $E_b$ of the potential determines the $\Delta E_v(\infty)$. The second term (= 0) is the force at equilibrium. Higher-order differentials corresponding to harmonic and nonlinear vibrations determine the shape of the u(r) only. External stimulus perturbs the crystal potential from $V_{cry}(r)$ to $V_{cry}(r)(1+\Delta_H) \cong E_b(1+\Delta_H)$ at equilibrium without modulating the wavefunction.

Because four lone pairs isolate a $H_2O$ molecule, the H-O bonding potential dominates the $V_{cry}(r)$. Therefore, the O 1s energy shift $\Delta E_{1s}(z)$ from that of an isolated oxygen atom $E_{1s}(0)$ is proportional to the H-O bond energy $E_H$ [161]. Any relaxation of the H-O bond shifts the $E_{1s}(z)$ away from the bulk reference $E_{1s}(\infty)$. The $\Delta E_{1s}(z)$ may be positive or negative, depending on the sources of perturbation, including bond length relaxation [162,163], charge polarization [163], Coulomb coupling, etc. $\Delta E_{1s}(z)$ in water ice follows the relation [161]:

$$\frac{\Delta E_{1s}(z)}{\Delta E_{1s}(\infty)} = \frac{E_{1s}(z) - E_{1s}(0)}{E_{1s}(\infty) - E_{1s}(0)} = \frac{E_H(z)}{E_H(\infty)} = [C_H(z)]^{-m}.$$

(19)

3.5.2    Elasticity–$\omega_x$–$\Delta E_{1s}$ correlation

From a dimensional viewpoint, the second-order derivative of the $u_x(r)$ at equilibrium is proportional to $E_x/d_x^2$ [125]. Approximating the vibration energy, $\mu_x\omega_x^2 x^2/2$, of an oscillator to the second differential of the Taylor series of its interaction potential, $u_x(r)$, yields the $\omega_x$ of the oscillator, where $\mu_x$ is the reduced mass of the oscillator and $x$ is the amplitude of vibration. The following expressions also approximate the elastic modulus by definition, and $\omega_x$ is correlated with the elasticity in the following relationships [3,5,164]:

$$\begin{cases} \Delta\omega_x = \omega_x(z) - \omega_x(0) \propto \left(\left.\frac{\partial^2 u_x(r)}{\mu_x \partial r^2}\right|_{r=d_x}/\mu_x\right)^{1/2} \propto \sqrt{E_x/\mu_x}\Big/d_x \\ Y_x \propto -V\left.\frac{\partial^2 u_x(r)}{\partial V^2}\right|_{r=d_x} \propto \frac{E_x}{d_x^3} \\ (\Delta\omega_x)^2 \cong Y_x d_x \end{cases},$$

where $\omega_x(0)$ is the reference point from which the $\omega_x(z)$ shifts; $V$ is the molecular volume; and $\Delta\omega_x(z)$ is a direct measure of the bond stiffness, which is the product of the Young's modulus $Y_x$ and the length $d_x$ of the bond. For liquid water, the elastic modulus is dominated by the weaker O:H nonbond ($x = L$), and polarization. Coulomb coupling contributes to the vibration by replacing the force constant $k_x$ with $(k_x + k_C)$ in the form of $\Delta\omega \propto [(k_x + k_C)/\mu_x]^{1/2}$. The following describe their relative shifts:

$$\begin{cases} \dfrac{\Delta\omega_x(z)}{\Delta\omega_x(\infty)} = \dfrac{\omega_x(z) - \omega_x(0)}{\omega_x(\infty) - \omega_x(0)} = [C_x(z)]^{-(1+m_x/2)} \\ \dfrac{Y_x(z)}{Y_x(\infty)} = [C_x(z)]^{-(3+m_x)} \end{cases}$$

(20)

$\Delta E_{1s}$ correlates with $\Delta\omega_H$ as follows:

$$\begin{cases} \Delta E_{1s} \propto E_H & (O_{1s} \text{ shift}) \\ \Delta\omega_H \propto d_H^{-1}\sqrt{E_H} & (\omega_H \text{ shift}) \\ (d_H \Delta\omega_H)^2 \cong \Delta E_{1s} & (\text{Correlation}) \end{cases}$$

(21)

This correlation means that $\Delta E_{1s}$ and the $\Delta\omega_H$ shift cooperatively in the same direction, but at different rates.

### 3.5.3 Critical temperatures versus bond energies

For other 'normal' materials, $T_C$ is proportional to the atomic cohesive energy, $T_C \propto zE_z$, where $z$ is the effective atomic CN and $E_z$ is the bond energy of the $z$-coordinated atom [122]. However, for water molecules, the $T_C$ is proportional to $E_H$ only, because of the 'isolation' of the H$_2$O molecule by its surrounding lone pairs. $E_L$ determines $T_C$ for evaporation $T_V$, as this process dissociates the O:H nonbond:

$$\dfrac{T_C(z)}{T_C(\infty)} = \begin{cases} \dfrac{E_H(z)}{E_H(z_b)} = [C_H(z)]^{-m_H} & (T_C < T_V) \\ \dfrac{E_L(z)}{E_L(z_b)} = [C_L(z)]^{-m_L} & (T_C = T_V) \end{cases}$$

(22)

### 3.6 Summary

All detectable properties vary functionally with the relaxation of either the H-O bond or the O:H nonbond. The relaxation of the H-O bond shifts the O 1s energy $\Delta E_{1s}$, the critical temperature $T_C$ for phase transition (except for evaporation), and the H-O phonon frequency $\omega_H$. The relaxation of the O:H nonbond contributes to polarization, $\omega_L$ frequency shift, elasticity, and molecular dissociation energy $E_L$. O:H-O relaxation dominates density change.

## 4 Compression: Proton centralization and ice regelation

- *Coulomb repulsion and O:H-O strength disparity differentiate H$_2$O from 'normal' materials in response to compression.*
- *Compression shortens the O:H bond and simultaneously lengthens the H-O bond towards identical O:H*

*and H-O lengths, lowering the compressibility of water ice.*
- *Compression lowers $T_C$ and $T_m$ by reducing $E_H$ from the bulk value of $E_H$ = 3.97 eV; O:H-O bond full recoverability dictates the regelation of ice.*
- *MD decomposition of measured V-P profile into $d_x$ –P curves results in universal $d_L$-$d_H$ cooperativity.*

4.1     Mysteries of compressed water ice
4.1.1   Regelation and low compressibility

Discovered by Faraday and Thomson in 1850's [165,166], regelation is the phenomenon of ice melting under pressure and freezing again when the pressure is relieved at temperatures around – 10 °C. 'Two pieces of thawing ice, if put together, adhere and become one; at a place where liquefaction was proceeding, congelation suddenly occurs. The effect will take place in air, in water, or *in vacuo*. It will occur at every point where the two pieces of ice touch; but not with ice below the freezing-point, i.e., with dry ice, or ice so cold as to be everywhere in the solid state' [165]. Looping a wire around a block of ice with a heavy weight attached to it can melt the local ice gradually until the wire passing through the entire block. The wire's track will refill as soon as pressure is relieved, so the ice block will remain solid even after wire passes completely through. Another example is that a glacier can exert a sufficient amount of pressure on its lower surface to lower the melting point of its ice, allowing liquid water flows from the base of a glacier to lower elevations when the temperature of the air is above the freezing point of water. The regelation is exceedingly interesting, because of its relation to glacial action under nature circumstances [167], in its bearing upon molecular action [168], and damage recovery of living cells.

It is usual in 'normal' substance that compression raises the critical temperature ($T_C$) at all phase transitions [129,169,170]. Compression stores energy into the substance by shortening all bonds and plastic deformation may occur when the pressure is relieved. However, the freezing temperature of liquid water is lowered to -22°C by applying 210 MPa pressure. Stretching ice (i.e. tensile, or negative, pressure) has the opposite effect — ice melts at +6.5°C when subjected to -95 MPa pressure [34]. Conversely, the $T_C$ for ice drops from 280 to 150 K at the transition from ordered ice-VIII to proton-disordered ice-VII phase when $P$ is increased from 1 to 50 GPa [62,171,172]. A MD study of a nanowire cutting through ice suggests that the transition mode and cutting rate depends on the wetting properties of the wire - hydrophobic and thicker wires cut ice faster [173]. A video clip shows that a copper wire cutting ice faster than a fishing wire because of thermal conductivity [174]. O:H-O bond has extraordinary ability of recovering fully its relaxation and damage [175].

Conversely, compression shortens the O-O distance but lengthens the H-O bond, resulting in the low compressibility of ice compared to 'normal' materials [176]. The compressibility of liquid water is slightly higher than that of ice.

4.1.2   Proton centralization

In 1972, Holzapfel [177] firstly predicted that, under compression, an O:H-O bond might be transformed from the highly asymmetrical O:H-O configuration to a symmetrical state in which the H proton lies midway between two $O^{2-}$ ions, leading to a non-molecular symmetrical phase of ice-X. Goncharov et al. [178] confirmed this prediction in 1998 using *in situ* high-pressure Raman spectroscopy. The proton centralization in the O:H-O bond of ice-VIII occurred at about 60 GPa and 100 K, and no further shift of phonon frequency was observed when they increased the pressure, since the O:H and H-O had both reached identical lengths (0.11 nm) [179,180]. Proton centralization also occurs in liquid $H_2O$ at 60 GPa and 85 K, and to liquid $D_2O$ at 70 GPa and 300 K [181].

Figure 12 shows the phase diagram for ice VII, VIII and X transitions obtained using path-integral MD calculations [182] and IR and Raman spectroscopic measurements [143,178,183-185]. The ice-X phase boundary at about 60 GPa changes insignificantly with temperature. Systematic studies [182,186] revealed that heating compensated for compression on the proton centralization, but the mechanism was unclear.

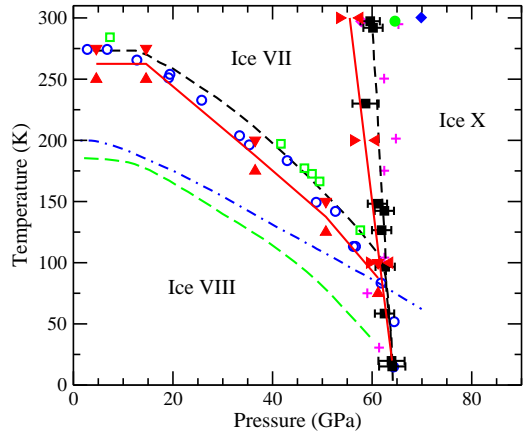

Figure 12. Phase diagram for ice VII, VIII, and X transition with the X phase boundary at 60 GPa varying insignificantly with temperature. Lines denote calculated values [182,187]; scattered symbols are experimental results [143,178,183-185]. (Reprinted with permission from [182].)

Compression-induced proton centralization for water has always been attributed to "translational proton quantum tunneling" [179,188,189] or to the extraordinary, yet unclear, behavior of the inter- and intramolecular bonds [190]. Teixeira [89] suggested that compression evolves the pairing potential wells into a single well located midway between $O^{2-}$ ions (see Figure 1b).

4.1.3    Phonon cooperative relaxation

Generally, compression stiffens all phonons of 'normal' materials such as carbon allotropes [125], ZnO [124], group IV [191], group III-V [192], and group II-VI [193] compounds without exception; however, for ice and water, compression stiffens the softer phonons ($\omega_L <$ 300 cm$^{-1}$) but softens the stiffer phonons ($\omega_H >$ 3000 cm$^{-1}$) [14,62,172,194,195].

Figure 13 shows the compression-induced $\omega_x$ cooperative relaxation of ice-VIII phase at 80 K [6, 103]. The $\omega_H$ softening also takes place in water at 23°C at 0.05–0.39 GPa pressure [195], which is accompanied by $\omega_L$ stiffening because of the O:H-O bond cooperativity. First principles and quantum Monte Carlo calculations suggested that the O:H interaction contributed positively, while the H-O bond lowered lattice energy as the pressure was increased [196].

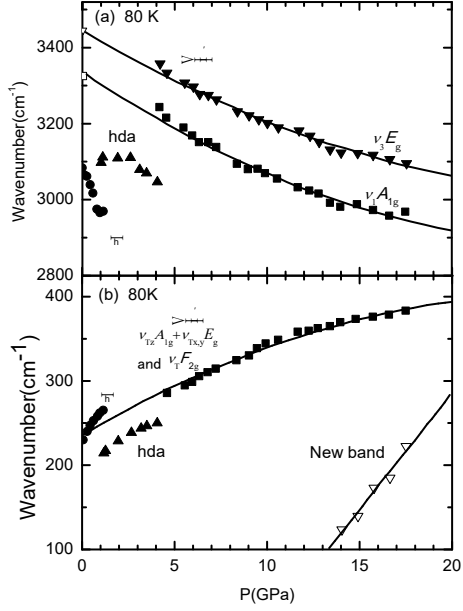

Figure 13. Pressure-dependence of (a) $\omega_H$ shift, and (b) $\omega_L$ shift. Experimental data fitted with solid lines. Compression softens the $\omega_H$ and stiffens the $\omega_L$ monotonically. (Reprinted with permission from [194].)

4.2   Bond–phonon–energy relaxation
4.2.1   Proton centralization and $d_x$ cooperative relaxation

Table 2 features the DFT-MD derivatives of the O:H-O bond segmental relaxation dynamics and the band gap change as a function of pressure. As shown in Figure 14a, compression shortens the O:H nonbond from 0.1767 to 0.1692 nm, meanwhile lengthens the H-O bond from 0.0974 to 0.1003 nm when the pressure is increased from 1 to 20 GPa. The $d_x$ relaxation agrees with the trends reported in [15,56,197]. The following polynomial formulates the relaxation, $d_x = d_{x0}[1 + \alpha_x(P - P_0) + \beta_x(P - P_0)^2]$, in which $P_0 = 1$ MPa is atmospheric pressure:

$$\begin{pmatrix} d_H / 0.975 \\ d_L / 1.768 \\ V / 1.060 \end{pmatrix} = \begin{pmatrix} 1 & 9.510 & 2.893 \\ 1 & -3.477 & -10.280 \\ 1 & -238.000 & 47.000 \end{pmatrix} \begin{pmatrix} P^0 \\ 10^{-4} P^1 \\ 10^{-5} P^2 \end{pmatrix}.$$

(23)

An MD numerical match to the $V$–$P$ profile of ice-VIII, measured by *in situ* high-pressure and low-temperature synchrotron XRD and Raman spectroscopy [14], is shown in Figure 14(a). This match leads to the equation of states, $V/V_0$ – $P$ curve, with $V_0 = 1.06$ cm$^3$/kg.

Encouragingly, the MD calculations convert the measured $V/V_0$ – $P$ profile into the $d_x/d_{x0}$ – $P$ curves, whose extrapolations meet at the exact point of proton centralization [178,179] (see Figure 14b). The proton centralization was calculated to occur at 58.6 GPa with an O-O distance of 0.221 nm, exactly consistent with the observed value of 0.220 nm at 59 GPa [179]. Therefore, it is evident that proton centralization arises from pressure-induced O:H-O bond asymmetrical relaxation rather than transitional H$^+$ proton quantum tunneling. Proton tunneling is unlikely because of the strong H-O bond of 3.97 eV energy. Constrained by measured proton centralization and the $V$–$P$ equation of states, the $d_x/d_{x0}$ – $P$ profiles describe the true situation irrespective of the probing technique or condition. Taking the P as a hidden parameter, the $d_x/d_{x0} - P$ profiles give the universal $d_L$–$d_H$ cooperativity.

Table 2. Pressure dependence of the mass density $\rho$, segmental lengths $d_x$, O-O distance $d_{O-O}$, and band gap $E_G$ of ice [2].

| | DFT | | | | MD | | |
|---|---|---|---|---|---|---|---|
| $P$ | $\rho$ (g/cm³) | $d_H$ (Å) | $d_L$ (Å) | $E_G$ (eV) | $d_H$ (Å) | $d_L$ (Å) | $d_{O-O}$ (Å) |
| 1 | 1.659 | 0.966 | 1.897 | 4.531 | 0.974 | 1.767 | 2.741 |
| 5 | 1.886 | 0.972 | 1.768 | 4.819 | 0.979 | 1.763 | 2.742 |
| 10 | 2.080 | 0.978 | 1.676 | 5.097 | 0.985 | 1.750 | 2.736 |
| 15 | 2.231 | 0.984 | 1.610 | 5.353 | 0.991 | 1.721 | 2.713 |
| 20 | 2.360 | 0.990 | 1.556 | 5.572 | 1.003 | 1.692 | 2.694 |
| 25 | 2.479 | 0.996 | 1.507 | 5.778 | – | – | – |
| 30 | 2.596 | 1.005 | 1.460 | 5.981 | | | |
| 35 | 2.699 | 1.014 | 1.419 | 6.157 | | | |
| 40 | 2.801 | 1.026 | 1.377 | 6.276 | | | |
| 45 | 2.900 | 1.041 | 1.334 | 6.375 | | | |
| 50 | 2.995 | 1.061 | 1.289 | 6.459 | | | |
| 55 | 3.084 | 1.090 | 1.237 | 6.524 | | | |
| 60 | 3.158 | 1.144 | 1.164 | 6.590 | | | |

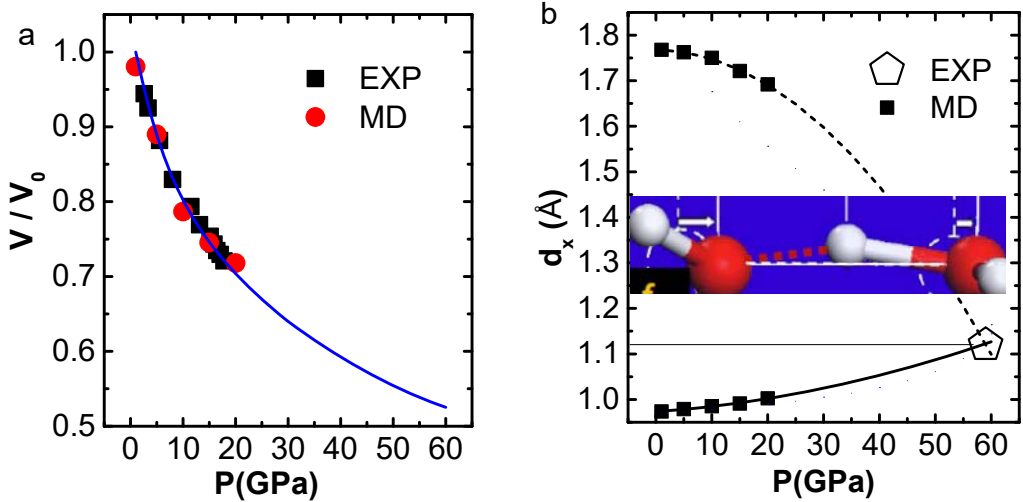

Figure 14. (a) MD conversion of measured $V$–$P$ profile [14] into (b) the $d_x$ relaxation curves that meet at the point of proton centralization occurring at 59 GPa and 0.22 nm [178,179]. (Reprinted with permission from [2].)

4.2.2 Phonon stiffness cooperative relaxation

Figure 15a features the MD-derived phonon spectra as a function of pressure, which agree with trends probed using Raman and IR spectroscopy from ice-VIII at 80 K [14,62,172]. Compression softens the $\omega_H$ from 3520 cm$^{-1}$ to 3320 cm$^{-1}$ and stiffens the $\omega_L$ from 120 to 336 cm$^{-1}$, disregarding the possible phase change and other supplementary peaks nearby. Figure 15b compares the calculated and a collection of the measured $\Delta\omega_x$ for ice.

Consistency in the pressure derived $\omega_x$ cooperativity in Figure 15b for both water [195] and ice [62,171,172] confirms that compression shortens and stiffens the O:H nonbond and relaxes the H-O bond to the opposite extent.

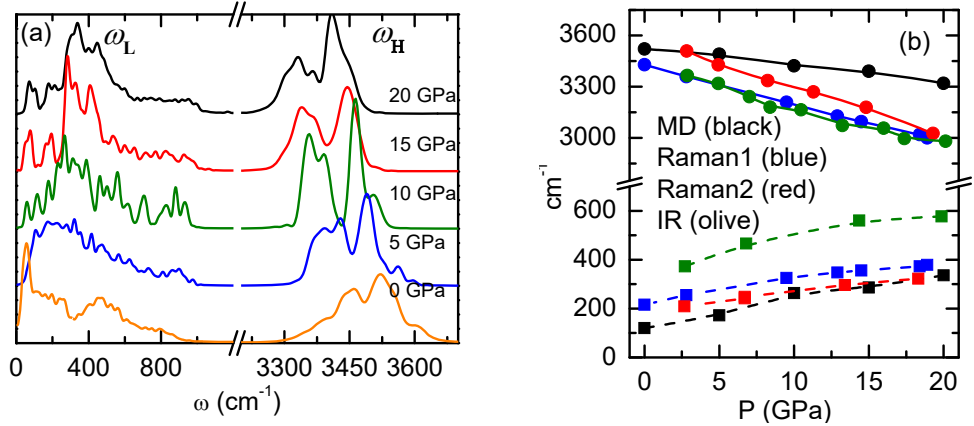

figure 15. (a) MD-derived $\omega_x$ relaxation; and (b) trend agreement between Raman and IR measurements and calculations for the ice-VIII phase at 80 K [14,62,172]. (Reprinted with permission from [2].)

4.2.3   Band gap expansion

Hermann and Schwerdtfeger [42] found that the onset of UV absorption by ice, as an indication of band gap, shifts positively with increasing pressure, making ice more transparent. They attributed this effect to an increase of the Stark shift in water caused by the electrostatic environment at smaller volumes. Generally, compression enlarges the optical band gap due to bond strength gain, since the band gap is proportional to the bond energy with involvement of electron–phonon coupling [124,198]. Band gap expansion of ice follows the same pressure trend but due to a different mechanism, because compression softens the H-O bond.

Figure 16a shows the DFT-derived DOS evolution of ice-VIII with pressure varying from 1 to 60 GPa. The bottom edge of the valence band shifts down from -6.7 eV at 1 GPa to -9.2 eV at 60 GPa, but its upper edge at the Fermi level remains unchanged. The conduction band shifts up from 5.0 to 12.7 eV at 1 GPa to 7.4–15.0 eV at 60 GPa. The band gap expands further at higher pressures, from 4.5 to 6.6 eV, as shown in Figure 16b, when the pressure increases from 1 to 60 GPa.

As currently understood, the band gap expansion in compressed ice arises not from the $E_H$ but is caused by a different mechanism. The energy shift of the DOS above $E_F$ results from polarization of the lone pair by the entrapped core electrons. The energy shift of the valence DOS below $E_F$ arises from entrapment of the bonding states of oxygen [97,116].

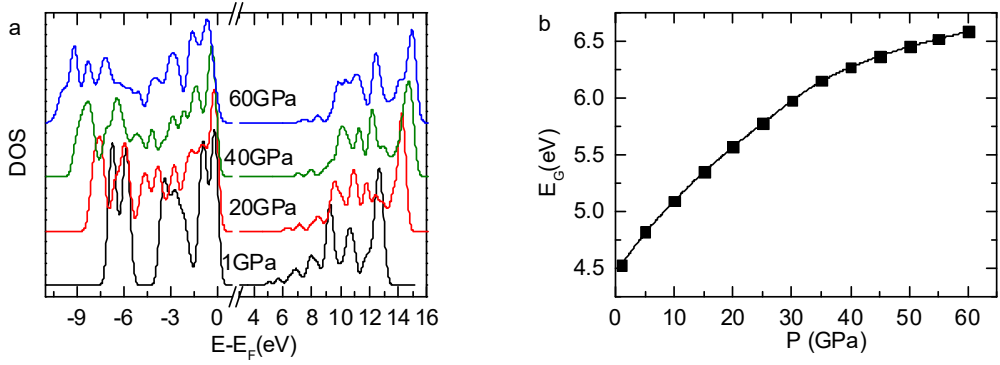

Figure 16. Compression widens the band gap of ice-VIII. (a) Compression entraps the valence DOS (bottom edge) and polarizes the conduction DOS, resulting in (b) band gap expansion. The upper edge of the valence band at the $E_F$ is conserved. (Reprinted with permission from [2].)

### 4.3 $E_H$ dominance of $T_C$ and $T_m$

The following proves that $E_H$ dictates the $T_C$ for ice VII–VIII phase transition, and $T_m$ for melting. Assuming that $T_C$ and $T_m$ changes with $E_x$ in the following relationship, Eq **Error! Reference source not found.**), but $x$ is as yet to be determined [129],

$$\frac{T_C(P)}{T_C(P_0)} = 1 - \frac{E_x(P)}{E_{x0}} = 1 - \frac{\int_{V_0}^{V} p\, dv_x}{E_{x0}} = 1 - \frac{S_H \int_{P_0}^{P} p\, \frac{d d_H}{dp}\, dp}{E_{H0}} < 1.$$

From Eq. (23), we have:

$$\frac{d d_x}{d p} = d_{x0}[\alpha_x + 2\beta_x(P - P_0)].$$

(24)

Reproduction of the measured $P$-dependent $T_C$ for the VII–VIII phase transition [62,171,172], and the $T_m$ for melting (-22 °C at 210 MPa; +6.5 °C at -95 MPa) [34] in requires that the integral must be positive. Only the $d_H$ in Eq. (23) and

Figure 17, meets this criterion ($\alpha_x > 0$, $\beta_x > 0$). Therefore, the H-O bond dominates the $T_C$.

Furthermore, matching the $T_C$–$P$ (panel a) and the $T_m$–$P$ (panel b) profiles using Eq. (24) yields an $E_H$ value of 3.97 eV at 1 MPa by taking the H atomic diameter of 0.106 nm as the diameter of the H-O bond [199]. This $E_H$ value agrees with the H-O dissociation energy of 4.66 eV for dissociating the H-O bond of water molecules deposited on a $TiO_2$ substrate with less than a monolayer coverage, and 5.10 eV for water monomers in the gaseous phase [106]. Molecular undercoordination differentiates values of 5.10, 4.66 and 3.97 eV for the H-O bond in various coordination environments [3].

Clearly, the relaxation of the H-O bond mediates the $T_m$ and the $T_C$, while $E_L$ is largely irrelevant to them. It is not surprising, therefore, that compression softens the H-O and hence lowers $T_C$ and $T_m$, while negative (tensile) pressure does the opposite by shortening and stiffening the H-O bond [34]. This finding clarifies that the status of the H-O bond relaxation governs the superheating and supercooling behavior of water and ice.

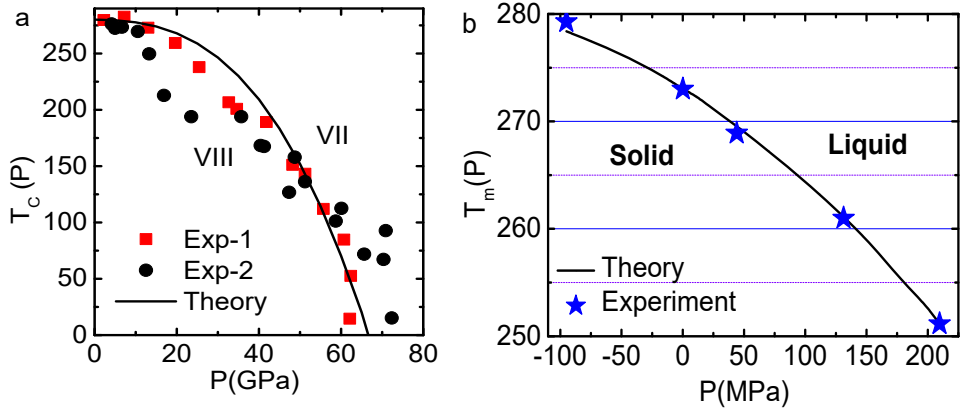

Figure 17. (a) Theoretical reproduction of the measured $T_C$–$P$ [62,171,172]; (b) $T_m$–$P$ (-22°C at 210 MPa; +6.5°C at -95 MPa) [34] profiles confirms that the $E_H$ dictates the $T_C$ and the $T_m$ with derivative of $E_H$ = 3.97 eV for bulk water and ice. (Reprinted with permission from [2,82].)

4.4    Regelation: O:H-O memory and quasi-solid phase boundary dispersion

It is known that molecular evaporation requires energy depending on $E_L$ only. If remove one $H_2O$ molecule from the bulk, one has to break four O:H nonbonds with energy of 0.38 eV per molecule [101]. Question may arise why the $E_H$ instead of the $E_L$ dominates the $T_C$ and $T_m$?

In order to clarify this paradox, let us look at the specific heat of water, shown in Figure 9. The shape and integral of the specific heat curve depend on the vibration frequency $\omega_x$ and the cohesive energy $E_x$ of the respective segment. The supposition of the two $\eta_x$ curves creates two intersecting points that correspond to the phase boundaries of the quasi-solid phase or the extreme density temperatures.

One can imagine what will happen to the crossing temperatures if one can depress the $\Theta_{DH}(\omega_H)$ and $E_H$ and meanwhile, elevate the $\Theta_{DL}(\omega_L)$ and $E_L$ by compression or the otherwise by tension. The $\eta_L$ will saturate slower and the $\eta_H$ quicker than they were. Compression ($\Delta P > 0$) will lower the upper intersecting point - the $T_m$ and, vice versa, as illustrated in Figure 18. Therefore, the $E_H$ determines approximately the $T_C$ through dispersing the specific heat crossing temperatures. Under any excitation the two $\omega_x$ always relax in opposite direction, which disperse the specific heat and hence superheating/supercooling occurs. Once the O:H-O bond breaks by cutting, oxygen atoms will find new partners for bonding to retain the sp³-orbital hybridization, which is the same to diamond oxidation and metal corrosion [97]. Therefore, O:H-O bond has the strong recoverability for O:H-O bond relaxation and dissociation without any plastic deformation, as well be shown in section 8.3. Compression sores energy into water ice by lowers the total bond energy, $E_L+E_H$, through inter electron pair compression [175].

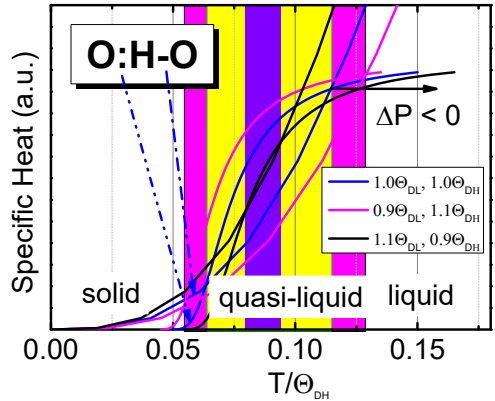

Figure 18. Superposition of the $\eta_x(T)$ curves defines two crossing points that correspond to the quasi-solid phase boundaries. The high temperature boundary corresponds to melting and the lower to freezing. Reversible compression ($\Delta P > 0$) and tension ($\Delta P < 0$) disperse the boundaries simultaneously by modulating the $\Theta_{Dx} \propto \omega_x$ and $E_x \propto \int_0^{T_{mx}} \eta_x(t)dt$, which depresses/elevates the $T_m$.

4.5    Summary

Consistency in numerical calculations and experimental observations verified the following:

1) Compression shortens and stiffens the O:H nonbond and lengthens and softens the H-O bond towards proton centralization, which lowers the compressibility of water ice and makes the O:H contribute positively, while the H-O contributes negatively to the lattice energy of compressed ice.
2) H-O bond relaxation determines $T_C$, the $T_m$ and the superheating or supercooling behavior of water and ice. Matching observations result in $E_H = 3.97$ eV for bulk water and ice.
3) Numerical conversion of the $V$–$P$ profile into the $d_x$–$P$ establishes $d_x$ cooperativity uniquely quantitatively.
4) Compression closes up the separation between boundaries of the quasi-solid phase and depresses the freezing temperature. Negative pressure does the opposite.
5) Unlike a 'normal' substance that gains energy with potential plastic deformation under compression, O:H-O bond demonstrates extreme recoverability of relaxation and dissociation. O:H-O tends to recover from its higher-energy state to initially lower state when stimulus is relieved. Findings may extend to damage recovery of living cells of which O:H-O bond dominates.

5    Molecular undercoordination: Supersolidity and slipperiness of ice

- *Molecular undercoordination shortens the H-O and lengthens O:H with a dual process of polarization that raises the hydrophobicity, viscoelasticity, and repulsion.*
- *H-O bond stiffening raises $\omega_H$, $T_m$, $E_{1s}$ and $E_H$ for H-O bond dissociation; O:H elongation lowers $\rho$, $\omega_L$ and $E_L$ for molecular dissociation.*
- *Water and ice share a common supersolid skin characterized by an identical $\omega_H$ of 3450 cm$^{-1}$.*
- *The less dense supersolid skin makes ice slippery and water skin hydrophobic and tough.*

5.1 Puzzles due to molecular undercoordination
5.1.1 Skin hydrophobicity, lubricity and thermal stability

Undercoordinated water molecules are referred to those with fewer than four neighbors as they occur in the bulk of water [18,107,108,200,201]. Molecular undercoordination occurs in the terminated O:H-O bonded networks, in the skin of a large volume of water, in hydration shells, and in the gaseous state. Such undercoordinated water molecules are even more fascinating than those that are fully coordinated [18,56,197,202-208]; for instance, water droplets encapsulated in hydrophobic nanopores [209,210] and ultrathin water films deposited on graphite, silica, and certain metals [71,139,202,211-216] behave like ice at room temperature – termed 'superheating', because they melt at a temperature well above the melting point (273 K) of ice in bulk. A monolayer of ice melts at 325 K according to MD calculations [138] compared to around 310 K of the skin of bulk water [4].

Superheating is more apparent at the curved skin. Sum frequency generation (SFG) spectroscopy reveals that two adjacent molecular layers are highly ordered at the hydrophobic contacts compared with those at a flat water–air interface [217] due to the presence of an air gap 0.5–1.0 nm thick in the hydrophobic contacts [218]. The air gap increases with the contact angle of the droplet or with the lowering of the effective CN of molecules at the skin.

However, when encapsulated in hydrophilic nanopores [219,220], or when wetted in hydrophilic topological configurations [221], water molecules perform in an opposite way and show attributes of supercooling, melting at temperatures below the bulk $T_m$. Altering the $H_2O/SiO_2$ interface from hydrophobic to hydrophilic by plasma sputtering raises the interface shear viscosity and changes the $\omega_H$ of the skin (3400 cm$^{-1}$) to the bulk value of 3200 cm$^{-1}$ [222].

The superheating of nanosized droplets is often accompanied by supercooling when freezing. Molecular undercoordination not only raises the $T_m$ but also lowers the least-density temperature $T_C$, for instance, from 242 K for a 4.4 nm droplet [134] to 205 K for a 1.4 nm droplet [133]. This observed size dispersion effect has been termed the 'extreme-density temperatures'.

More interestingly, theoretical calculations and measurements have both demonstrated that a monolayer film of water at room temperature manifests 'quasi-solid' behavior, and a hydrophobic nature that prevents it from being wetted itself by a water droplet [139,223]. A video clip [224] shows that a water droplet bounces repeatedly and continuously on water, evidence that the skins of both the bulk water and the droplet are elastic and hydrophobic. Water droplets also dance on solid surfaces, regardless of substrate temperatures and materials (-79°C $CO_2$; 22°C hydrophobic surface, and 100°C Al plate) [225], which indicates that the skin of the droplet is not only elastic but also thermally stable.

Amazingly, molecular undercoordination makes ice the slipperiest and the skin of water hydrophobic and toughest of ever known. The hydrophobicity and high surface tension, 72 mN/m at 25 °C , entitles water skin with numerous fascinations [226]. For instances, small insects such as a strider can walk and glide freely. If carefully placed on the surface, a small needle floats on the water even though its density is times higher than that of water. If the surface is agitated to break up the tension, then the needle will sink quickly. Skin tension helps seeds bury themselves by causing awns to coil and uncoil. It enables a floating fern to maintain an air layer, even when submerged. It also makes a beetle fly in two dimensions, not three. Surface tension also allows human and agricultural pathogens to travel long distances in tiny, buoyant droplets. The hardly noticed skin tension does play a big role in life at large.

## 5.1.2 Electron $\Delta E_{1s}$ versus phonon $\Delta\omega_H$

Following the same trend as 'normal' materials, molecular undercoordination imparts to water local charge densification [204,205,227-230], binding energy entrapment [203,227,231,232], and nonbonding electron polarization [229]. For instance, the O 1s level shifts more deeply from the bulk value of 536.6 eV to 538.1 eV and 539.7 eV when bulk water is transformed into skin or into gaseous molecules [233,234], as shown in Figure 19a [235].

Atomic undercoordination lowers the atomic cohesive energy and the thermal stability of 'normal' materials in general. However, molecular undercoordination raises the $T_m$ in water, as discussed above. The energy necessary for dissociating a $(H_2O)_N$ cluster into $(H_2O)_{N-1} + H_2O$ increases, conversely, when the cluster size is reduced to a trimer (Figure 19b) [236], which conflicts with the traditional understanding of 'normal' material behavior.

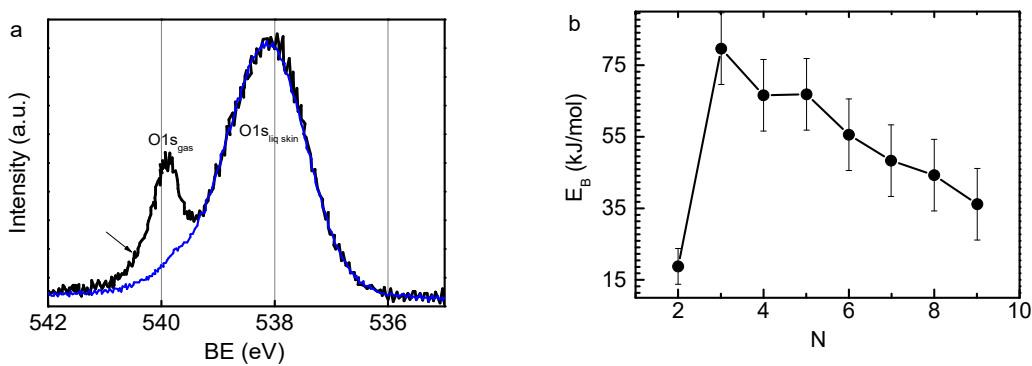

Figure 19. XPS O1s spectra of (a) water containing emission from the liquid skin at 538.1 eV and from the gaseous phase at 539.9 eV (reprinted with permission from [235]); (b) energy required for dissociating a $(H_2O)_N$ cluster into $(H_2O)_{N-1} + H_2O$ (1 kJ/mol = 0.02 eV/molecule). (Reprinted with permission from [236].)

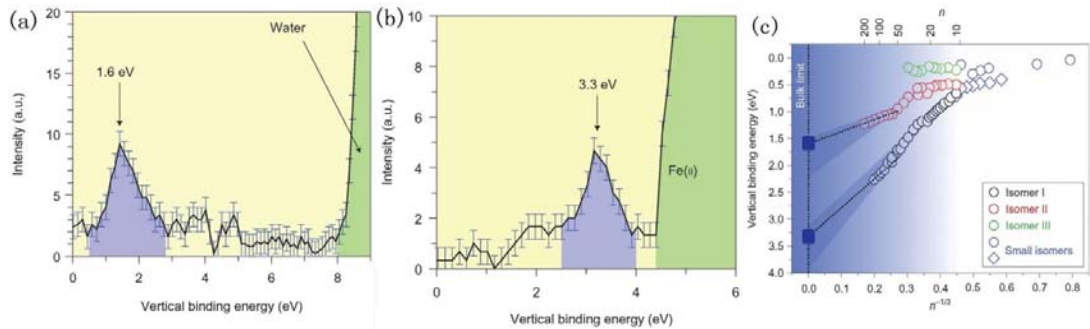

Figure 20. Molecular undercoordination polarizes nonbonding electrons. The vertical bound energy for solvated electrons was measured to be (a) 1.6 eV in the skin, and (b) 3.3 eV in the bulk of the liquid water. (c) The bound energy reduces further with the number $n$ of $(H_2O)_n$ clusters. (Reprinted with permission from [229].)

However, an ultrafast liquid-jet UPS [229], shown in Figure 20, resolved the vertical bound energies (being equivalent to work function) of 1.6 eV and 3.3 eV for the solvated electrons in the skin and in the bulk center of the solution, respectively. In addition, the bound energy decreases with the number *n* of the (H$_2$O)$_n$ clusters, evidence that undercoordination substantially enhances nonbonding electron polarization [3].

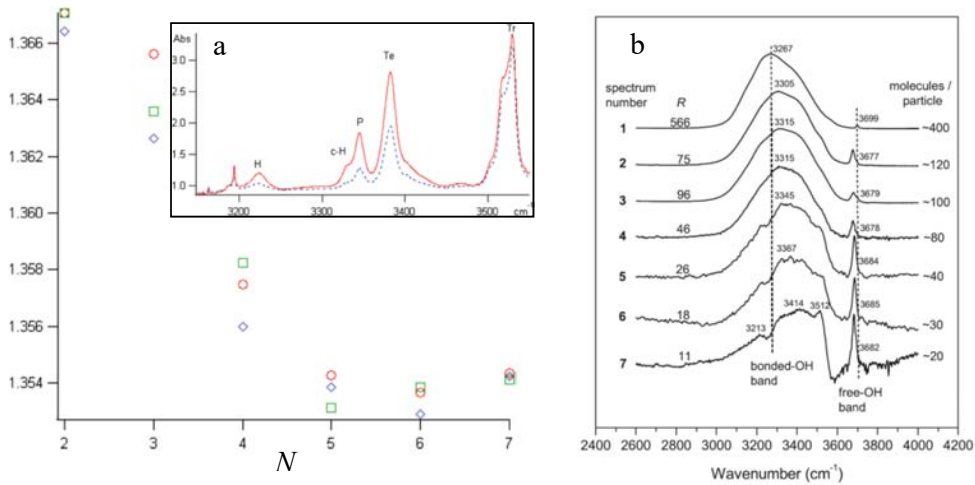

Figure 21. Size-dependent $\omega_H$ of (a) (H$_2$O)$_N$ clusters (in the frequency ratio of H-O/D-O) and (b) large clusters. Line (2) corresponds to a dimer [237], (3, Tr) to a trimer [238], (4, Te) to a tetramer, (5, P) to a pentamer, (6, c-H) corresponds to a cyclic hexamer, (7, H) corresponds to a cage hexamer. Red circles correspond to He matrix, green squares correspond to Ar matrix, and blue diamonds corresponds to p-H$_2$ measured at 2.8 K. Inset (a) denotes the sharp $\omega_H$ peaks for the small clusters. Size-reduction stiffens the H-O bonds with little disturbance to the dangling H-O bonds at 3700 cm$^{-1}$ in (b). (Reprinted with permission from [57,239].)

Normally, the loss of neighbor atoms softens the phonons of 'normal' materials such as diamond and silicon except for the G mode (1550 cm$^{-1}$) for graphene [240] and the A$_1$ mode (141 cm$^{-1}$) for TiO$_2$ [241] as these two modes are independent of atomic CN. However, water molecular undercoordination stiffens the stiffer $\omega_H$ significantly [242,243]. The $\omega_H$ has a peak centered at 3200 cm$^{-1}$ for bulk water, and at 3450 cm$^{-1}$ for the skins of water ice [78]. The $\omega_H$ for gaseous molecules is around 3650 cm$^{-1}$ [57,65,113,239]. The $\omega_H$ shifts from 3200 to 3650 cm$^{-1}$ when the N of the (H$_2$O)$_N$ cluster drops from 6 to 1 (Figure 21(a)) [48,239,244]. Encapsulation by Kr and Ar matrices changes the $\omega_H$ slightly by 5–10 cm$^{-1}$ due to the involvement of interface interaction [48]. Size-induced $\omega_H$ stiffening also occurs in large molecular clusters [57] (see Figure 21(b)). When N drops from 475 to 85, $\omega_H$ transits from the dominance of the 3200 cm$^{-1}$ component (bulk attribute) to the dominance of the 3450 cm$^{-1}$ component (skin attribute) [245]. The high frequency at approximately 3700 cm$^{-1}$ corresponds to the vibration of the dangling H-O bonds, with possible charge transportation in the skin of water and ice [207,246].

### 5.1.3 Density loss by $d_{O-O}$ elongation

Bonds in the skins of metals, alloys, semiconductors, insulators and nanostructures contract globally with respect to the underlying bulk material. The spacing between the first and the second atomic layers of these systems contracts by 10–14% relative to the bulk geometry. For nanostructures, the extent of contraction is greater; the relaxation extends radially into deeper layers [128,247]. The skin bond contraction takes the full responsibility for the unusual behavior of the undercoordinated adatoms, point defects and nanostructures of different shapes, including the nanoscale size-dependence of the known bulk properties and the size-induced emergence of properties that the bulk parent never shows [122].

However, the skin O-O distance for water expands by 5.9% or more, compared to a 4.6% contraction of the skin O-O distance of liquid methanol [248], which differentiates the surface tension of 72 mN/m for water from 22 mN/m for methanol. The O-O distance in the skin and between a dimer is about 3.00 Å; the O-O distance in the bulk varies from 2.70 [249] to 2.85 Å [76]. Besides, the volume of water confined in the 5.1 and 2.8 nm sized $TiO_2$ pores expands by 4% and 7.5% respectively, with respect to the bulk water [250]. MD calculations also reveal that the $d_H$ contracts from 0.9732 Å at the center to 0.9659 Å at the skin of a free-standing water droplet containing 1000 molecules [251]. The O-O elongation results in a density loss in the water skin to a value as low as 0.4 g·cm$^{-3}$ (corresponding to $d_{O-O}$ = 3.66 Å) [252,253].

5.1.4 Does a liquid skin cover ice or a solid skin form on water?

Whether a solid skin forms on water or a liquid skin covers ice has long been a paradox in this field of research. Correlation between the coordination environment and the skin anomalies of water and ice remain unclear despite extensive investigation since 1859 when Faraday [254] first proposed that a quasi-liquid skin appears to lubricate ice, even at temperatures below freezing [255,256]. The slipperiness of ice has also been perceived as pressure-promoted melting [257] or friction-induced heating [258], while the extraordinary hydrophobicity and toughness of water skin are commonly attributed to the presence of a layer of molecules in the solid state [202,213].

In fact, ice remains slippery while one is standing still on it without involvement of pressure or friction [60,78,259]. According to the phase diagram of ice, pressure induces solid–liquid transition only at -22°C and above [257,260], but ice remains slippery at temperatures far below this. This fact rules out the mechanism of pressure melting. On the other hand, if a liquid lubricant exists on ice, the vibration amplitudes of molecules in the liquid skin should be greater than in the bulk ice, but interfacial force microscopy measurements have excluded this occurrence [259]. The skin layer is, however, somewhat viscoelastic at temperatures between -10 and -30°C, which is evidence of the absence of a liquid skin at such low temperatures.

Furthermore, water skin is ice-like at ambient temperature. SFG spectral measurements and MD calculations suggested that the outermost two layers of water molecules have an 'ice-like' order at room temperature [261]. At ambient temperature, ultrathin films of water perform like ice with a hydrophobic nature [139,202]. These facts exclude the possibilities of skin pre-melting. Water at temperatures of 7, 25, and 66°C at atmospheric pressure has an ordered skin 0.04–0.12 nm thick [262]. Does a liquid skin form on ice? Does a solid skin cover water? These apparent paradoxes are solved from the perspective of skin O:H-O bond relaxation using the currently employed theoretical and experimental strategies.

5.2 Skin bond–electron–phonon attributes
5.2.1 Bond relaxation of $(H_2O)_N$ clusters

Figure 22 and Table 3 show O:H-O bond relaxation as a function of N for the $(H_2O)_N$ clusters derived from calculations using the PW and the OBS algirithms. Consistent with results reported in [18,200], the following are confirmed:

1) Molecular undercoordination shortens the H-O bond and lengthens the O:H nonbond consistently; $E_H$ increases and $E_L$ drops when N is reduced, as the BOLS-NEP notation predicts.
2) The H-O bond shortening (lengthening) is always coupled with the O:H lengthening (shortening), irrrespective of the algorithm, which is evdence of the expected O:H-O bond cooperativity.

3) The non-monotonic change of $d_x$ results from the effective CN that varies not only with the number of molecules N of the $(H_2O)_N$ cluster but also with the geometrical configuration of the cluster. The effective CN of a ring-like cluster is smalller than that of a cage for the same N value.
4) The inconsistent results from different algorithns suggest that the focus should be more on the trend and the natural origin than on the accuracy of the derived values. Numerical derivatives serve as useful references for concept verification.

Table 3. DFT-derived segmental length $d_x$, total electronic binding energy $E_{Bind}$ (-eV), and segmental $E_x$, of $(H_2O)_N$ clusters*. References contain experimental results.

|  | N | $d_H$ (Å) | $d_L$ (Å) | θ (°) | $d_{OO}$ (Å) | $E_{Bind}$ (eV) | $E_H$ (eV) | $E_L$ (eV) |
|---|---|---|---|---|---|---|---|---|
| Monomer | 1 | 0.969 |  |  |  | 10.4504 | 5.2252 | - |
| Dimer | 2 | 0.973 | 1.917 | 163.6 | 2.864 | 21.0654 | 5.2250 | 0.1652 |
| Trimer | 3 | 0.981 | 1.817 | 153.4 | 2.730 | 31.8514 | 5.2238 | 0.1696 |
| Tetramer | 4 | 0.986 | 1.697 | 169.3 | 2.672 | 42.4766 | 5.2223 | 0.1745 |
| Pentamer | 5 | 0.987 | 1.668 | 177.3 | 2.654 | 53.1431 | 5.2208 | 0.1870 |
| book | 6 | 0.993 | 1.659 | 168.6 | 2.640 | 63.8453 | 5.2194 | 0.2020 |
| Cages | 6 | 0.988 | 1.797 | 160.4 | 2.748 | – | – | – |
|  | 8 | 0.992 | 1.780 | 163.6 | 2.746 | – | – | – |
|  | 10 | 0.993 | 1.748 | 167.0 | 2.725 | – | – | – |
| Clusters | 12 | 0.992 | 1.799 | 161.7 | 2.758 | – | – | – |
|  | 20 | 0.994 | 1.762 | 165.4 | 2.735 | – | – | – |
| Bulk | Ih | 1.010 | 1.742 |  |  |  | 3.97 [6] | 0.095[101] |

*The total energy of a cluster is that required to excite all the electrons to the vacuum level; the binding energy $E_{Bind}$ is the energy required to combine atoms together to form a cluster: $E_{Bind} = E_{cluster} - \sum E_{atom} = \sum E_{bond}$. For N = 1, there are two H-O bonds only. Thus, the H-O bond energy is one-half of the $E_{Bind}$: $E_H(1) = E_{Bind}(1)/2$. For N = 2 to 6: $E_H(N) = E_H(1) - \alpha(d_H(N) - d_H(1))^2$, where $\alpha$ is the elastic constant.

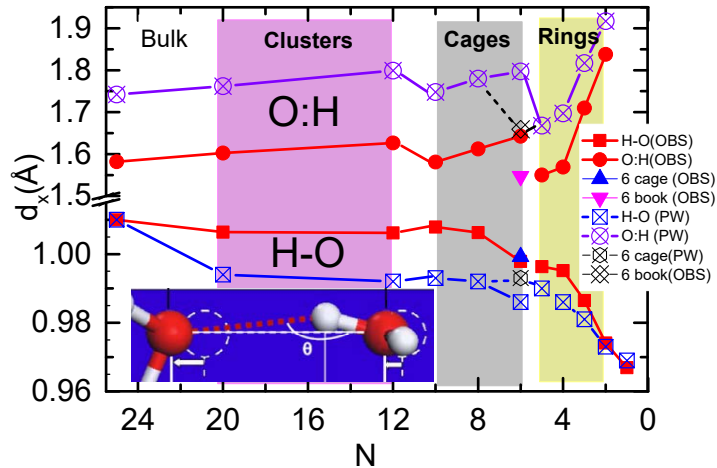

Figure 22. Cluster size-dependence of $d_x$ in the $(H_2O)_N$ clusters optimized using the PW [147] and the OBS [148] methods. N = 6 gives the 'cage', 'book' and 'prism' hexamer structures, with almost identical binding energies. The non-monotonic tread stems from the effective molecular CN that also changes with geometrical

configuration. (Reprinted with permission from [206].)

As N is reduced from 24 (an approximation to the bulk) to two (a dimer), the H-O bond contracts by 4% from 0.101 nm to 0.097 nm, and the O:H bond expands by 17% from 0.158 to 0.185 nm, according to the OBS derivatives. This cooperative relaxation expands the O-O distance by 13% and lowers the density by 30% for the dimer. The monotonic relaxation profiles for the $d_x$ at N ≤ 6 will be discussed in the subsequent discussions without rendering the generality of the conclusion.

### 5.2.2 Skin mass density

Figure 23 features particularly the residual length spectra (RLS) for the MD-derived $d_x$ of ice. Subtracting the length spectrum calculated using the 360-molecular unit cell without skin from that with a skin resulted in the RLS. The RLS indicates that the $d_H$ contracts from the bulk value of about 1.00 to about 0.95 Å at the skin, while the $d_L$ elongates from about 1.68 to about 1.90 Å, with high fluctuation. This cooperative relaxation lengthens the O-O by 6.8% and is associated with an 18% density loss. The peak of $d_H$ = 0.93 Å even corresponds to the undercoordinated H-O radicals, whose vibration frequency is around 3650 cm$^{-1}$ [3].

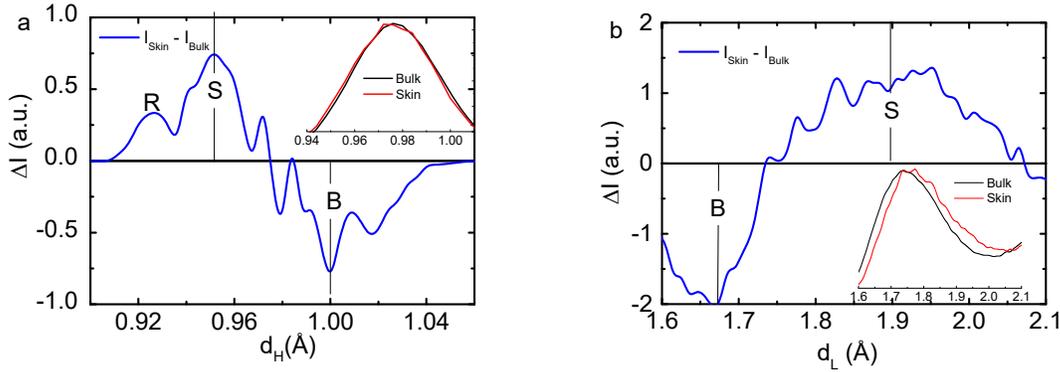

Figure 23. MD-derived RLS reveals that (a) $d_H$ contracts from the bulk value (B) of ≈ 1.00 to 0.95 Å for the skin (S) and to 0.93 Å for the H-O free radicals (R), which is coupled with (b) $d_L$ elongation from the bulk value (B) of ≈ 1.68 to ≈ 1.90 Å, with high fluctuation. The insets show the raw spectra of the unit cell with skin (denoted 'skin') and without skin (denoted 'bulk'). (Reprinted with permission from [4].)

The skin O:H-O relaxation in fact lengthens the O-O distance $d_{O-O}$ and lowers the mass density $\rho$ [38,45,56,248,251]. With the measured $d_{O-O}$ of 2.965 Å [248] as input, segmental lengths of $d_H$ = 0.8406 Å and $d_L$ = 2.1126 Å are derived, which correspond to a 0.75 g·cm$^{-3}$ skin mass density [4]. In comparison, the MD derivatives in Figure 23 are 0.82 g·cm$^{-3}$ density compared to a density of 0.70 g·cm$^{-3}$ for a dimer. These values, 0.70–0.82 g·cm$^{-3}$, are much lower than 0.92 g·cm$^{-3}$ for bulk ice.

### 5.2.3 Cluster-size-dependence of $\Delta\omega_x$

Figure 24(a) shows the cluster-size-dependence of the calculated vibration spectra of $(H_2O)_N$ with respect to the ice-Ih phase. As expected, N reduction stiffens the $\omega_H$ from 3100 to 3650 cm$^{-1}$ and meanwhile softens the $\omega_L$ from 250 to 170 cm$^{-1}$ as the bulk water turns into dimers. The ∠O:H-O bending mode $\omega_{B1}$ (400–1000 cm$^{-1}$) shifts to a slightly lower value, but the ∠H-O-H libration mode $\omega_{B2}$ (≈1600 cm$^{-1}$) remains unchanged [263].

N-reduction-stiffened $\omega_H$ in Figure 24b is consistent with spectroscopic measurements (Figure 21a). For instance, reduction of the $(H_2O)_N$ cluster from N = 6 to 1 stiffens the $\omega_H$ from 3200 to 3650 cm$^{-1}$ [244]. The skin $\omega_H$ of 3450 cm$^{-1}$ corrsponds to an effective cluster size of N = 2–3. Indeed, molecular undercoordination shortens and stiffens the H-O bond, and lengthens and softens the O:H nonbond consistently, which confirms the proposal of O:H-O bond cooperativity and the BOLS notation.

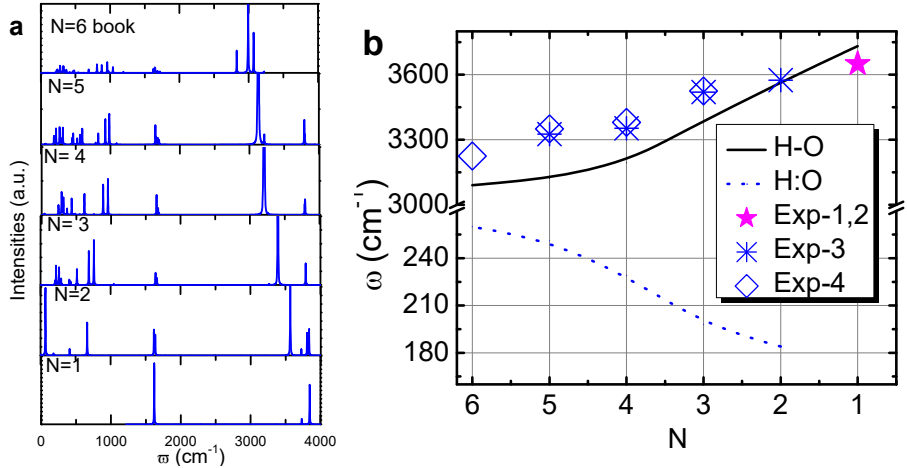

Figure 24. Cluster size dependence of (a) the $\omega$ relaxation in comparison with (b) measurements (scattered data) shown as Exp-1 [113], Exp-2 [244], Exp-3 [239], and Exp-4 [48] for (H$_2$O)$_N$ clusters. Measurements of the $\omega_L$ redshift have not been frequently carried out due to experimental sensitivity limitations. (Reprinted with permission from [3].)

5.2.4  Identical $\omega_H$ for the skins of water and ice

Figure 25 (a), (b) features particularly the RPS for ice in comparison to (c) the measured $\omega_H$ RPS for both water and ice [78]. The valleys of the RPS represent the bulk feature, while peaks show the skin attributes. Appropriate offset of the calculated RPS is necessary, as the MD code overestimates intra- and intermolecular interactions [5]. As expected, $\omega_L$ undergoes a redshift, while the $\omega_H$ undergoes a blueshift with three components. The $\omega_H$ blueshift results from the stiffening of the skin H-O bonds (S) and the free H-O radicals (R). The $\omega_L$ redshift arises from O-O repulsion and polarization. The polarization in turn screens and splits the intramolecular potential, which adds another $\omega_H$ peak (denoted P as polarization) with frequency being lower than that of the bulk valley (B).

Most strikingly, the measured RPS shows that the skins of water and ice share the same $\omega_H$ value of 3450 cm$^{-1}$, which indicates that the H-O bond in both skins is identical in length and energy, since $\omega_H \propto (E_H/d_H^2)^{1/2}$. The skin $\omega_L$ of ice may deviate from that of liquid water because of the extent of polarization, although experimental data is absent at this moment. Nevertheless, the skin $\omega_H$ stiffening agrees with the DFT-MD derivatives that the $\omega_H$ shifts from ≈ 3250 cm$^{-1}$ at 7 Å depth to ≈ 3500 cm$^{-1}$ of the 2 Å skin of liquid water [264]. Therefore, it is neither the case that an ice skin forms on water nor that a liquid skin covers ice. Rather, an identical supersolid skin covers both. The concept of supersolidity is adapted from the superfluidity of solid $^4$He at mK temperatures. The skins of $^4$He fragments are highly elastic and frictionless with repulsion between them when in motion [265].

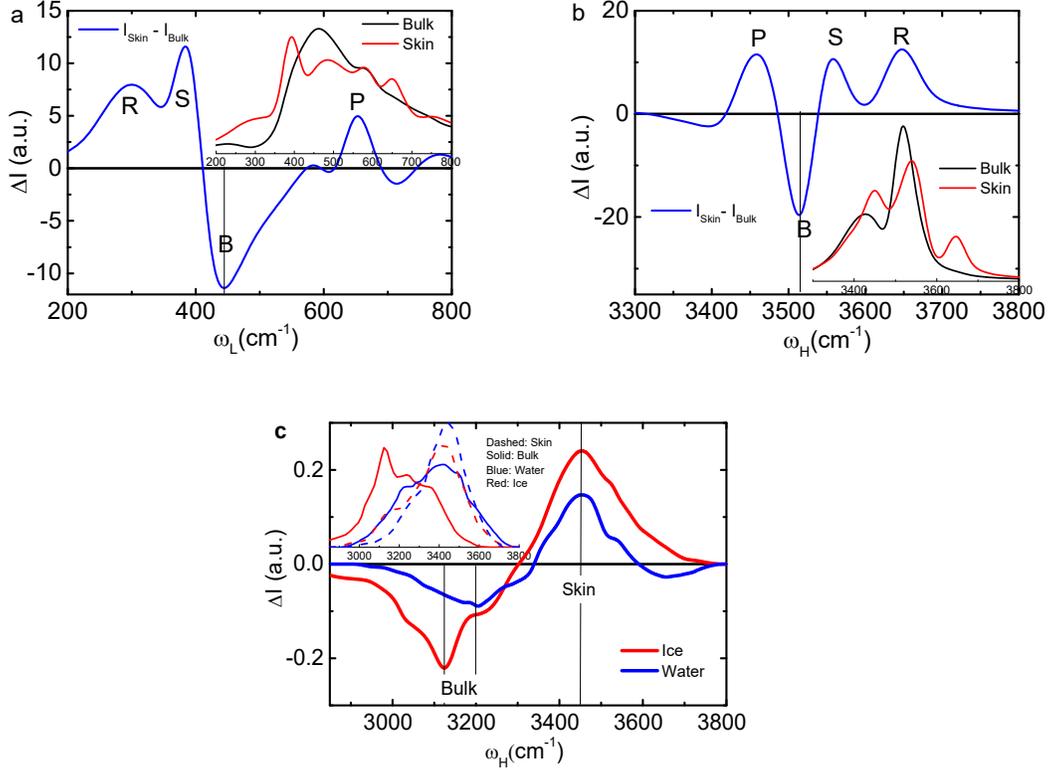

Figure 25. RPS of the MD-derived (a) $\omega_L$ and (b) $\omega_H$ of ice, and (c) the measured $\omega_H$ of water (at 25°C) and ice (at -20 and -15°C) [78]. The insets in (a) and (b) show the raw spectra of calculation. The broken lines of inset (c) were collected at 87°C and the solid lines were collected at 0°C with respect to the surface normal of ice (red) and water (blue). Calculations show that the $\omega_L$ undergoes a redshift, while the $\omega_H$, splitting into three, undergoes a blueshift. Features S corresponds to the skin H-O bond; R corresponds to the free H-O radicals; the P component arises from the screening and splitting of the crystal potential by the polarized nonbonding electrons. The skins of water and ice share the same $\omega_H$ of 3450 cm$^{-1}$. The peak intensity changes with the scattering from ice and water.

5.2.5   Skin electron entrapment versus H-O bond energy

Table 4 features the DFT-derived Mulliken charge accumulation at the skin and in the bulk of water. O increases its net charge from the bulk value of -0.616 to -0.652 e for the skin. The net charge of a water molecule increases from 0.022 to -0.024 e correspondingly.

Table 4. DFT-derived charge localization at the skin and in the bulk of ice and derivatives (in bold) based on the referenced data using Eq. (25). Negative sign represents net electron gain.

|  | Skin | Bulk | $(H_2O)_1$ | O atom |
| --- | --- | --- | --- | --- |
| $q_O$ | -0.652 | -0.616 | – | – |
| $q_H$ | 0.314 | 0.319 | – | – |
| Net $q$ of $H_2O$ | -0.024 | 0.022 | – | – |
| $E_{1s}$ (eV) [233-235] | 538.1 | 536.6 | 539.7 | **525.71** |
| $E_H$ (eV) | **4.52/4.66** | 3.97 [1] | 5.10 [106] | – |
| $T_m$ (K) | **311/320** | 273 | – | – |

The skin H-O bond energy $E_H$(Skin) and the atomic O 1s energy $E_{1s}(0)$, as listed in Table 4, can be estimated using the referenced data and the formulae in section 2.3 [122]:

$$\frac{\Delta E_{1s}(N)}{\Delta E_{1s}(\infty)} = \frac{E_{1s}(N) - E_{1s}(0)}{E_{1s}(\infty) - E_{1s}(0)} = \frac{E_H(N)}{E_H(\infty)} = \frac{T_C(N)}{T_C(\infty)} = \left(\frac{d_H}{d_{H0}}\right)^{-m}.$$

(25)

The $E_H$(Skin) = 3.97 × (538.1/536.6) = 4.52 eV is compatible with the value of 4.66 eV for breaking the H-O bond of $H_2O$ molecules deposited on a $TiO_2$ surface in less than a monolayer coverage using laser excitation [106]. The deviation $\Delta E_H$(Skin) = 0.14 eV (about 3%) arises mainly from molecular undercoordination in these two situations — one is the water skin and the other is the even less coordinated molecules on the $TiO_2$ surface, which indicates that a weak interaction may exist between water molecules and the hydrophobic $TiO_2$ surface; an air gap may be present.

With the known values of $(d_H, E_H)_{Skin}$ = (0.84 Å, 4.52 eV) and $(d_H, E_H)_{Bulk}$ = (1.0 Å, 3.97 eV) and the $E_H(1)$ = 5.10 eV, the bond nature index is estimated as $m$ = 0.744 and the $d_H(1)$ = 0.714 Å of a monomer. The densely and locally entrapped core electrons of the undercoordinated water molecules polarize in a dual-process the nonbonding electrons. Therefore, molecular clusters, surface skins, and ultrathin films manifest strong polarization, as shown in Figure 20. Polarization enhancement of the undercoordinated water molecules [18,266] arise not only from the O-O elongation but also from charge-gain according to DFT optimization in Table 4.

5.3    Skin thermo–mechano–dynamics
5.3.1  Melting point elevation

The $T_m$(Skin) is estimated based on the correlation between the $T_C(N)$ and the $\Delta E_{1s}(N)$ from Eq. (25):

$$\frac{T_C(Skin)}{T_C(\infty)} = \frac{T_m(Skin)}{273} = \frac{E_H(Skin)}{E_H(Bulk)} = \frac{4.52 \sim 4.66}{3.97},$$

which yields $T_m(Skin) = 273 \times (4.52 - 4.66)/3.97 = 311 - 320$ K. The skin of the bulk water melts at temperatures in the range 315 ± 5 K. It is not surprising that water skin performs like ice or glue at room temperature (298 K) and that the monolayer water melts at about 325 K [138]. The $T_m$ increases with the curvature of the skin. Therefore, molecular clusters, surface skins, and ultrathin films of water exhibit the attribute of superheating.

Defects also raise the temperature of melting ice. MD simulations suggested that freezing preferentially starts in the subsurface of water instead of the outermost layer, which remains ordered during freezing [267]. The subsurface accommodates better than the bulk the increase of volume connected with freezing. Furthermore, bulk melting is mediated by the formation of topological defects which preserve the coordination of the tetrahedral network. Such defect clusters form a defective region involving about 50 molecules with a surprisingly long lifetime [268]. These findings verify the BOLS-NEP expectations that the undercoordinated water molecules are indeed thermally stable. Therefore, a liquid layer never forms on ice [267] or surrounding defects [268].

5.3.2  Curvature-enhanced thermal stability
Water droplets encapsulated in hydrophobic nanopores [210] and point defects [267,268] are thermally even more stable than the bulk water even because of the undercoordinated molecules in the curved skin and the presence of the air gap at contacts. Curvature-enhanced $T_m$ explains the curvature-dependence of the critical temperature for transiting the initial contact angle of a droplet to zero. As shown in Figure 26, such transition happens at 185, 234, and 271°C for water droplets at the respective initial contact angles of 27.9°, 64.2° and 84.7° on quartz, sapphire and graphite substrates [269]. Likewise, a water droplet on a roughened Ag surface (with nanocolumnar structures) having a greater contact angle and higher curvature, freezes 62 s later than on a smooth Ag surface at -4°C [270].

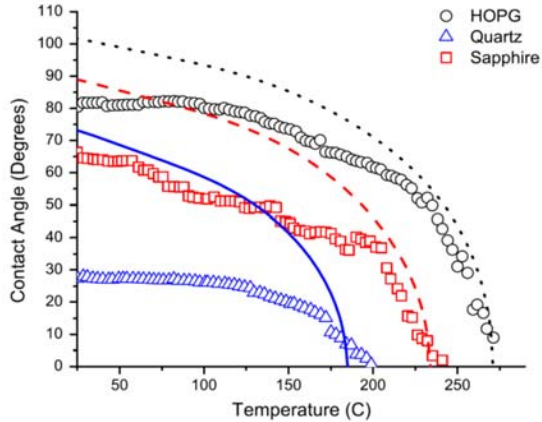

Figure 26. Evolution of water contact angle on quartz, sapphire and graphite as a function of temperature (°C). Lines are model fitting. (Reprinted with permission from [269].)

5.3.3  Viscoelasticity, repulsion, and hydrophobicity

The polarization of molecules enhances the skin repulsion and viscoelasticity. The high viscoelasticity and the high density of skin dipoles are essential to the hydrophobicity and lubricity at contacts [271]. According to the BOLS-NEP notation, the local energy densification stiffens the skin and the densely and tightly entrapped bonding charges polarize nonbonding electrons to form anchored skin dipoles [265].

Table 5 features the MD-derived thickness-dependent $\gamma$, $\eta_s$ and $\eta_v$ of ice films. Reducing the number of molecular layers increases them all. The O:H-O cooperative relaxation and associated entrapment and polarization enhances the surface tension to reach the value of 73.6 mN/m for five layers, which approaches the measured value of 72 mN/m for water skin at 25°C. Generally, the viscosity of water reaches its maximum at a temperature around the melting point $T_m$ [272].

Table 5. Thickness-dependent surface tension $\gamma$ and viscosity $\eta$.

| Number of layers | 15 | 8 | 5 |
|---|---|---|---|
| $\gamma$ (mN/m) | 31.5 | 55.2 | 73.6 |
| $\eta_s$ ($10^{-2}$mN·s/m$^2$) | 0.007 | 0.012 | 0.019 |
| $\eta_v$ ($10^{-2}$mN·s/m$^2$) | 0.027 | 0.029 | 0.032 |

The negative charge gain and the nonbonding electron polarization provide electrostatic repulsive forces lubricating ice. Measurements, shown in Figure 27, in fact verified the presence of the repulsive forces between a hydrated mica substrate and the tungsten contacts at 24°C [212]. Such repulsive interactions appear at 20% – 45% relative humidity (RH). The repulsion corresponds to an elastic modulus of 6.7 GPa. Monolayer ice also forms on a graphite surface at 25% RH and 25°C [273]. These findings and the present numerical derivatives evidence the presence of the supersolidity with repulsive forces because of bonding charge densification, surface polarization and $T_m$ elevation.

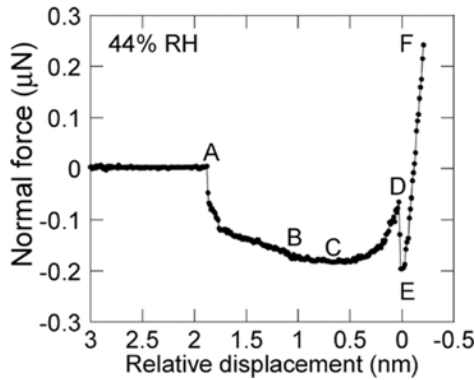

Figure 27. Normal force profiles between mica and tungsten tip at 44% RH. Point A is the initiation of water nucleation and condensation; B and C are the formation of a complete water bridge; D is the maximum attractive force before the tip–substrate contact; E denotes the sudden drop of force; and F indicates the tip–substrate contact repulsive force. (Reprinted with permission from [212].)

5.4    Skin supersolidity slipperizing ice and toughening water skin

As justified above, the skin of water and ice form an extraordinary supersolid phase that is elastic [78], hydrophobic [139,223], polarized [207,229] and thermally stable [138], with densely entrapped bonding electrons [231,233-235] and ultra-low-density [248]. The fewer the molecular neighbors there are, the smaller the water molecule size is, the greater the molecular separation is, and therefore the greater the supersolidity will be.

The supersolid skin is responsible for the slipperiness of ice and the hydrophobicity and toughness of water skin. This understanding of the slipperiness of ice may extend to the superfluidity of $^4$He [265] and water droplet flow in carbon nanotubes [274]. Atomic undercoordination-induced local strain and the associated entrapment and polarization might rationalize $^4$He superfluidity — elastic and repulsive between locked dipoles at contacts. It is understandable now why the rate of the pressure-driven water flow through carbon nanotubes is orders higher in magnitude and faster than is predicted from conventional fluid-flow theory [275]. It is within expectation that the narrower the channel diameter is, the faster the flow of the fluid will be [274,276], because of the curvature-enhanced supersolidity of the water droplet.

The mechanism of slipperiness of ice is analogous to the self-lubrication of metal nitride [129,277] and oxide [278] skins with electron lone pairs coming into play. Nano-indentation measurements have revealed that the elastic recovery of TiCrN, GaAlN and $\alpha$-Al$_2$O$_3$ surfaces could reach 100% for a critical indentation frictional load (e.g., < 5 N for carbon nitride) at which the lone pair breaks. Figure 28 shows that both ice and nitrides share the comparatively low friction coefficients of 0.1. The polarization of the lone pairs slipperize the skins of nitrides and oxides.

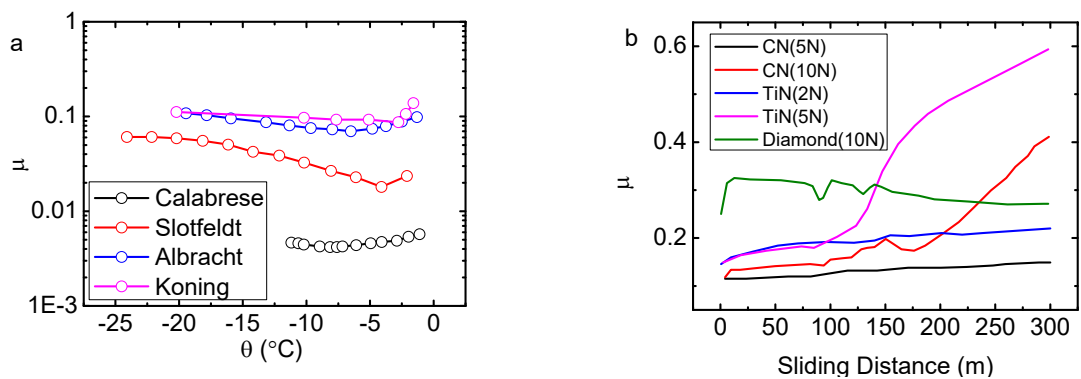

Figure 28. Friction coefficients of (a) ice and (b) nitrides under different loads in comparison to that of a diamond film. Lowering the operating temperature from the ambient (b) may reduce the coefficient. The abrupt increase of the coefficient indicates the presence of the critical load at which the lone pair nonbond breaks (Reprinted with permission from [60,277] and references therein.)

5.5   Summary

Undercoordination-induced O:H-O bond relaxation and the associated binding electron entrapment and the dual-process of nonbonding electron polarization clarify the anomalous behaviour of water molecules with fewer than four neighbours — in particular, the skin supersolidity of water and ice. Agreement between numerical calculations and experimental observations verified the following:

1) Undercoordination-induced O:H-O relaxation results in the supersolid phase that is elastic, hydrophobic, thermally more stable with density loss, which dictates the unusual behaviour of water molecules at the boundary of the O:H-O networks or in the nanoscale droplet.
2) H-O bond contraction densifies and entraps the core and bonding electrons; H-O bond stiffening shifts positively the O1s energy, the $\omega_H$ and the $T_m$ of molecular clusters, surface skins, and ultrathin films of water.
3) A dual-process of polarization makes the skins hydrophobic, viscoelastic, and frictionless.
4) Neither a liquid skin forms on ice nor a solid skin covers water; rather, a common supersolid skin covers both. The supersolid skin causes slippery ice and toughens water skin.
5) These understanding may extend to the superfluidity of $^4$He and the lubricity of water droplet flow in carbon nanotubes.

6   Thermal excitation: Ice floating, mass-density and phonon-frequency oscillation

- *Specific-heat disparity and inter-oxygen repulsion drive the mass-density and phonon-stiffness oscillation over the full temperature range, with the transition at 277, 258, and 80 K.*
- *In the liquid and solid phase, the $d_L$ contracts more than the $d_H$ elongates, resulting in the 'regular' process of cooling densification.*
- *During freezing, the $d_H$ contracts less than the $d_L$ elongates, leading to O-O elongation and density loss; at T < 80 K, both segments are conserved but the ∠O:H-O angle stretches, resulting in slight volume expansion.*
- *The O-O distance is longer in ice than in water, resulting in a lower density, causing ice to*

*float.*

6.1 Thermal anomalies of water ice

6.1.1 Density oscillation in the full temperature range

$H_2O$ density anomalies, in particular the floating of ice, continues to puzzle the community despite extensive investigation conducted over past decades [52,85,108,133,136,139,279-286]. When water freezes, its volume increases by up to 9% at atmospheric pressure. By contrast, the volume of liquid argon shrinks by 12% on freezing [176]. The $\rho(T)$ profiles for water droplets 1.4 nm [133] and 4.4 nm [134] in size (see Figure 29a) show that, in the liquid (I, bulk) phase and in the solid (III) phase, $H_2O$ exhibits the normal process of cooling densification at different rates; at the freezing transition phase (II), volume expansion occurs; at temperatures below 80 K (IV), ice volume again increases slightly [135]. The critical temperature for the lowest density is strongly droplet-size-dependent.

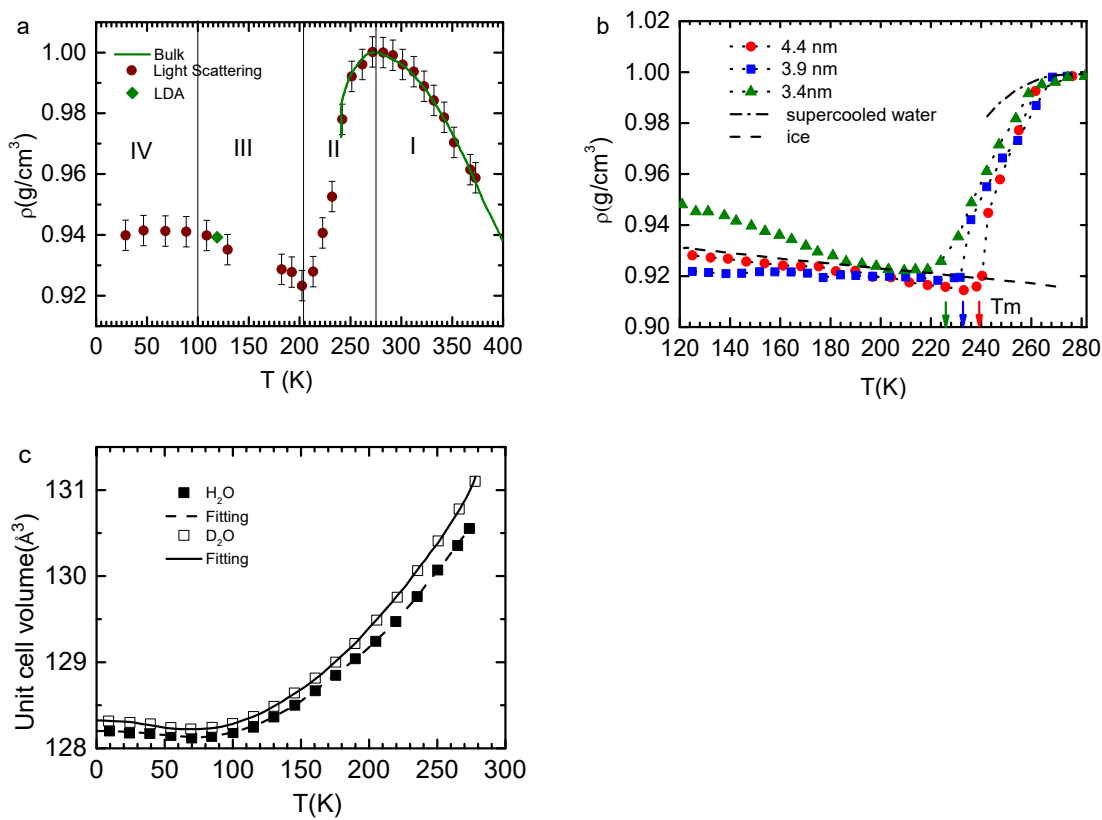

Figure 29. Density $\rho(T)$ profiles of water droplets measured using (a) Raman and FTIR (1.4 nm), and (b) small-angle X-ray scattering. Cooling densification takes place in the liquid (I: $T \geq 277$ K for bulk) and to the solid phase (III: $T \geq 80$ K with size-induced dispersion of the least-density temperature from 202 to 242 K) while cooling expansion occurs at the transition point (II: $273 \geq T \geq 205$ (244) K). (c) Slight volume expansion happens to $H_2O$ and $D_2O$ at $T \leq 80$ K (IV). (Reprinted with permission from [133-135].)

6.1.2 Available mechanisms for density anomalies

Theoretical studies on the physical anomalies of water ice have mainly focused on the density change in the freezing region in terms of supercooling. The mechanism behind the 'regular' process of cooling densification in the liquid and solid phases has attracted little attention. The following mechanisms have addressed freezing expansion:

1) The mixed-phase scheme [85,133,136,281-283,285,286] suggests that a competition between the 'ice-like' nanoscale fragments or the ring- or chain-like low-density liquid (LDL), and the tetrahedrally structured high-density liquid (HDL) fragments dictates the volume expansion at freezing [90,283]. Cooling increases the proportion of the LDL phase, and then ice floats. The many-body electronic structure and the non-local vdW interactions could be possible forces driving volume expansion [109].
2) The monophase notation [20,52,108,279,280] explains that water contains a homogeneous, three-dimensional, tetrahedrally coordinated structured phase with thermal fluctuation that is not quite random [108,284]. Unlike the mixed-phase scheme, the monophase model attributes freezing expansion to the O:H-O bond relaxation in length and angle in a fixed but as yet unclear manner.
3) The linear correlation model [287] rationalizes the fact that the local density changes homogeneously with the length, angle and network topology of the O:H-O bond. O:H-O bond elongation is responsible for thermal expansion, while the angular distortion causes contraction. Therefore, the competition between the O:H-O bond angle and its length relaxation determines the density anomalies of water ice.
4) The model of two kinds of O:H-O bond [288,289] suggests that one kind of stronger and another kind of weaker O:H-O bond coexist randomly in the ratio of about 2:1. By introducing these two types of O:H-O bond, Tu and Fang [289] reproduced a number of the anomalies of water, particularly the thermodynamic properties in the supercooled state. They found that the exchange between the strong and the weak O:H-O bonds enhances the competition between the open and the collapsed structures of liquid water.

6.1.3    Phonon stiffness oscillation and O 1s thermal entrapment

Phonons of 'normal' materials undergo heat softening because of the thermal lengthening and softening of all bonds involved [191,192,240,290-294]. However, as shown in Figure 30, heating stiffens the stiffer $\omega_H$ phonons and softens the softer $\omega_L$ phonons of liquid water [25,113,295-299] and ice at temperatures above 80 K [194,297,300-302]. Free-standing water droplets in the supercooling state also demonstrates thermal $\omega_H$ stiffening in temperatures between -34.6 and 90.0°C [303]. Marechal [304] observed that thermal $\omega_H$ stiffening and $\omega_L$ softening happen simultaneously, not only in liquid $H_2O$ but also to liquid $D_2O$, despite an offset in the characteristic peak. However, the $\omega_H$ transition at 0°C from thermal stiffening to thermal softening, as shown in Figure 30d [297], has hitherto received little attention.

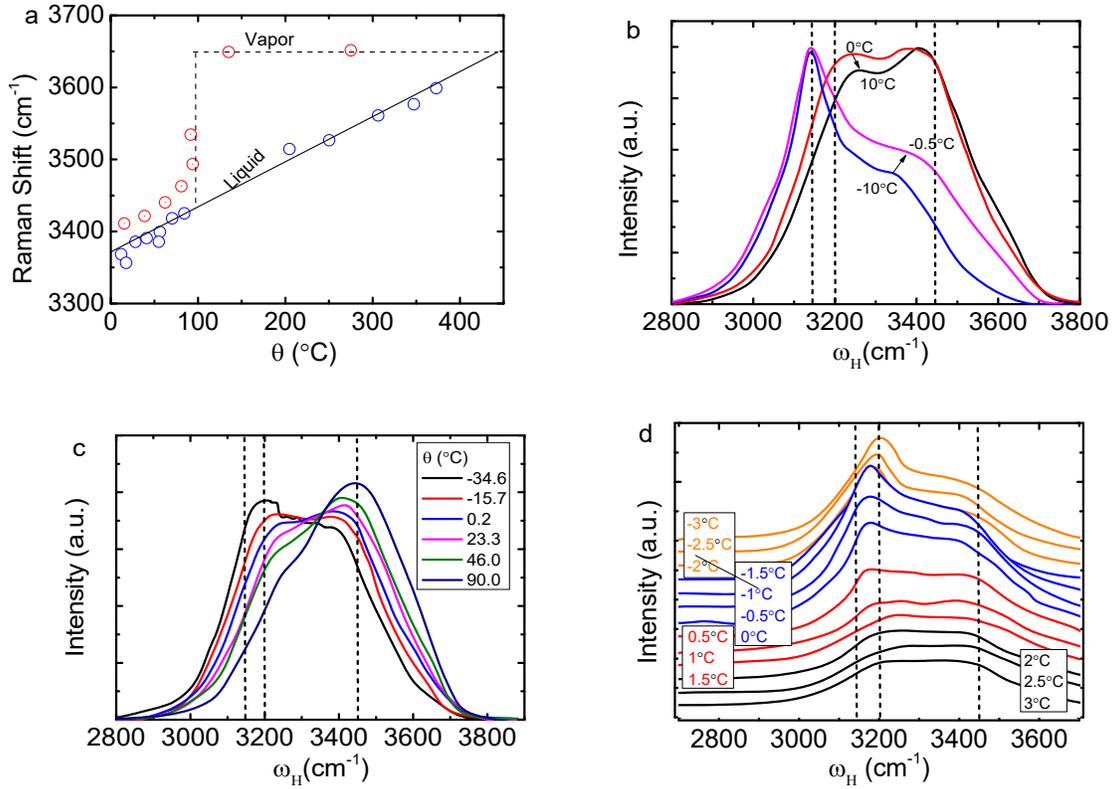

Figure 30. Thermal $\omega_H$ stiffening for bulk water in the temperature ranges from (a) 0 to 300°C, and (b) -10 to 10°C. (c) Thermal $\omega_H$ stiffening for supercooled water droplets in the range -34.6 to 90°C at ambient pressure. (d) The $\omega_H$ transitions at 0°C from thermal-stiffening to cooling-stiffening. The $\omega_H$ in (a) is unchanged in the vapor phase (half-filled circles) at 3650 cm$^{-1}$; empty circles show linear dependence of $\omega_H$ on temperature in the superheating liquid. (Reprinted with permission from [113,297,303,305].) Hatched vertical lines indicate $\omega_H$ at 3450 cm$^{-1}$ for the skin, 3200 cm$^{-1}$ for bulk water, and 3150 cm$^{-1}$ for bulk ice.

However, thermal annealing of low-density amorphous ice from 80 to 155 K softens $\omega_H$ from 3120 to 3080 cm$^{-1}$[306], which is counter to the trend of $\omega_H$ heating stiffening in ice-VIII crystals. Thermal relaxation increases the structural order of the amorphous state on more extended length scales as the average O-O distance becomes shorter with narrower distribution.

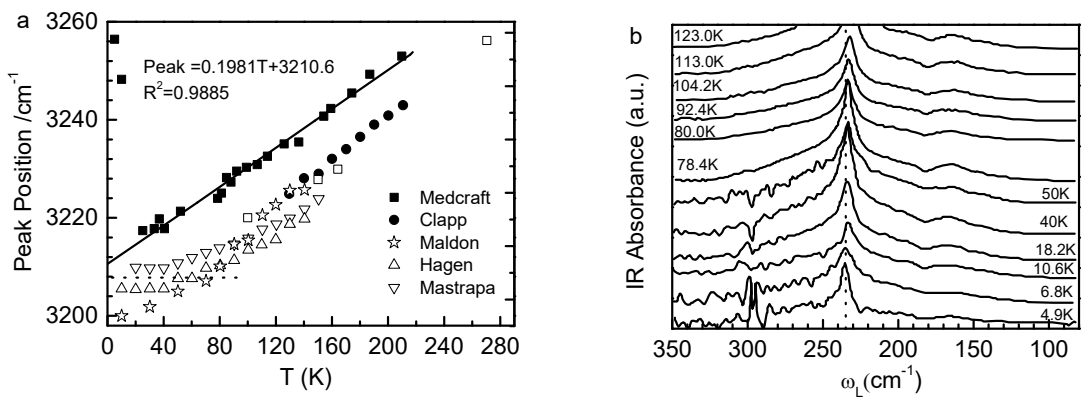

Figure 31. Insignificant shift of (a) the $\omega_H$ and (b) the $\omega_L$ at $T \leq 60$ K (reprinted with permission from [307,308] and references therein). This indicates that $\eta_L \cong \eta_H \cong 0$ almost silences the O:H-O bond in this temperature regime [5]. Broken lines guide viewing.

Figure 31 shows that both the $\omega_H$ and the $\omega_L$ approach almost an almost constant value at $T < 60$ K [307]. Using IR spectroscopy, Medcraft et al. [308] measured the size- and temperature-dependence of $\omega_L$ in the temperature range 4–190 K. They found that heating softened the $\omega_L$ at $T > 80$ K but the $\omega_L$ remains almost unchanged below 60 K. Earlier Raman spectroscopy had revealed that the $\omega_L$ for bulk ice drops monotonically with temperature down to 25 K with relatively lower resolution [309].

Heating also deepens the O 1s energy (see Figure 32) towards that of gaseous molecules [310,311]. The O 1s thermal entrapment has always been explained as the consequence of the mixed-phase configuration, that is, ordered tetrahedral and distorted O:H-O bonded networks [51,90].

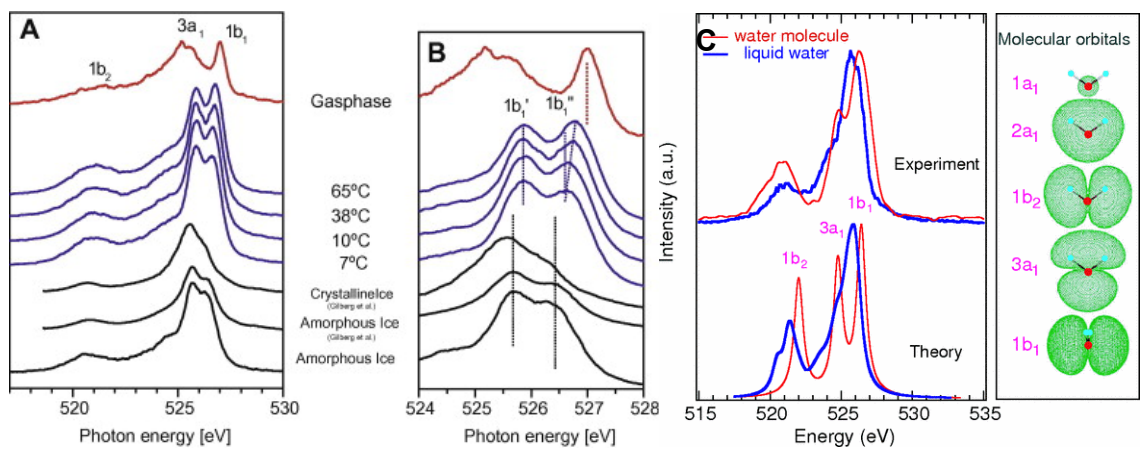

Figure 32. O 1s spectra of vapor, liquid water, and amorphous and crystalline ice at different temperatures, with an energy scale displaying the full spectrum (A) or only the $1b_1$ orbital (B). The peak ($1b_1$) splits into a double of peaks ($1b_1'$ and $1b_1''$ in B). Panel C shows the O 1s orbital energies of molecules and liquid water; the side panel illustrates the molecular orbitals. (Reprinted with permission from [310,311].)

6.1.4  Emerging challenges

Phonon and electron spectroscopic and diffraction studies to date have mainly focused on the structural phases composition and transition under compression and thermal excitation [14,183,194,300]. Determination of the following attributes is beyond the scope of available models focusing on the phase composition in the supercooling states. Solving the emerging challenges from the perspective of O:H-O bond relaxation could be possible:

1) Thermal oscillation dynamics of $\omega_x$ and $\rho$ in the full temperature range
2) The 'regular' process of cooling densification in the liquid and solid phases
3) Slight density drop and $\omega_x$ conservation at extremely low temperatures
4) O 1s thermal entrapment and phonon cooperative relaxation
5) Size dispersion of the extreme-density temperatures for water droplet — superheating and supercooling in the transition phase

6.2  Bond–electron–phonon relaxation
6.2.1  Bond angle–length relaxation and density oscillation

Figure 33 features the MD-derived relaxation of the segmental lengths $d_x$, the ∠O:H-O angle $\theta$, snapshots of the MD trajectory, and the O-O distance as a function of temperature. Figure 33a shows that the shortening of the master segments (denoted with arrows) is always coupled with a lengthening of the slaves during cooling. In the liquid region I and in the solid region III, the O:H bond contracts more than the H-O bond elongates, resulting in a net loss of the O-O length. Thus, cooling-driven densification of $H_2O$ takes place in both the liquid and the solid phases. This mechanism differs completely from that conventionally adopted for the standard cooling densification of other 'normal' materials in which only one kind of chemical bond is involved [130].

In contrast, in the transition phase II, the master and the slave exchange roles. The H-O bond contraction is less than the O:H bond expansion, producing a net gain in the O-O length and resulting in density loss. Calculations reveal no region IV below 80 K as observed, due to the limitation of the algorithm used. Encouragingly, the entire process of relaxation follows the modeling prediction based on the specific-heat disparity, as discussed in section 2.2.3.

Figure 33b shows that widening of the angle $\theta$ contributes to volume change. In the liquid phase I, the mean $\theta$ value of 160° remains almost constant. However, the snapshots of the MD trajectory in Figure 33c and the MD video in [5] show that the V-shaped H-O-H motifs remain intact at 300 K over the entire duration of recording. This configuration is accompanied by large fluctuations of $\theta$ and $d_L$ flashing in this regime, which indicates the preference for a tetrahedrally-coordinated structure of water molecules [107], even for a single molecule at the low temperature of 5 K [96].

In region II, cooling widens $\theta$ from 160° to 167°, which contributes a maximum of +1.75% to the O:H-O bond elongation and about 5% to the volume expansion. In phase III, $\theta$ increases from 167° to 174° and this trend results in a maximal value of -2.76% to the volume contraction in ice. An extrapolation of the θ widening in Figure 33b results in the O—O distance lengthening in region IV, which explains the slight drop in density and the steady $\omega_L$ ($d_L$ and $E_L$) observed at temperatures below 80 K [308,309]. Therefore, the angle relaxation contributes only positively to the mass density loss in phase II without apparent influence on relevant physical properties such as $T_C$, $E_{1s}$, $\omega_x$ etc., although molecular polarization may mediate the angle.

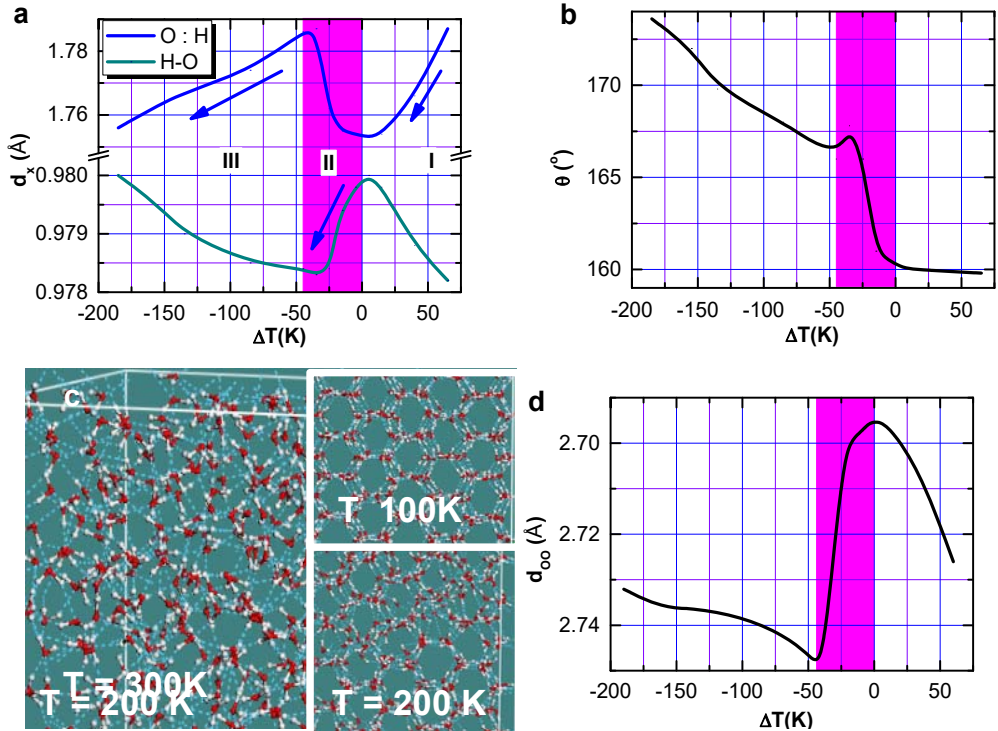

Figure 33. MD-derived $d_x$ and ∠O:H-O angle $\theta$ relaxation dynamics. (a) Arrows indicate the masters that contract in different regions accompanied with elongation of the other parts as a slave. $\Delta T = T - T_{max}$ where $T_{max} = 277$ K is the maximal density temperature for bulk. (b) Cooling widening of $\theta$ contributes positively to freezing expansion and negatively to cooling densification in the solid phase while it remains almost constant in liquid. (c) The snapshots of MD trajectory show that the V-shaped $H_2O$ motifs remain intact at 300 K because of the stronger H-O bonding (3.97 eV/bond). (d) The calculated O-O distance oscillation agrees with the density change of water ice [133]. (Reprinted with permission from [5].)

The MD-video in [5] shows that, in the liquid phase, the H and the O attract each other in the O:H, but the O-O repulsion prevents this occurrence. The intact H-O-H motifs move ceaselessly like a complex pendulum because of the high fluctuation and frequent switching the O:H interaction on and off.

The evolution of the O-O distance shown in Figure 33d agrees well in trend with the measured density evolution in the relevant temperate range [133]. In ice, the O-O distance is always longer than in water — hence ice floats, without necessary involvement of the mixed-phase. Therefore, the entire process of density oscillation arises from O:H-O bond relaxation subject to the specific-heat disparity.

6.2.2    Phonon frequency $\omega_x$ thermal oscillation

As expected, the Raman spectra of water cooled using programmed liquid nitrogen, in Figure 34, show three regions: $T > 273$ K (I), $273 \geq T \geq 258$ K (II), and $T < 258$ K (III) [5]:

1)  In liquid phase I, $T \geq 273$ K, cooling stiffens $\omega_L$ abruptly from 75 to 220 cm$^{-1}$ and softens $\omega_H$

from 3200 to 3140 cm$^{-1}$ with indication of ice forming at 273 K. The cooperative $\omega_x$ shift indicates that cooling shortens and stiffens the O:H bond but lengthens and softens the H-O bond in the liquid phase, which confirms again that the O:H bond cooling contraction dominates O:H-O relaxation in liquid water, as predicted by calculations.

2) In the freezing phase II, 273–258 K, the situation reverses. Cooling stiffens $\omega_H$ from 3140 to 3150 cm$^{-1}$ and softens $\omega_L$ from 220 to 215 cm$^{-1}$ (see the shaded areas). Consistent with the Raman $\omega_H$ shift measured at temperatures around 273 K [301,305], the cooperative shift of $\omega_x$ confirms the switching of the master and the slave roles of the O:H and H-O during freezing; H-O contraction dominates.

3) In the solid phase III, T ≤ 258 K, the master-slave role reverts to its behavior in the liquid region, albeit with a different relaxation rate. Cooling from 258 to 98 K stiffens $\omega_L$ from 215 to 230 cm$^{-1}$ and softens $\omega_H$ from 3150 to 3100 cm$^{-1}$. Other supplementary peaks at about 300 and 3450 cm$^{-1}$ change insignificantly; the skin $\omega_H$ of about 3450 cm$^{-1}$ in water and ice is indeed thermally insensitive. The cooling softening of $\omega_H$ agrees with that measured using IR spectroscopy of ice clusters of 8–150 nm size [307]. When the temperature drops from 209 to 30 K, $\omega_H$ shifts from 3253 to 3218 cm$^{-1}$.

4) Figure 31 shows that both $\omega_H$ and $\omega_L$ remain almost constant at $T < 70$ K, which indicates that neither the length nor the energy of the two segments changes with temperature because of their extremely low specific heat (Figure 9).

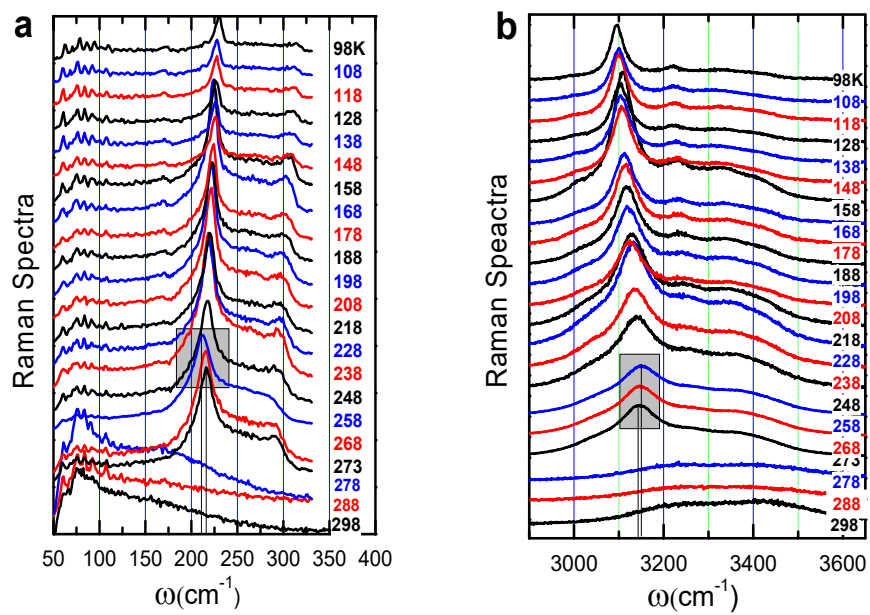

Figure 34. Temperature-dependent Raman shifts of (a) $\omega_L$ < 300 cm$^{-1}$ and (b) $\omega_H$ > 3000 cm$^{-1}$ in the temperature regions of $T > 273$ K, $273 \geq T \geq 258$ K, and $T < 258$ K, confirming the expected O:H-O length and stiffness cooperative oscillation over the full temperature range (reprinted with permission from [5]).

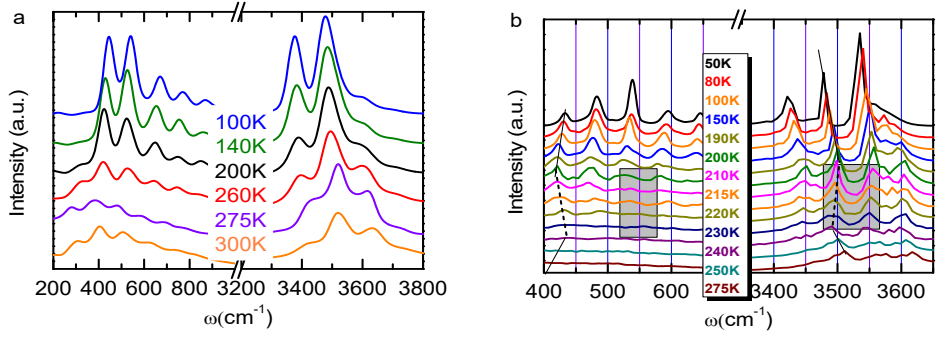

Figure 35. Temperature-dependent power spectra of $H_2O$. (a) Splitting of the high-frequency peaks at 260 K indicates the transition from water to ice at 200–260 K. (b) Phonon oscillation (indicated with hatched lines) holds the same trend as that of Raman measurements in Figure 34.

Figure 35 shows the *T*-dependent power spectra of $H_2O$ derived from MD calculations. The splitting of the high-frequency peaks at 260 K indicates the transition from water to ice at 200–260 K. The three-region phonon thermal oscillation is the same as the measurements in Figure 34. Figure 36 compares the measured and the calculated phonon thermal relaxation dynamics. As expected, $\omega_L$ stiffening (softening) always couples with $\omega_H$ softening (stiffening) in all three regions. Offsets of the calculated $\omega_L$ by -200 cm$^{-1}$ and $\omega_H$ by -400 cm$^{-1}$ compared to experiments suggest the presence of artifacts in the MD algorithm that deal inadequately with the ultra-short-range interactions.

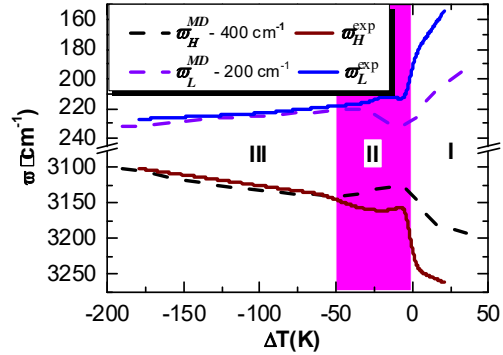

Figure 36. Comparison of the measured (solid lines) and the calculated (broken lines) phonon relaxation dynamics. Indicated 200/400 cm$^{-1}$ offsets of the calculated $\omega_x$ match observations. (Reprinted with permission from [5].)

It is now clear why heating softens the $\omega_H$ of amorphous ice [306] rather than stiffening it, as occurs in crystalline ice. The $\omega_H$ redshift indicates H-O bond lengthening. Molecular undercoordination shortens the H-O bond, which is distributes randomly in the amorphous phase. Annealing removes the defect and relaxes the $\omega_H$ towards crystallization. Therefore, $\omega_H$ redshift occurs in amorphous ice upon annealing, which is within the BOLS expectation.

### 6.2.3  O 1s energy shifting versus $\omega_H$ stiffening

The correlation $(d_H \Delta\omega_H)^2 \cong \Delta E_{1s}$ in Eq. (21) indicates that both the $\Delta E_{1s}$ and the $\Delta\omega_H$ shift are always in the same direction, at different rates, when the specimen is excited. Therefore, the $1b_1''$ peak corresponds to the skin $\omega_H$ at 3450 cm$^{-1}$, and 1b$_1'$ to the bulk $\omega_H$ at 3200 cm$^{-1}$ for water (see Figure 32). The O 1s goes deeper, and the $\omega_H$ shift is consistently higher at heating, because heating shortens and stiffens the H-O bond. It is expected that the $\Delta E_{1s}$ also undergoes thermal oscillation but its measurement in ultra-high vacuum is very difficult.

6.3   Isotope effect on $\omega_x$ thermal relaxation

Figure 37 shows that the isotope (D) has two effects on the IR spectrum of ordinary H$_2$O [304]. One is the intensity attenuation of all peaks and the other is the general phonon softening [304]. However, $\omega_x$ maintains the shift trend due to heating — $\omega_H$ stiffening and $\omega_L$ softening. The mechanism of this isotope effect remains unclear.

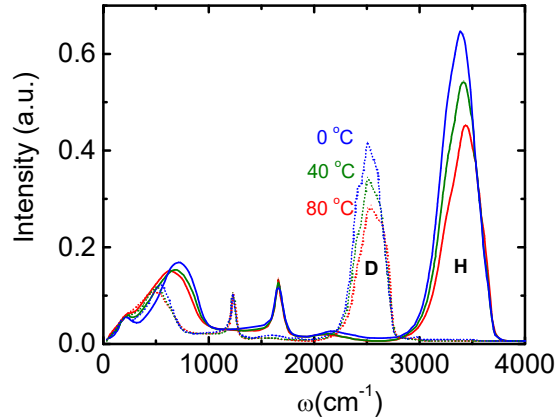

Figure 37. IR spectra of 1 μm-thick ordinary water (H) and heavy water (D) reveal that isotope D attenuates the intensity and softens all phonons of the H$_2$O in general, although the thermal stiffening of $\omega_H$ and thermal softening of $\omega_L$ remain. (Reprinted with permission from [304].)

This puzzle may be explored from the perspective of effective mass reduction. The isotope contributes only to reducing the $\mu(m_1,m_2) = m_1 m_2 / (m_1 + m_2)$ in the expression $\omega_x \propto (E_x/\mu_x)^{1/2}/d_x$. Considering the mass difference, both vibration modes shift their relative frequencies compared to ordinary water in the following manner:

$$\frac{\Delta\omega_{xH}}{\Delta\omega_{xD}} \cong \left(\frac{\mu_D(m_1,m_2)}{\mu_H}\right)^{1/2} = \begin{cases} \left[\dfrac{\mu_D(2,16)}{\mu_H(1,16)}\right]^{1/2} = (17/9)^{1/2} = 1.374 \quad (\omega_H) \\ \left[\dfrac{\mu_D(20,20)}{\mu_H(16,16)}\right]^{1/2} = (5/4)^{1/2} = 1.180 \quad (\omega_L) \end{cases},$$

(26)

where, for intramolecular H-O vibration, $m_1$ is the mass of H (1 atomic unit) or D (2 atomic units); and $m_2$ is the mass of O (16 units); and for intermolecular H$_2$O:H$_2$O vibration, $m_1 = m_2$ is the mass of 2H + O (18 units) or 2D + O (20 units). Measurements shown in Figure 37 yield the following:

$$\frac{\Delta \omega_{xH}}{\Delta \omega_{xD}} \approx \begin{cases} 3400/2500 = 1.36 & (\omega_H) \\ 1620/1200 = 1.35 & (\omega_B) \\ 750/500 = 1.50 & (\omega_L) \end{cases}.$$

(27)

The difference between the numerical derivatives in Eq. (26) and the measurements in Eq. (27) arises mainly from Coulomb coupling, particularly for $\omega_L$. It is thus justifiable to adopt first-order approximation as helpful for describing the isotopic effect on the phonon relaxation dynamics of $\omega_x$. Therefore, the addition of the isotope softens all the phonons by mediating the effective mass of the coupled oscillators in addition to the quantum effect that may play some role. The peak intensities in the isotope are also lower because of the enhanced scattering by the low-frequency vibrations.

6.4   Summary

Consistency in the four-region oscillation of $\rho$, $d_x$, $\omega_x$ and the proposed specific-heat evidence that the coupled O:H-O bond oscillators describe adequately the true situation of water ice when cooling. Agreement between calculations and the measurements verify the following:

1) Inter-oxygen repulsion and the segmental specific-heat disparity of the O:H-O bond govern the change in the angle, length and stiffness of the segmented O:H-O bond, and the oscillation of the mass density and the phonon-frequency of water ice over the full temperature range.
2) The segment with relatively lower specific heat contracts and drives the O:H-O bond cooling relaxation. Cooling stretching of the O:H-O angle contributes positively to volume expansion in the freezing phase but it contributes negatively to cooling densification in the solid phase. Angle relaxation has no direct influence on the physical properties, with the exception of mass density.
3) In the liquid and solid phases, the O:H bond contracts more than the H-O bond elongates, resulting in the cooling densification of water and ice. This mechanism is completely different from the process experienced by other 'normal' materials when only one type of chemical bond is involved.
4) In the freezing transition phase, H-O bond contracts less than the O:H bond lengthens, resulting in volume expansion during freezing. Stretching of the O:H-O bond angle lowers the density slightly at $T < 80$ K as the length and energy of the O:H-O are conserved.
5) The O-O distance is larger in ice than it is in water, and therefore ice floats.

7   Multiple fields coupling

- *Stimuli jointly relax the O:H-O bond in a superposition manner.*
- *Compression enhances the effect of liquid and solid cooling.*
- *Molecular undercoordination has the opposite effect of compression.*
- *Molecular undercoordination disperses the extreme-density temperatures by specific-heat modulation through $\omega_x$ cooperative relaxation — supercooling and superheating coexist.*

7.1   Undercoordination enhances the effect of liquid and solid heating

Cluster size reduction lowers the skin molecular CN and raises the fraction of the undercoordinated molecules. Undercoordination enhances the effect of liquid and solid heating on $d_H$ contraction. Medcraft et al. [307] confirmed this trend by examining the joint effects of size reduction and thermal excitation on the $\omega_H$ frequencies of ice nanoparticles 3–200 nm in diameter over a temperature range of 5–209 K. They found that reducing particle size below 5 nm stiffens the $\omega_H$ by some 40 cm$^{-1}$. The size effect is not apparent for particles larger than 8 nm because of the proportion of undercoordinated skin molecules lost. They also found that the peak $\omega_H$ shifts up by 35 cm$^{-1}$ from 3218 cm$^{-1}$ at 30 K to 3253 cm$^{-1}$ at 209 K.

MD calculations [263] shown in Figure 38 also confirm this coupling effect. The $\omega_H$ of molecules at the polymer proxy undergoes a further 35 cm$^{-1}$ blueshift at 310 K compared to bulk water at the same temperature; instead, the $\omega_L$ of bulk water undergoes a redshift upon heating. In fact, cluster size reduction lengthens the O:H bond and softens $\omega_L$; heating enhances this size trend on $\omega_L$ softening. The joint effect results in the MD-derived trends in Figure 38. These observations confirm that size reduction and heating have the same effect on $\omega_H$ stiffening and $\omega_L$ softening.

However, at $T < 60$ K, the $\omega_x$ shows almost no change [307,308], except for a slight increase in volume [135] if the particle size remains unchanged. Although O:H-O angle stretching increases the volume, the specific heat $\eta_x \approx 0$ relaxes neither the length nor the stiffness of the O:H or the H-O bond, conserving the $\omega_x$.

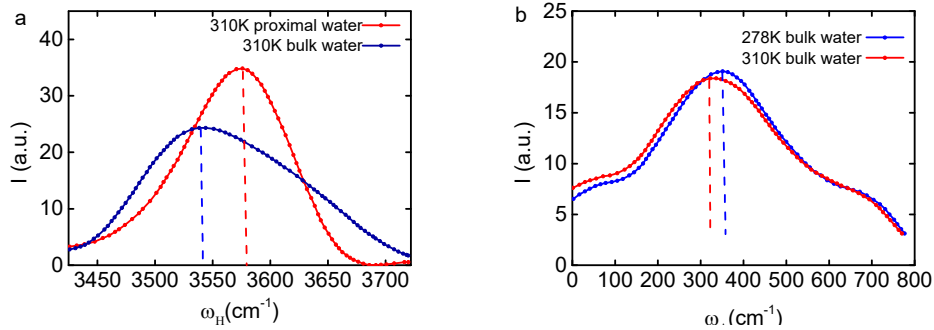

Figure 38. (a) Cluster size reduction (proximal) stiffens $\omega_H$ at the same temperature; and (b) heating softens the $\omega_L$ of bulk water (reprinted with permission from [263]) because both molecular undercoordination and heating shorten and stiffen the H-O bond, and lengthen and soften the O:H bond.

7.2   Compression has an opposite effect to undercoordination

Figure 39 shows the joint effect of undercoordination and compression on $\Delta E_{1s}$ and the valence band shift for water clusters of different sizes [46]. Except for the O 1s peak at 539.7 eV for gaseous molecules, size growth and compression jointly shift the O 1s energy from 539.7 to 538.2 eV towards the component centered at 538.1 eV for the skin of bulk water [233,234]. UPS reveals that the entire valence band of a molecule subjected to 7.5 kPa pressure shifts up and expands in width when the cluster size grows [46].

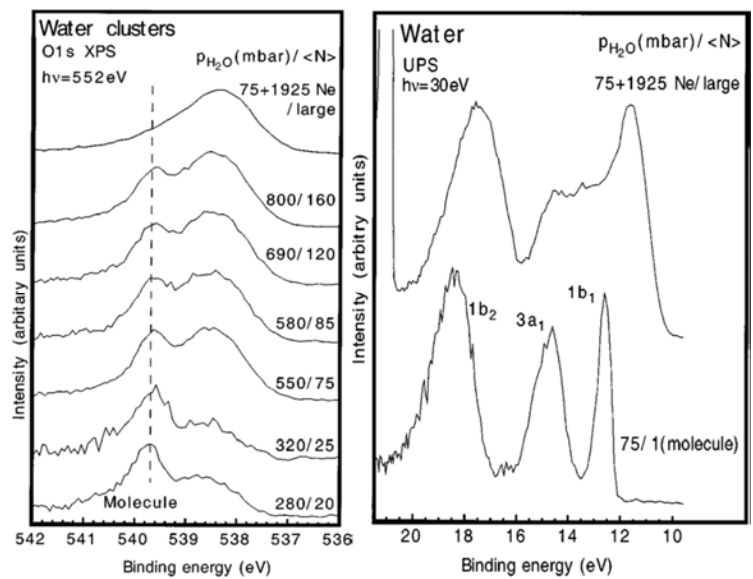

Figure 39. Joint effect of compression and undercoordination on $\Delta E_{1s}$ (left) and the valence band (right) for water clusters (10 mbar = 1 kPa). The vertical broken line denotes the $E_{1s}$ at 539.7 eV for gaseous molecules. The $1b_1$, $1b_2$, and $3a_1$ orbitals of water molecules (monomer) are shown. Compression enhances the effect of size growth in raising the $E_{1s}$ and the valence band. (Reprinted with permission from [46].)

The coupling effect of compression and size growth on the O 1s and valence band follows the BOLS notation — that is, the amount of energy shift is proportional to $E_H$. Compression softens but molecular undercoordination stiffens the H-O bond. Therefore, size growth enhances the effect of compression on the binding energy shift in all energy bands.

Systematic studies [182,186] have also revealed that cooling enhances the effect of compression on the structure phase transition and dipole moment of ice. In the solid phase, both cooling and compression shorten the O:H bond and lengthen the H-O bond.

7.3     Size dispersivity of extreme-density temperatures

One important fact is that the extreme-density temperatures change with droplet size. The least-density temperature varies from 205 to 242 to 258 K when the water droplet increases in size from 1.4 [133] to 4.4 nm [134] and to the bulk water [5] (see Figure 29). The melting temperature of the skin is as high as 315 K. Much attention has been paid to supercooling related to liquid-freezing or superheating related to ice-melting without knowing why they happen and what the correlation is between superheating and supercooling.

As discussed in section 2.3, size reduction increases the curvature and the fraction of undercoordinated molecules in a droplet, which stiffens the H-O bond and softens the O:H nonbond. Size reduction raises the $\omega_H(\theta_{DH})$ and lowers the $\omega_L(\theta_{DL})$. The droplet-size-induced $\theta_{Dx}(\omega_x)$ relaxation mediates the specific heat and hence disperses the extreme-density temperatures. Reducing the droplet size stretches the $\eta_H$ outwardly and compresses the $\eta_L$ inwardly along the $T$-axis (see Figure 9), which results in supercooling at freezing and superheating at melting. Thus it is not surprising that the extreme-density temperatures of droplets change with droplet size.

8     Potential paths for the O:H-O bond at relaxation

- *Lagrangian dynamics provides the best means solving the motion of the coupled O:H-O oscillators with particularly asymmetrical and short-range interactions.*
- *Solution then enables mapping of the potential paths for the O:H-O bond at relaxation by transforming the ($d_x$, $\omega_x$) into the ($k_x$, $E_x$) at each state of quasi-equilibrium.*
- *Both oxygen atoms move in the same direction along the potential paths by different amounts under excitation of cooling, compressing, and clustering.*
- *The Lagrangian derived CN trend agrees with measured $E_H$ of 3.97, 4.66, and 5.10 eV for bulk, skin, and monomer of water.*

8.1     Lagrangian oscillating dynamics
8.1.1   Harmonic approximation of the coupled oscillator

The potential energy $V$ of the coupled O:H-O oscillators contains 3 terms (Eq. (2) and Figure 6b):

$$\begin{cases} V_L(r_L) = V_L(d_{L0} - u_L) & (vdW-like) \\ V_H(r_H) = V_H(d_{H0} + u_H) & (Exchange) \\ V_C(r_C) = V_C(d_{C0} - u_L + u_H) = V_C(d_C - u_C) & (C-repulsion) \end{cases} \quad (28)$$

where $d_{C0} = d_{L0} + d_{H0}$ is the distance between adjacent oxygen ions at equilibrium without contribution of Coulomb repulsion; $d_C = d_L + d_H$ denotes this distance at quasi-equilibrium with involvement of the repulsion; the displacement $u_C = u_L + \Delta_L - u_H + \Delta_H$ is the change of distance between neighboring oxygen ions at quasi-equilibrium; $\Delta_x$ is the dislocation caused by repulsion. Displacements $u_L$ and $u_H$ take opposite signs because the O:H and H-O dislocate in the same direction [2]. A harmonic approximation of the potentials at each quasi-equilibrium site, by omitting the higher-order terms in their Taylor series, yields:

$$\begin{aligned} V &= V_L(r_L) + V_H(r_H) + V_C(r_C) \\ &= \sum_n \left\{ \frac{d^n V_L}{n! dr_L^n}\bigg|_{d_{L0}} (-u_L)^n + \frac{d^n V_H}{n! dr_H^n}\bigg|_{d_{H0}} (u_H)^n + \frac{d^n V_C}{n! dr_C^n}\bigg|_{d_C} (-u_C)^n \right\}, \\ &\approx \left[ V_L(d_{L0}) + V_H(d_{H0}) + V_C(d_C) \right] - V'_C u_C + \frac{1}{2}\left[ k_L u_L^2 + k_H u_H^2 + k_C u_C^2 \right] \end{aligned}$$
(29)

where $V_x(d_{x0})$, commonly denoted $E_{x0}$, is the potential well depth ($n = 0$ terms) of the respective segment at equilibrium. Noting that the Coulomb potential never reaches equilibrium, and that the repulsion force is always positive, the other potentials are expanded at their quasi-equilibrium points in the Taylor series. As will be shown shortly, an on-site harmonic approximation ensures sufficient accuracy of the derived potential paths for O:H-O relaxation without high-order approximation.

In the Taylor series, the $n = 1$ term equals zero for the L and the H segment potentials at ideal equilibrium without Coulomb repulsion, $V'_x(d_x) = 0$. At quasi-equilibrium, the sum of $V'_x(d_x) + V'_C(d_C) = 0$, or $V''_x \cdot u_x + V''_C \cdot u_C = 0$, due to the presence of Coulomb repulsion. Here, $V_C' = dV_C/dr_C |_{d_C}$ denotes the first-order derivative at the quasi-equilibrium position. Terms of $n = 2$, or the curvatures of the respective potentials, denote the force constants, i.e., $k_x = V'' = d^2V_x/dr_x^2 |_{d_{x0}}$ for the harmonic oscillators.

Substituting Eq. (28) into Lagrangian Eq. (12) yields the O:H-O bond oscillating dynamics, with $f_P$ being the non-conservative force due to compression [6]:

$$\begin{cases} m_L \dfrac{d^2 u_L}{dt^2} + (k_L + k_C) u_L - k_C u_H + k_C (\Delta_L - \Delta_H) - V_C' - f_P = 0 \\ m_H \dfrac{d^2 u_H}{dt^2} + (k_H + k_C) u_H - k_C u_L - k_C (\Delta_L - \Delta_H) + V_C' + f_P = 0 \end{cases}$$

(30)

### 8.1.2 General solution

A Laplace transformation of (30) yields the general solution,

$$\begin{cases} u_L = \dfrac{A_L}{\gamma_L} \sin \gamma_L t + \dfrac{B_L}{\gamma_H} \sin \gamma_H t \\ u_H = \dfrac{A_H}{\gamma_L} \sin \gamma_L t + \dfrac{B_H}{\gamma_H} \sin \gamma_H t \end{cases}$$

(31)

The vibration angular frequency $\gamma_x$ depends on the force constants and the reduced masses of the oscillators. The $\gamma_x$ and its combination with $A_x$ is the amplitude of the respective oscillator. This general solution indicates that the O:H and the H–O segments share the same form of eigenvalues of stretching vibration. The force constant $k_x$ is correlated to $\omega_x$ as follows:

$$k_{H,L} = 2\pi^2 m_{H,L} c^2 (\omega_L^2 + \omega_H^2) - k_C \\ \pm \sqrt{\left[2\pi^2 m_{H,L} c^2 (\omega_L^2 - \omega_H^2)\right]^2 - m_{H,L} k_C^2 / m_{L,H}},$$

(32)

where $c$ is the velocity of light *in vacuo*. Omitting the Coulomb repulsion decouples the oscillators into the classically isolated H–O and H$_2$O:H$_2$O oscillator with the respective angular frequency of vibration $\omega_x \propto \sqrt{k_x / m_x}$.

### 8.2 Specific solutions
### 8.2.1 Short-range interactions at equilibrium

With the known Coulomb potential, the measured segmental $d_x$ (or density [1]), and $\omega_x$ [2], parameters in the L-J ($E_{L0}$, $d_{L0}$) and in the Morse ($E_{H0}$, $\alpha$) potentials can be mathematically obtained. Table 6 lists the 0-th, first, second and third derivatives of the Taylor series. Table 7 lists the corresponding energies as functions of pressure. The harmonic approximation is indeed valid because the third derivative is much smaller than the second derivative.

Table 6. Derivatives of the L-J and Morse potentials at equilibrium.

| Derivatives | L-J potential | Morse potential | Results |
|---|---|---|---|
| $V_{x0}(E_{x0})$ | $E_{L0}$ | $E_{H0}$ (3.97 eV)[3] | $E_{x0}$ |
| $V_x' = 0$ | 0 | 0 | $d_{x0}$ |
| $V_x' + V_c' = 0$ | 0 | 0 | $d_x = d_{x0} + u_x$ |
| $V_x'' = k_x$ | $72E_{L0}/d_{L0}^2$ | $2\alpha^2 E_{H0}$ | $\alpha$ |
| $V_x'''$ | $-1512E_{L0}/d_{L0}^3$ | $-6\alpha^3 E_{H0}$ | |

Table 7. Pressure dependence of the first four terms of the Taylor series of the L-J and Morse potentials. Contribution of the third term is negligibly small. (Reprinted with permission from [6].)

| P (GPa) | $E_x$ (eV) | | | | | | | |
|---|---|---|---|---|---|---|---|---|
| | L-J potential | | | | Morse potential | | | |
| | 0th | 1st | 2nd($10^{-3}$) | 3rd($10^{-3}$) | 0-th | 1st | 2nd($10^{-3}$) | 3rd($10^{-5}$) |
| 0 | 0.0625 | 0 | 16.8102 | 10.1750 | 3.9700 | 0 | 0.7465 | 1.02 |
| 5 | 0.1063 | | 8.2883 | 2.7002 | 3.6447 | | 0.6387 | 0.85 |
| 10 | 0.1458 | | 4.7185 | 0.9904 | 3.3859 | | 0.5300 | 0.66 |
| 15 | 0.1755 | | 2.9185 | 0.4391 | 3.1875 | | 0.4247 | 0.49 |
| 20 | 0.1919 | | 1.9033 | 0.2212 | 3.0450 | | 0.3271 | 0.34 |
| 30 | 0.2477 | | 0.6599 | 0.0397 | 2.6290 | | 0.1880 | 0.16 |
| 40 | 0.2498 | | 0.2432 | 0.0089 | 2.1285 | | 0.1022 | 0.07 |
| 50 | 0.2165 | | 0.0967 | 0.0024 | 1.6465 | | 0.0581 | 0.03 |
| 60 | 0.1605 | | 0.0697 | 0.0017 | 1.1595 | | 0.0626 | 0.05 |

### 8.2.2 Force constants versus vibration frequencies

If $\omega_x$ and $k_C$ are given, it is possible to obtain the force constants $k_x$, the potential well depths $E_{x0}$, and the cohesive energy $E_x$, at each quasi-equilibrium site of the O:H and H-O segments. The force constant due to Coulomb repulsion is $k_C = q_O^2 / (2\pi \varepsilon_r \varepsilon_0 d_C^3)$ at quasi-equilibrium. Applying the known $\varepsilon_r = 3.2$, $\varepsilon_0 = 8.85 \times 10^{-12}$ F/m, $q_O = -0.634$ e (average of skin and bulk in Table 4), and $d_C = 2.743$ Å (section 4.1) results in $k_C = 0.17$ eV/Å².

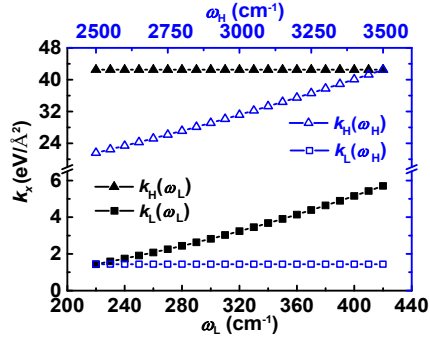

Figure 40. $\omega_x$-dependence of the force constants $k_x$ with $k_C = 0.17$ eV/Å$^2$ (colors differentiate the H from the L segment). The $k_L(\omega_L)$ and the $k_H(\omega_H)$ are much more sensitive to their respective frequency than the crossing terms of $k_L(\omega_H)$ and $k_H(\omega_L)$ that remain almost constant. (Reprinted with permission from [6].)

Figure 40 shows the functional dependence of $k_x$ on $\omega_x$, derived from Eq. (32). The $k_L$ increases from 1.44 to 5.70 eV/Å$^2$ while $k_H$ increases from 21.60 to 42.51 eV/Å$^2$ with their respective frequency shifts. The cross-terms $k_L(\omega_H)$ and $k_H(\omega_L)$, however, remain almost constant. Therefore, Eq. (32) may be simplified as,

$$k_x = 4\pi^2 c^2 m_x \omega_x^2 - k_C$$

or,

$$\omega_x = (2\pi c)^{-1} \sqrt{\frac{k_x + k_C}{m_x}}$$

(33)

With the measured $\omega_L = 237.42$ cm$^{-1}$, $\omega_H = 3326.14$ cm$^{-1}$ and the known $k_C = 0.17$ eV/Å$^2$ for the ice-VIII phase under atmospheric pressure [62,172,194], Eq. (33) yields $k_L = 1.70$ eV/Å$^2$ and $k_H = 38.22$ eV/Å$^2$. With the known $d_L = 1.768$ Å and $d_H = 0.975$ Å experiencing Coulomb repulsion [2], the free $d_{L0}$ is obtained to be 1.628 Å, and $d_{H0}$ to be 0.969 Å without involvement of the Coulomb repulsion. Coulomb repulsion lengthens the O-O distance from 2.597 to 2.733 Å by 0.136 Å at the 5% level.

With the derived values of $k_L = 1.70$ eV/Å$^2$, $k_H = 38.22$ eV/Å$^2$, and the known value of $E_{H0} = 3.97$ eV, all the parameters in the potentials and the force fields may be determined (Eq. (2)) at ambient pressure:

$$\begin{cases} k_L = 72 E_{L0}/d_{L0}^2 = 1.70 \text{ eV}/\text{A}^2 \\ k_H = 2\alpha^2 E_{H0} = 38.22 \text{ eV}/\text{A}^2 \end{cases}$$

or

$$\begin{cases} E_{L0} = 1.70 \times 1.628^2/72 = 0.062 \text{ eV} \\ \alpha = (38.22/3.97/2)^{1/2} = 2.19 \text{ A}^{-1} \end{cases}$$

(34)

8.2.3  Pressure-dependent $(d_x, \omega_x)$ and $(k_x, E_x)$

The $d_x(P)$ in Figure 14b [2] and $\omega_x(P)$ in Figure 13b provide the input for obtaining the specific solution

$(k_x, E_x)$:

$$\begin{pmatrix} d_H/0.9754 \\ d_L/1.7687 \\ \omega_H/3326.140 \\ \omega_L/237.422 \end{pmatrix} = \begin{pmatrix} 1 & 9.510\times 10^{-2} & 0.2893 \\ 1 & -3.477\times 10^{-2} & -1.0280 \\ 1 & -0.905 & 1.438 \\ 1 & 5.288 & -9.672 \end{pmatrix} \begin{pmatrix} P^0 \\ 10^{-2}P^1 \\ 10^{-4}P^2 \end{pmatrix}.$$

(35)

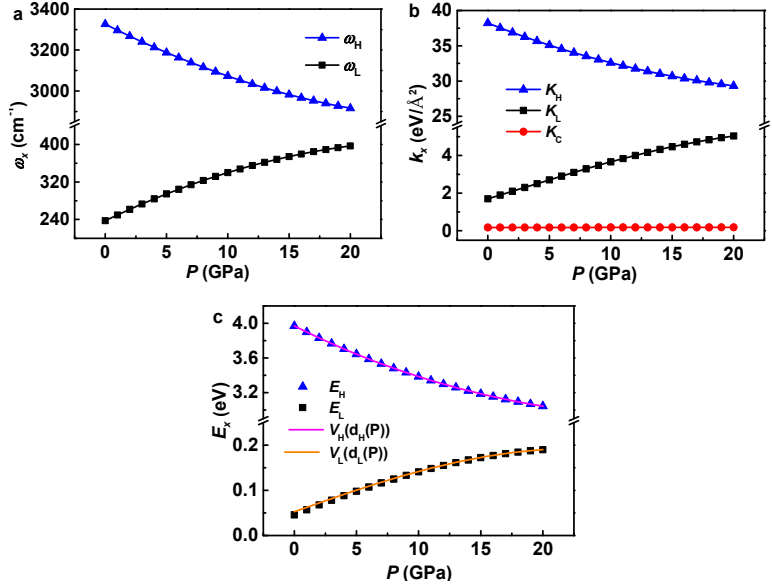

Figure 41. The known pressure-dependence of (a) $\omega_x(P)$ [14,62,172,194] and $d_x(P)$ (Figure 14(b)) yield (b) the $k_x(P)$ and (c) the $E_x(P)$ for the respective segment of the O:H-O bond in compressed ice. The $k_C$ remains almost constant (b) but couples the two segments. Agreement between the scattered data of harmonic approximation at each quasi-equilibrium point and the plotted continuum functions $V_x(d_x)$ in (c) verifies the reliability of the on-site harmonic approximation for the inharmonic system. (Reprinted with permission from [6].)

Lagrangian solution transforms the measured $(d_x, \omega_x)$ into $(k_x, E_x)$ for the segmented O:H-O bond. The following formulate the Lagrangian derivatives (Figure 41b and c):

$$\begin{pmatrix} k_H/38.223 \\ k_L/1.697 \\ E_H/3.970 \\ E_L/0.046 \end{pmatrix} = \begin{pmatrix} 1 & -1.784 & 3.113 \\ 1 & 13.045 & -15.258 \\ 1 & -1.784 & 3.124 \\ 1 & 25.789 & -49.206 \end{pmatrix} \begin{pmatrix} P^0 \\ 10^{-2}P^1 \\ 10^{-4}P^2 \end{pmatrix}.$$

(36)

Results shown in Figure 41b indicate that the $k_C$ (curvature of the Coulomb potential) keeps almost constant under compression if the $q_O$ and the $\varepsilon_r$ are conserved. The $k_L$ increases more rapidly than $k_H$ decreases because of the interplay of the mechanical compression, the Coulomb repulsion, and the strength disparity of the two segments. Figure 41c indicates that increasing the pressure from 0 to 20 GPa strengthens the O:H bond ($E_L$) from 0.046 to 0.190 eV, while softening the H–O bond ($E_H$) from 3.97 eV to 3.04 eV because of repulsion. This trend agrees with the numerical derivation, indicating that O:H contributes positively and H-O contributes negatively to lattice energy [196]. As given in Table 8, when the pressure is increased to 60 GPa, $k_L$ reaches 10.03 and $k_H$ to 1.16 eV/Å$^2$.
$E_L$ increases to a maximum of 0.25 at 40 GPA, and recovers to 0.16 at 60 GPa.

8.3   Potential paths for O:H-O bond relaxation
8.3.1   Proton centralization of compressed and cooled ice

The results in Table 8 confirm that compression shortens and stiffens the O:H bond, which lengthens and softens the H–O bond through Coulomb repulsion, which shortens the O-O distance towards proton centralization. As $d_L$ shortens by 4.3% from 0.1768 to 0.1692 nm, $d_H$ lengthens by 2.8% from 0.0975 to 0.1003 nm when the pressure is increased from 0 to 20 GPa [2]. When the pressure increases further to 60 GPa, the length of the O:H equals the length of the H–O at 0.11 nm, forming a symmetrical O:H-O bond in the ice-X phase.

However, $E_H$ (1.16 eV) remains higher than $E_L$ (0.16 eV) at 60 GPa, which means that the $sp^3$-hybridization of O is retained during the symmetrization. The nature of the respective short-range interaction persists even though the length and force constants approach equality. This means that it is unlikely to that the $sp^3$-hybridized orbits of oxygen will be dehybridized by compression. Therefore, phase X remains the H$_2$O molecular structure despite identical length scales. The counters of the potential paths follow each respective potential curve during symmetrization.

Most strikingly, unlike the 'normal' substance that gains energy with possible plastic deformation under compression [312], O:H-O bond always losses energy, instead. The O:H-O bond always tends to recover its higher energy state at relaxation to lower initial state without any plastic deformation, exhibiting recoverability of relaxation and damage.

Table 8. Pressure-dependence of the O:H-O segmental cohesive energy $E_x$ and force constant $k_x$, and the deviated displacement ($\Delta_x$) from the equilibrium position. (Reprinted with permission from [6].)

| P (GPa) | $E_L$ (eV) | $E_H$ (eV) | $\Delta E_{H+L}$ (eV) | $k_L$ (eV/Å$^2$) | $k_H$ (eV/Å$^2$) | $\Delta_L$ (10$^{-2}$ nm) | $\Delta_H$ (10$^{-4}$ nm) |
|---|---|---|---|---|---|---|---|
| 0 | 0.046 | 3.97 | 0 | 1.70 | 38.22 | 1.41 | 6.25 |
| 5 | 0.098 | 3.64 | -0.278 | 2.70 | 35.09 | 0.78 | 6.03 |
| 10 | 0.141 | 3.39 | -0.485 | 3.66 | 32.60 | 0.51 | 5.70 |
| 15 | 0.173 | 3.19 | -0.653 | 4.47 | 30.69 | 0.36 | 5.26 |
| 20 | 0.190 | 3.04 | -0.786 | 5.04 | 29.32 | 0.27 | 4.72 |
| 30 | 0.247 | 2.63 | -1.139 | 7.21 | 25.31 | 0.14 | 3.85 |
| 40 | 0.250 | 2.13 | -1.636 | 8.61 | 20.49 | 0.08 | 3.16 |
| 50 | 0.216 | 1.65 | -2.15 | 9.54 | 15.85 | 0.05 | 2.71 |
| 60 | 0.160 | 1.16 | -2.696 | 10.03 | 11.16 | 0.04 | 3.35 |

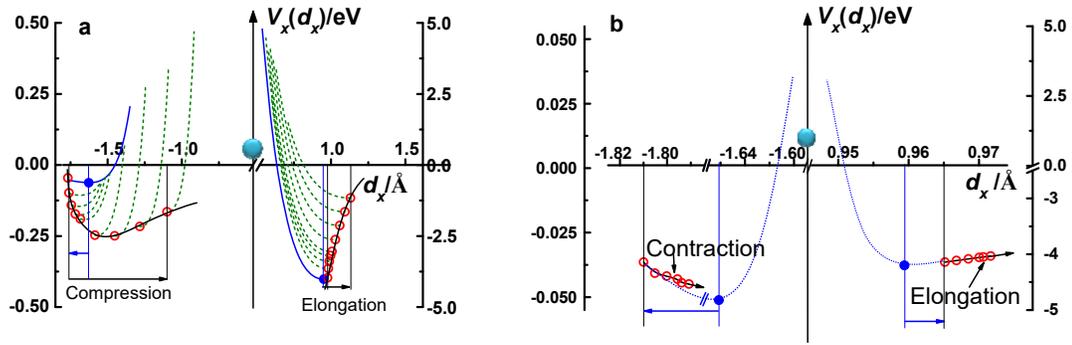

Figure 42. Potential paths $V_x(r)$ for the O:H-O bond relaxation in ice: (a) when compressed (l. to r.: P = 0, 5, 10, 15, 20, 30, 40, 50, 60 GPa); and (b) for cooling (l. to r.: 238–98 K) [5]. Small blue circles represent the intrinsic equilibrium (length and energy) of $O^{2-}$ without Coulomb repulsion or compression ($V'_x = 0$, $V_C = 0$, $P = 0$). Open red circles denote the quasi-equilibrium caused by both the Coulomb repulsion and the pressure/temperature (left-hand ends: $V'_x + V'_C = 0$, $P = 0$; otherwise: $V'_x + V'_C + f_{P,T} = 0$, where $f_{P,T}$ is the driving force). Broken lines denote potentials at quasi-equilibrium in (a). Thick solid lines through all the open circles are the $V_x(r)$ paths of the O:H-O bond at relaxation under compressing or cooling. Note the vertical energy scale difference between the two segments in one panel and the lateral scale difference between these two panels. The sphere at the coordinate origin is a $H^+$ proton. (Reprinted with permission from [6].)

The potential paths pertaining to the O:H-O relaxation dynamics have thus been resolved. Figure 42 shows the $V_x(r)$ paths for the O:H–O bond in compressed and cooled ice. Sun et al. [5] provided the $d_x(T)$ and $\omega_x(T)$ input for the $V_x(r)$ paths of the O:H-O bond in ice at cooling.

The potential paths show the relaxation dynamics. During relaxation, Coulomb repulsion pushes both $O^{2-}$ ions moving firstly outwardly from their ideal equilibrium position. Upon being compressed or cooled, both $O^{2-}$ ions in ice move to the right along the potential paths. The intrinsic equilibrium position of the $O^{2-}$ forming the H-O bond almost superposes on its quasi-equilibrium position, with a displacement of only $6.25 \times 10^{-4}$ nm at first. However, for the O:H, the displacement is $1.41 \times 10^{-2}$ nm, evidence that the O:H bond is much softer than the H-O bond. Coulomb repulsion and external stimulus relax $E_x$ along the respective potential path step-by-step. These are amounts beyond the scope of any available physical techniques, but the Lagrangian solution provides the values. It is also impossible to detect the potential directly without appealing to Lagrangian-Laplace mechanics.

8.3.2   Molecular undercoordination

Using the MD derivatives of $d_x$ and $\omega_x$ as input (Figure 22 and Figure 24), the Lagrangian solution then derived the $k_x$, $E_x$ potential paths for the O:H-O bond relaxation with the $(H_2O)_N$ size reduction, as shown in Figure 43. Numerically, the least $E_L = 4.3$ meV for a dimer seemed not so reasonable because the MD artifacts estimate the short-range interactions inadequately. Encouragingly, the trend of $E_H$ change agrees with the measured vales of 3.97 eV for bulk water ice, 4.52–4.66 eV for the skin, and 5.10 eV for gaseous monomers, which verify the BOLS-NEP expectation and the physical origin for undercoordination-induced O:H-O bond relaxation.

Table 9. N-dependence of $d_x$ and $E_x$ for the O:H-O bond in $(H_2O)_N$.

| N | $E_L$ (meV) | $E_H$ (eV) | $k_L$ (eV/Å²) | $k_H$ (eV/Å²) | $\Delta_L$ (10⁻¹ Å) | $\Delta_H$ (10⁻³ Å) |
|---|---|---|---|---|---|---|
| 6 | 76.53 | 3.763 | 2.222 | 35.92 | 1.074 | 6.641 |
| 5 | 51.31 | 3.974 | 1.646 | 38.22 | 1.441 | 6.207 |
| 4 | 44.92 | 4.033 | 1.503 | 38.85 | 1.576 | 6.099 |
| 3 | 11.38 | 4.411 | 0.733 | 42.83 | 3.214 | 5.499 |
| 2 | 4.30 | 4.542 | 0.525 | 44.19 | 4.477 | 5.324 |

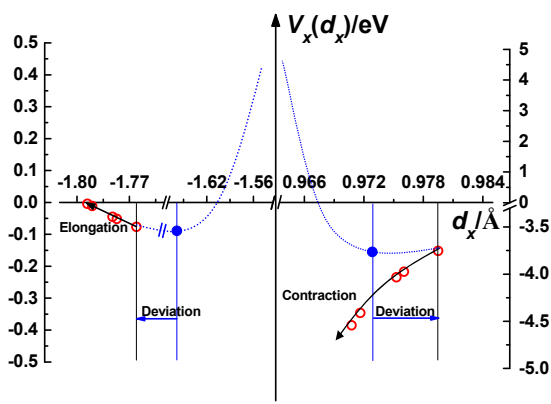

Figure 43. Potential paths (red circles) for the O:H-O bond relaxation as a function of N (l. to r.: N = 2, 3, 4, 5, 6) in the $(H_2O)_N$ clusters. Blue circles are the ideal equilibrium without inter-oxygen repulsion. These indicate that undercoordination reduces the $H_2O$ size ($d_H$) but increases their separations ($d_{O-O}$) with H-O bond stiffening and O:H nonbond softening.

8.4    Summary

The Lagrangian-Laplace solution to the motion dynamics of the coupled O:H-O oscillators transforms the measured ($d_x$, $\omega_x$) into ($k_x$, $E_x$) and thus enables the potential paths to be probed for the O:H-O bond at relaxation, which is beyond the scope of currently available experimental means. Results consistently provide evidence of the persistence and significance of the asymmetrical short-range interactions and Coulomb repulsion in the flexile, polarizable O:H-O bond. The Lagrangian strategy is useful for solving strongly correlated systems with multiple short-range interactions. The asymmetrical, local, short-range potentials and their switches at the atomic site are the most meaningful information, and the long-range interactions serve as the average background in general.

9    Uniqueness of the structural solution

- *Mass density unifies the size, separation, and geometry of molecular packing in water ice.*
- *Solution reconciles the documented proton centralization, skin $d_{O-O}$ and $d_x$ cooling relaxation.*

- *Water prefers a fluctuated, tetrahedrally coordinated structure with a supersolid skin.*
- *Bi-phase exits to water nanodroplet and nanobubble in a core-shell configuration.*

9.1    Challenge: Geometric and scale uncertainties

In all previous sections, we have persevered with the tetrahedral structure of water ice, as only this notation provides the O:H-O bond configuration. However, the structure order and the length scale of molecular packing in water and ice remain open to debate. Traditionally, independent and instantaneous accuracy is sought for one of the strongly correlated parameters, providing fuel for endless argument. For instance, the separation between adjacent oxygen atoms ($d_{O-O}$) has been reported to vary from 2.70 to 3.00 Å [56,76,248,249,313-321], and molecular size ($d_H$) from 0.970 to 1.001 Å [322]. The molecular CN varies from two [90] to four or even greater [323]. Certainty is needed regarding the geometrical structure of liquid $H_2O$, and whether mono- or mixed-phase order.

As an important yet often-overlooked quantity for water ice, the mass density $\rho$, which can easily be determined with certainty, does unify uncertain issues such as geometric configuration and length scale. Based on the essential rule of $sp^3$-orbit hybridization of oxygen [88,97] and the O:H-O bond cooperativity [2,3,5], it should be possible to resolve these uncertainties without the need for assumptions or approximations.

9.2    Density–geometry–length scale correlation

The packing order in Figure 5c defines that each cube of $a^3$ volume accommodates only one $H_2O$ molecule on average. With the known mass of a $H_2O$ molecule consisting of 8 neutrons, 10 protons and 10 electrons, $M = (10 \times 1.672621 + 8 \times 1.674927 + 10 \times 9.11 \times 10^{-4}) \times 10^{-27}$ kg, and the known density $\rho = M/a^3 = 1$ (g·cm$^{-3}$) at 4°C at atmospheric pressure, this tetrahedrally-coordinated structural order defines unambiguously the molecular separation, $d_{O-O}$. Furthermore, plotting $d_L(P)$ against $d_H(P)$ (see Eq. (23) and Figure 14b), straight away yields the $d_x$ length (unit in Å) cooperativity, free from probing conditions or methods. The $d_{O-O}$, $d_x$ and $\rho$ are correlated as follows [1]:

$$\begin{cases} d_{OO} = d_L + d_H = 2.6950\rho^{-1/3} & (Molecular\ separation) \\ d_L = \dfrac{2d_{L0}}{1+exp[(d_H - d_{H0})/0.2428]}; & (d_{h0} = 1.0004, d_{L0} = 1.6946) \end{cases}.$$

(37)

Thus, given any known density change, it is possible to scale the size $d_H$ and the separation $d_{O-O}$ of $H_2O$ molecules with the given molecular structure of tetrahedron. If the relaxation of $d_x$ matches the value of detection, the structure in Figure 5c and the $d_x$ cooperativity derived herein are justified as true and unique.

Figure 44a shows the conversion of the measured $\rho(T)$ into the $d_{O-O}(T)$ for water droplets of different sizes [133,134]. The $d_{O-O}$ values of 2.70 Å measured at 25°C, and 2.71 Å at -16.8°C [249] exactly match the derivative for the larger droplet, and testify to the truth of both Eq. (37) and the tetrahedral structure as representing the actual condition of water and ice. Furthermore, the data reported in [249] is essentially accurate and correct.

From the referenced data, Eq. (37) may be used to make conversions into other relevant quantities.

Table 10 summarizes the $d_{O-O}$, $d_x$, $\rho$ and $\omega_x$ for the skin and the bulk of water and ice, and compares them to those of ice at 80 K, and water dimers. For example, from the known $d_{O-O} = 2.965$ Å [248], $d_H = 0.84$ Å and $\rho = 0.75$ g·cm$^{-3}$ are derived for the supersolid skin of water and ice.

Table 10. Experimentally derived quantities ($\omega_x$, $d_x$, $d_{O-O}$, $\rho$) of water and ice.
(Reprinted with permission from [4].)

|  | Water (298 K) | | Ice (253 K) | Ice (80 K) | Vapor |
|---|---|---|---|---|---|
|  | bulk | skin | bulk | Bulk | Monomer |
| $\omega_H$ (cm$^{-1}$) | 3200 [78] | 3450 [78] | 3125 [78] | 3090 [5] | 3650 [65] |
| $\omega_L$ (cm$^{-1}$) [5] | 220 | ~180 [3] | 210 | 235 | – |
| $d_{O-O}$ (Å) [1] | 2.700 [249] | 2.965 [248] | 2.771 | 2.751 | 2.980 [248] |
| $d_H$ (Å) [1] | 0.9981 | 0.8406 | 0.9676 | 0.9771 | 0.8030 |
| $d_L$ (Å) [1] | 1.6969 | 2.1126 | 1.8034 | 1.7739 | ≥ 2.177 |
| $\rho$ (g·cm$^{-3}$) [1] | 0.9945 | 0.7509 | 0.92 [285] | 0.94 [285] | ≤ 0.7396 |

9.3  Uniqueness of structure solution

One can obtain the $d_L$, $d_H$, $d_{O-O}$ and $\rho$ by solving the following simple equation with any one of these parameters as input [1]:

$$d_L - 2.5621 \times \left[1 - 0.0055 \times exp\left((d_{OO} - d_L)/0.2428\right)\right] = 0.$$

Agreeing with the MD derivatives given in Figure 33a, decomposition of the $d_{O-O}(T)$ into the $d_x(T)$ in Figure 44 confirms the following predictions:

1) Cooling shortens the O:H nonbond in the liquid ($T > 277$ K) and in the solid phase ($T < 205/242$ K), which lengthens the H-O bond slightly and lowers the density.
2) In the freezing transition phase, the relaxation process reverses, leading to the O-O length gain and density loss.
3) At $T \leq 80$ K, $d_x$ remains almost thermally stable because the specific heat $\eta_L \approx \eta_H \approx 0$ in this regime [5].
4) Dispersion of the least-density temperature from 205 K (for a 1.4 nm-sized droplet) to 241 K (4.4 nm-sized droplet) and to 258 K (for bulk water [5]) follows the expectations of Figure 9. Droplet size reduction stretches the $\eta_H$ and compresses the $\eta_L$ along the temperature axis, resulting in supercooling and superheating.

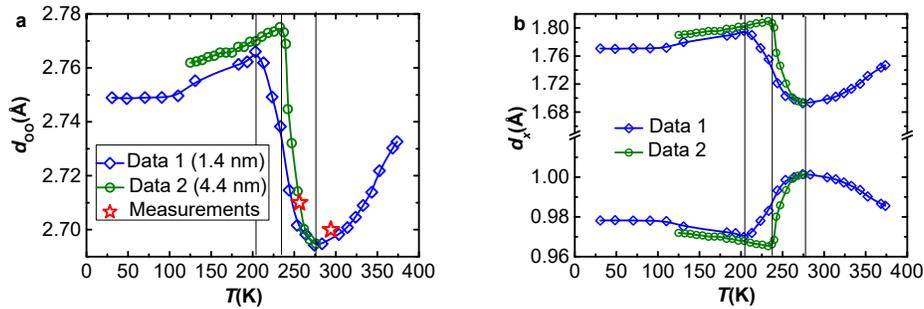

Figure 44. (a) $\rho(T) \approx d_{O-O}(T)$; and (b) $d_{O-O}(T) \approx d_x(T)$ conversion for water droplets ($T < 273$ K) and water bulk ($T > 273$ K) [133,134]. Matching the $d_{O-O}(T)$ profile to the measured $d_{O-O}$ at 25°C

and -16.8°C in panel (a) [249] not only verifies the validity of the uniqueness of the tetrahedral structure and the $d_x$ cooperativity, but also the accuracy and reliability of the data reported in [249]. (Reprinted with permission from [1].)

Figure 45 shows the solution consistency that unifies the ($d_H$, $d_L$, $d_{O-O}$, $\rho$) and the structural order pertaining to: (i) compressed ice [14], (ii) cooling water and ice [133,134], and, (iii) water skin and molecular monomers [248,249]. The currently derived value $d_H = 1.0004$ Å at unit density lies within observed values ranging from 0.970 to 1.001 Å [322]. The $d_{O-O}$ values [90,109,283] greater than the ideal value of 2.6950 Å at $\rho = 1$ (g·cm$^{-3}$) correspond to the skin that exists only in sites of water ice composed of molecules with fewer than four neighbors [3].

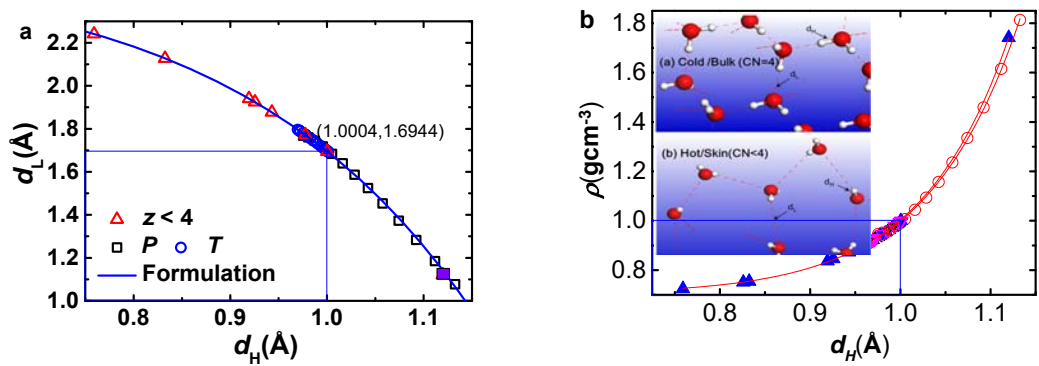

Figure 45. (a) $d_L$–$d_H$, and (b) $\rho$–$d_H$ profiles for H$_2$O molecules, which match (a) (i) ice under compression ($d_H > 1.00$ Å) [14]; (ii) water ice at cooling (0.96 < $d_H$ < 1.00 Å) [133,134]; and (iii) water skin and monomer ($d_H < 1.00$ Å) [76,249,313,314,318-321]. The value $d_H = 1.0004$ Å is the standard at $\rho = 1$ g·cm$^{-3}$. For $d_H$ shorter than 0.96 Å, it matches (b) in the skin or clusters [3,56,248]. The inset in (b) illustrates a change of molecular size and separation with temperature and molecular CN. (Reprinted with permission from [1].)

9.4    Nanobubble and nanodroplet bi-phase configuration

Findings herewith indicate that bulk water prefers uniquely the mono-phase of fluctuating tetrahedral structure with the addition of low-density supersolid skin. The skin is 0.04–0.12 nm thick, according to small-angle X-ray scattering and TIP4P/2005 force-field calculations at 7, 25 and 66°C water and atmospheric pressure [262]. However, the skin contains at least two molecular layers that are subject to molecular undercoordination, according to the current understanding. In the skin, the geometrical configuration of water molecules remains the bulk attribute, but the length scale changes with the CN loss — the size of a molecule shrinks but the separation between molecules expands. H-O free radicals or dangling bonds are present at the surface; these are even shorter and stiffer than the skin H-O bond, as phonon spectroscopy has probed at 3700 cm$^{-1}$.

Bulk water holds the mono-phase of tetrahedral structure as the volume of the skin is negligible; however, for a sufficiently small water droplet, the volume competition between the skin and the core is no longer negligible. Therefore, small droplets hold a bi-phase structure in a core–shell order. Therefore, it is now clear why some studies have reported a mono-phase structure, and some have argued for a mixed-phase structure in water droplets of different size. The bi-phase structures are never distributed either homogeneously or randomly in liquid water, though this cannot be resolved experimentally at the present time.

Gas bubbles at the nanometer scale have far-reaching physical, chemical and biological effects [82]. They are difficult to destroy, and are thermally much more stable than bubbles at the millimeter scale [324]. A bubble is the inversion of a droplet; a soap bubble, for example, contains two skins — an inner and outer skin — both of which are in the supersolid phase. The volume proportion of the entire liquid shell volume of such supersolid phases is much greater than in a droplet. Therefore, bubbles demonstrate the supersolid state more significantly: elasticity, hydrophobicity and thermal stability, which makes bubbles mechanically stronger and thermally more stable than droplets. Further investigation of the thermo–mechanico–dynamics and bond–electron–phonon cooperative relaxation in bubbles is in active progress.

10  Mpemba paradox: O:H-O bond memory and water–skin supersolidity

- *Heating stores energy into water by simultaneously stretching the O:H and shortening the H-O.*
- *Cooling has the opposite effect, emitting energy at a rate of history dependence.*
- *Heating and supersolidity jointly elevate skin thermal diffusivity, favoring outward heat flow.*
- *Being sensitive to the source volume, skin radiation and drain temperature, the Mpemba effect proceeds only in a strictly non-adiabatic 'source–path–drain' cycle system for heat 'emission–conduction–dissipation'.*

10.1   Why does hotter water freeze faster?

The Mpemba effect [325-329] is the assertion that hot water freezes quicker than cold water, even though it must pass through the same lower temperature on the way to freezing. This puzzle has baffled thinkers such as Francis Bacon, René Descartes and Aristotle [325], who first noted: "The water has previously been warmed contributes to its freezing quickly: for so it cools sooner". Hence, many people, when they want to cool water quickly, begin by putting it in the sun.

However, a commonly accepted explanation or numerical reproduction of this phenomenon remains challenging despite efforts made since the age of Aristotle. Proposed factors explaining this effect include evaporation [330], thermal convection [331,332], solutes [333], frosting [334], supercooling [334,335], etc. According to the winner [336] of a competition held in 2012 by the Royal Society of Chemistry, thermal convection provides the best possible rationalization of the energy 'emission–conduction–dissipation' dynamics in the 'source–path–drain' cycle system in which the Mpemba paradox takes place. However, little attention has yet been paid to the nature and the relaxation dynamics of the O:H-O bond [337] that is the primary constituent of the liquid. The liquid serves as both the source and the path in this event.

10.2   Numerical solution: Water-skin supersolidity
10.2.1 Fourier thermal–fluid equation

Figure 46 illustrates an adiabatically-walled, open-ended, one-dimensional tube cell for solving the Fourier fluid transport equation by the finite element method [7]. Water at an initial temperature $\theta_i$ in the cell is divided along the *x*-axis into two regions: the bulk (B, from $-L_1 = -9$ mm to 0) and the skin (S, from 0 to $L_2 = 1$ mm). The tube is cooled in a drain of constant temperature $\theta_f$, which is subject to variation to allow it to be examined for sensitivity.

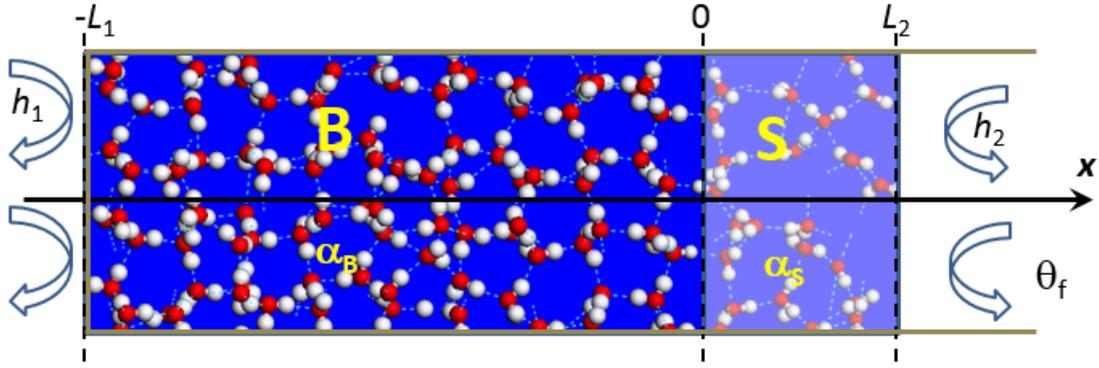

Figure 46. The water in the adiabatically-walled, open-ended, one-dimensional tube cell at initial temperature $\theta_i$ is cooled in the drain of constant temperature $\theta_f$. The liquid source is divided along the x-axis into B, the bulk ($-L_1 = -9$ mm, 0) and S, the skin (0, $L_2 = 1$ mm), with respective thermal diffusivities $\alpha_B$ and $\alpha_S$. The mass densities are $\rho_S/\rho_B = 3/4$ [1,3] in the respective region. The bulk-skin interface is at $x = 0$; $h_j$ is the heat transfer (radiation) coefficient at the two ends of the tube, with the absence ($j = 1$, left-hand end) and presence ($j = 2$, right-hand end) of the skin.

The time-dependent gradient of temperature change at any site ($x$), which follows the function and the initial- and boundary conditions:

$$\frac{\partial \theta(x)}{\partial t} = \nabla \cdot \left( \alpha(\theta(x), x) \nabla \theta(x) \right) - v \cdot \nabla \theta(x)$$

$$\begin{cases} \alpha(\theta, x) = \dfrac{\kappa_B(\theta, x)}{\rho_B(\theta, x) C_{pB}(\theta, x)} \times \begin{cases} 1 & (Bulk) \\ \approx \rho_B/\rho_S (= 4/3) & (Skin) \end{cases} \\ v_S = v_B = 10^{-4} (m/s) \end{cases},$$

$$\begin{cases} \theta = \theta_i & (t = 0) \\ \theta(0^-) = \theta(0^+); \theta_x(0^-) = \theta_x(0^+) & (x = 0) \\ h_i(\theta_f - \theta) \pm \kappa_i \theta_x = 0 & (x = -l_1; l_2) \end{cases}.$$

(38)

The first term describes thermal diffusion and the second term describes thermal convection in the Fourier transport equation, where $\alpha$ is the thermal diffusivity and $v$ is the convection rate. The known temperature-dependence of the thermal conductivity $\kappa(\theta)$, the mass density $\rho(\theta)$, and the specific heat under constant pressure $C_p(\theta)$, given in Figure 4, determines the thermal diffusivity $\alpha_B$ of the bulk water. The skin supersolidity [3] contributes to $\alpha_S$ in the form $\alpha_S(\theta) \approx 4/3 \alpha_B(\theta)$, because the skin mass density 0.75 gcm$^{-3}$ is 3/4 times the standard density at 4°C. $\alpha_S(\theta)$ is subject to optimization as the skin supersolidity may modify the $\kappa(\theta)/C_p(\theta)$ value in a yet unknown manner.

The boundary conditions represent that the temperature $\theta$ and its gradient $\theta_x = \partial \theta / \partial x$ continue at the interface ($x = 0$) and the thermal flux $h(\theta_f - \theta)$ is conserved at each end for $t > 0$. The velocity of convection $v$ takes the bulk value of $v_S = v_B = 10^{-4}$ m/s, or zero for examination. As the heat transfer (through radiation) coefficient $h_j$ depends linearly on the thermal conductivity $\kappa$ in each respective region [338], the standard value of $h_1/\kappa_B = h_2/\kappa_S = 30$ w/(m$^2$K) [99] is necessary for solving the problem. The $h_2/\kappa_S$ term contains heat reflection by the boundary. The ratio $h_2/h_1 > 1$ describes the possible effect of skin thermal radiation.

### 10.2.2 Roles of convection, diffusion, and boundaries

The computer reads in the digitized $\rho(\theta)$, $\kappa(\theta)$ and $C_p(\theta)$ in Figure 4 to compose the $\alpha_B(\theta)$ before each iteration of calculating the partitioned elemental cells. Besides the thermal diffusivity and the convection velocity in the Fourier equation, systematic examination of all possible parameters in the boundary conditions is essential. The results in Figure 47 and Figure 48 reveal the following:

1) Characterized by the crossing temperature, the Mpemba effect happens only in the presence of the supersolid skin ($\alpha_S/\alpha_B > 1$) irrespective of thermal convection.
2) Complementing skin supersolidity, thermal convection distinguishes only slightly between the skin and bulk temperature$\Delta\theta$, and raises the crossing temperature negligibly.
3) The Mpemba effect is sensitive to the source volume, the $\alpha_S/\alpha_B$ ratio, the radiation rate $h_2/h_1$ and the drain temperature $\theta_f$.
4) The bulk/skin thickness ($L_1$:$L_2$) ratio and the thermal convection have little effect on observations.

For instance, increasing the liquid volume may annihilate the Mpemba effect because of the non-adiabatic process of heat dissipation. It is understandable that cooling a drop of water (1 mL) needs shorter time than cooling one cup of water (200 mL) at the same $\theta_i$ and under the same conditions. Higher skin radiation $h_2/h_1 > 1$ promotes the Mpemba effect. Therefore, conditions for the Mpemba effect are indeed very critical, which explains why it is not frequently observed.

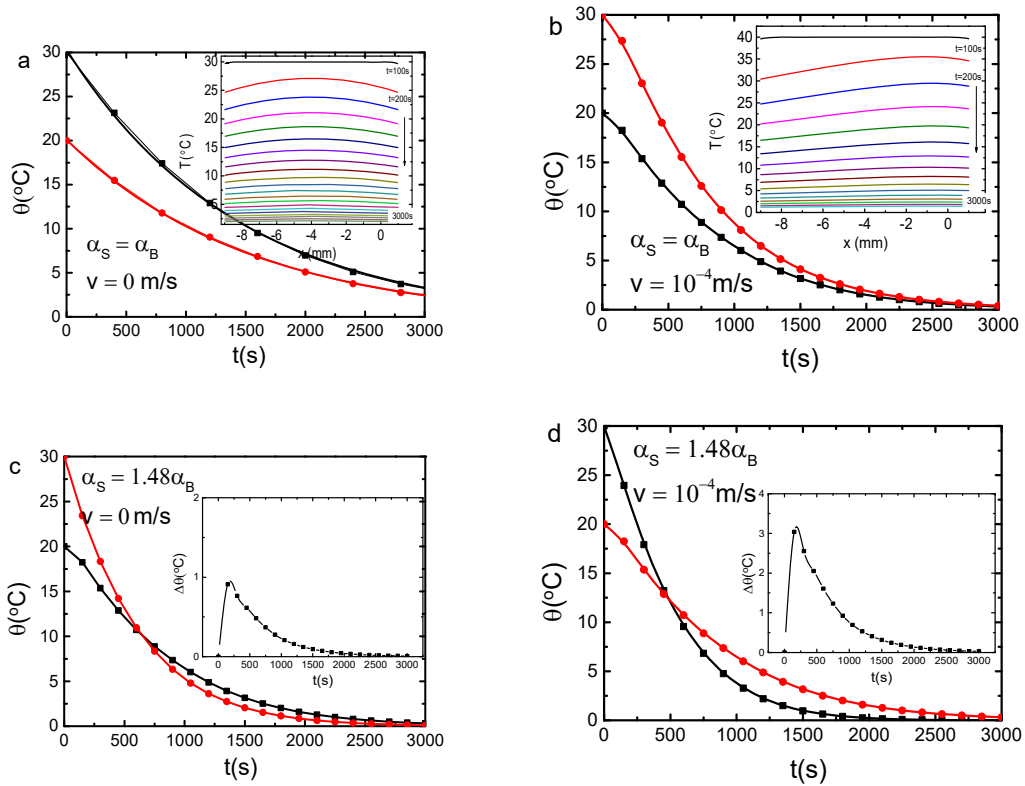

Figure 47. Thermal relaxation curves $\theta(\theta_i, t)$ at $x = 0$: (a), (b) with supersolid skin absent ($\alpha_S/\alpha_B = 1$); and (c), (d) with supersolid skin present (optimized at $\alpha_S/\alpha_B = 1.48$); and (a), (c) with thermal convection absent ($v_S = v_B = 0$); and (b), (d) with thermal convection present ($v_S = v_B = 10^{-4}$ m/s). The Mpemba effect is characterized by the crossing temperature which occurs only in the presence of the skin supersolidity, irrespective of the thermal convection. The insets in (a) and (b) show the time-dependent thermal field in the tube. Supplementing the skin supersolidity, convection only slightly raises $\Delta\theta$ and the crossing temperature — as the insets in (c) and (d) show. (Reprinted with permission from [7].)

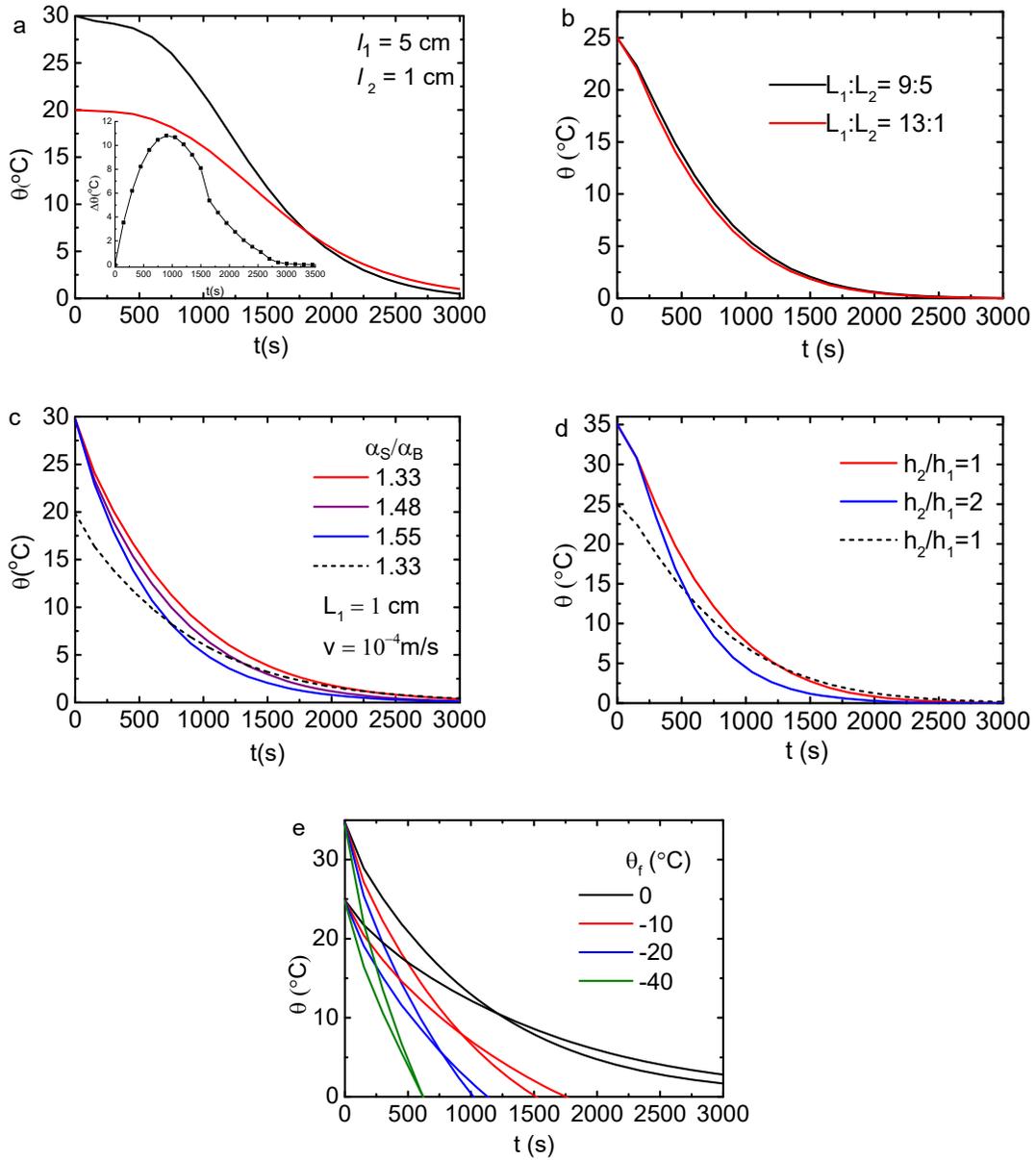

Figure 48. Sensitivity of the Mpemba effect (crossing temperature) to: (a) the source volume; (b) the bulk/skin thickness ratio ($L_1:L_2$); (c) the supersolidity ratio $\alpha_S/\alpha_B$; (d) the radiation rate $h_2/h_1$; and (e) the drain temperature $\theta_f$. Volume inflation (from 1 to 5 cm) in (a) prolongs the time until the crossing temperature is reached, and raises the skin temperature (see inset). (b) The $L_1:L_2$ ratio has little effect on the relaxation curve. Increasing (c), the $\alpha_S/\alpha_B$ and (d) the $h_2/h_1$ ratio promotes the Mpemba effect. (e) Lowering the $\theta_f$ shortens the time until the crossing temperature is reached. The sensitivity examination is conducted based on the conditions of $\alpha_S/\alpha_B = 1.48$, $v_S = v_B = 10^{-4}$ m/s, $\theta_f = 0°C$, $L_1 = 10$ mm, $L_2 = 1$ mm, $h_1/\kappa_B = h_2/\kappa_S = 30$ w/(m²K) unless indicated. (Reprinted with permission from [7].)

### 10.2.3 Mpemba attributes reproduction

Figure 49 shows the numerical reproduction of the observed Mpemba attributes (insets) [326,336], which confirm the following:

1) Hot water freezes faster than cold water under the same conditions.
2) The liquid temperature $\theta$ drops exponentially with cooling time ($t$) until the transition of water into ice, with a relaxation time $\tau$ that drops as $\theta_i$ is increased
3) The water skin is warmer than sites inside the liquid and the skin of hotter water is even warmer throughout the course of cooling.

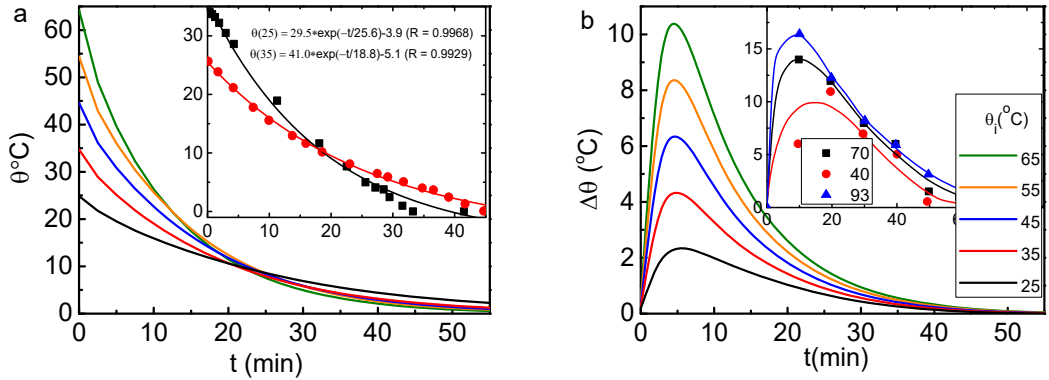

Figure 49. Numerical reproduction of the measured data (see insets): (a) thermal relaxation $\theta(\theta_i, t)$; and (b) skin-bulk temperature difference $\Delta\theta(\theta_i, t)$ curves [326,336] for water cooling from different $\theta_i$. Results were obtained using the conditions given in Figure 48. (Reprinted with permission from [7].)

### 10.3 Experimental revelation: O:H-O bond memory
### 10.3.1 O:H-O bond relaxation velocity

The following formulates the decay curve $\theta(\theta_i, t)$ shown in Figure 49a [336]:

$$\begin{cases} d\theta = -\tau_i^{-1}\theta dt & \text{(decay function)} \\ \tau_i^{-1} = \sum_j \tau_{ji}^{-1} & \text{(relaxation time)} \end{cases}$$

(39)

where the $\theta_i$-dependent relaxation time $\tau_i$ is the sum of $\tau_{ji}$ over all possible $j$-th process of heat loss during cooling.

It is encouraging that a combination of the documented experimental profiles of the $\theta(\theta_i, t)$ (Figure 49a inset) and the $d_H(\theta)$ (Figure 50a converted form Figure 4a using Eq. (37)) directly reveals the memory of the O:H-O bond without any assumption or approximation being needed. The $\theta(\theta_i, t)$ curve provides the slope of $d\theta/dt = -\tau_i^{-1}\theta$ and the $d_H(\theta) = 1.0042 - 2.7912 \times 10^{-5} \exp[(\theta+273)/57.2887]$ (Å) formulates the measured $\theta$ dependence of the H-O bond relaxation. Multiplying both slopes immediately yields the linear velocity of $d_H$ relaxation at cooling.

As the O:H nonbond and the H-O bond are correlated by Eq. (37), the relaxation velocities of their lengths and energies are obtained, since $E_x = k_x(\Delta d_x)^2/2$ approximates the energy storage with the known $d_H$ velocity. For simplicity and conciseness, the focused will be on the instantaneous velocity of $d_H$ during relaxation:

$$\frac{d(d_H(\theta))}{dt} = \frac{d(d_H(\theta))}{d\theta}\frac{d\theta}{dt} = -\tau_i^{-1}\theta\frac{\Delta(d_H(\theta))}{57.2887},$$

where:

$$\Delta(d_H(\theta)) = -2.7912 \times 10^{-5} \exp[(\theta+273)/57.2887].$$

(40)

Figure 50b plots the $\theta_i$-dependence of the $d_H$ linear velocity, which confirms that the O:H-O bond indeed possesses memory. Although passing through the same temperature on the way to freezing, the initially shorter H-O bond at higher temperature remains highly active compared to its behavior otherwise when they meet on the way to freezing.

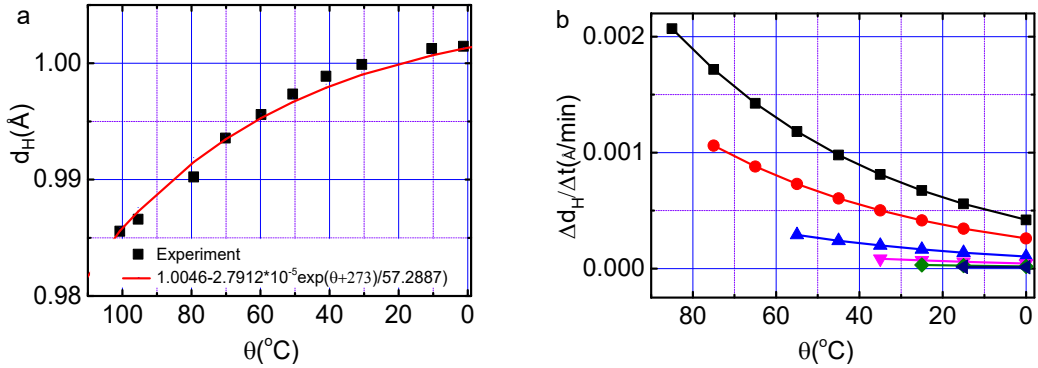

Figure 50. (a) Measured $d_H(\theta)$ (the solid line fitted to scattered data); and (b) derived $\theta_i$ (the starting point of each line) and its dependence on the $d_H$ velocity during relaxation at cooling. The velocity of the initially shorter H-O bond at higher $\theta_i$ always remains higher than otherwise when they meet. (Reprinted with permission from [7].)

10.3.2 Relaxation time versus initial energy storage

Solving the decay function Eq. (40) yields the relaxation time $\tau_i(t_i, \theta_i, \theta_f)$:

$$\tau_i = -t_i \left[ Ln\left(\frac{\theta_f + b_i}{\theta_i + b_i}\right) \right]^{-1}.$$

(41)

An offset of the $\theta_f (= 0°C)$ and the $\theta_i$ by a constant $b_i$ is necessary to ensure $\theta_f + b_i \geq 0$ in the solution ($b_i = 5$ was taken with reference to the fitted data in Figure 49a). Using the measured $t_i$, $\theta_i$ and $\theta_f$ given in Figure 51a (scattered data) as input, the respective $\tau_i$ may be found, shown by the solid line. The derived $\tau_i$ drops exponentially with the increase of the $\theta_i$ (Figure 51a), or with the increase of initial energy storage, or the initial vibration frequency measurements [113], as Figure 51b shows.

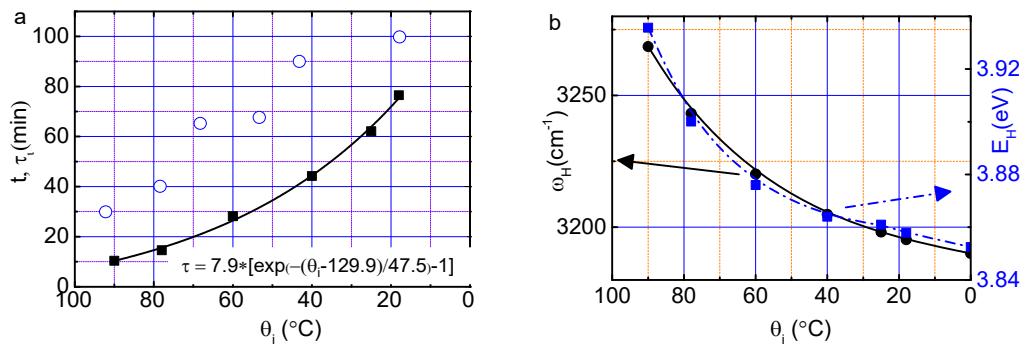

Figure 51. (a) Cooling time $t$ (scattered circles) dependent of relaxation time $\tau_i$ (fitted solid line) is correlated to (b) the initial energy $E_H$ (solid black line) and vibration frequency $\omega_H$ (broken blue line) of measurements [113] for the liquid source cooling from different $\theta_i$. (Reprinted with permission from [7].)

10.4    Heat 'emission–conduction–dissipation' dynamics
10.4.1  Liquid source and path: Heat emission and conduction

Figure 52 illustrates the cooperative relaxation of the O:H-O bond in water under thermal cycling. An interplay of the O:H vdW-like force, the H-O exchange interaction, the O--O Coulomb repulsion, the specific-heat disparity between the O:H, and the H-O bond, always dislocate O atoms in the same direction along the respective potential paths [2].

Generally, heating stores energy in a substance by stretching all bonds involved. However, heating excitation stores energy in water by lengthening the O:H nonbond. The O:H expansion weakens the Coulomb interaction, which shortens the H-O bond by shifting the $O^{2-}$ towards the $H^+$ (red line linked spheres in Figure 52 are in the hot state). Cooling does the opposite (blue line liked spheres), analogous to suddenly releasing a pair of coupled, highly deformed springs, one of which is stretched and the other compressed. This emits energy at a rate that depends on the deformation history (i.e., how much they were stretched or compressed). Energy storage and emission of the entire O:H-O bond occurs mainly through H-O relaxation, since $E_L$ (about 0.1 eV) is only 2.5% of $E_H$ (about 4.0 eV) [3]. The O:H-O bond memory and the unusual way of energy emission yield the history-dependent H-O bond relaxation velocity at cooling, as shown in Figure 50b.

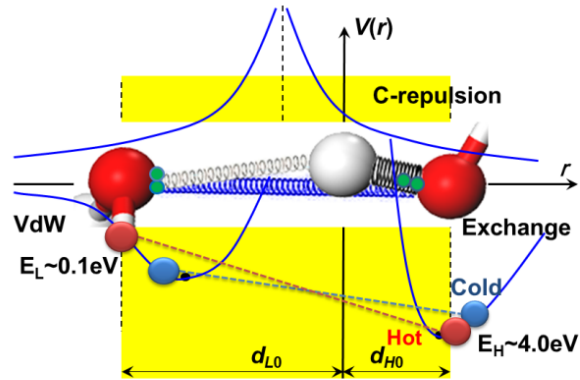

Figure 52. O:H-O bond cooperative relaxation under thermal cycling [339,340]. The $d_{H0}$ and $d_{L0}$ are the respective references at 4°C. Indicated are the O:H vdW-like nonbond interaction ($E_L$ ~0.1 eV; left-hand side), the H-O bond exchange interaction ($E_H$ ~ 4.0 eV; right-hand side), and the inter-oxygen electron-pair repulsion (paired green dots). A combination of these interactions and the specific-heat disparity between the O:H and the H-O dislocate O atoms in the same direction by different amounts when cooling. The relaxation proceeds along the O:H-O bond potentials from hotter (red line linked spheres, labeled 'hot') to colder state (blue line linked spheres, labeled 'cold').

During liquid heating, molecular undercoordination has the same effect as above on O:H-O bond relaxation. Heating and molecular undercoordination are mutually enhanced during O:H-O bond relaxation and the associated thermal diffusivity in the skin region. Mass density is lowered, raising the thermal diffusivity, which favors outward heat diffusion in the conduction path.

10.4.2  Source–drain interface: Nonadiabatic cycling

It is necessary to emphasize that the Mpemba effect happens only in circumstances where the water temperature drops abruptly from $\theta_i$ to $\theta_f$ at the source–drain interface. Examination has indicated that the Mpemba crossing temperature is sensitive to the volume of the liquid source (Figure 48a). Larger liquid volumes may prevent this effect by heat-dissipation hindering. As confirmed by Brownridge [334], any spatial temperature decay between the source and the drain could prevent the Mpemba effect. Examples of such decay might include sealing the tube ends, an oil film covering, a vacuum isolating the source–drain chamber, muffin-tin-like containers connecting, or multiple sources contributing to a limited volume. Conducting experiments under identical conditions is necessary to minimize artifacts such as radiation, source/drain volume ratio, exposure area, container material, etc.

10.4.3  Other factors: Supercooling and evaporating

It has been clear that the $E_H$ determines the critical temperature for phase transition [2]. Superheating at melting and supercooling at freezing is associated with the shorter H-O bond in water molecules with fewer than four neighbors, such as those that form the skin, a monolayer film, or a droplet on a hydrophobic surface or hydrophobically confined [218]. Supercooling is also associated with the longer H-O bond being in contact with hydrophilic surface [219] or with the longer H-O bond being compressed [2]. Supercooling of the colder water in the Mpemba process [334] is evidence that the initially longer H-O bonds in the colder water reacts more slowly to the relaxation at the freezing point than the bonds in the warmer water, because of the lower momentum of relaxation–memory effect.

The involvement of ionic solutes or impurities [341,342] mediates Coulomb coupling because of the alternation of charge quantities and ion volumes [343]. Salting affects the H-O phonon blueshift in the same way as heating [121,344,345], since it is expected to enhance the velocity of heat ejection at cooling. Mass loss due to evaporation of the liquid source [327] has no effect on the relaxation rate of O:H-O bonds since mass loss for cooling is negligible; it has been confirmed that the mass loss is only 1.5% or less in experiments freezing water at 75°C down to -40°C.

10.5    Summary

Reproduction of observations revealed the following pertaining to Mpemba paradox:

1) O:H-O bonds possess memory whose thermal relaxation defines intrinsically the rate of energy emission. Heating stores energy in water by O:H-O bond deformation. The H-O bond is shorter and stiffer in hotter water than it is in colder water. Cooling does the opposite, emitting energy with a thermal momentum that is history-dependent.
2) Heating enhances the skin supersolidity that elevates the skin thermal diffusivity with a critical ratio of $\alpha_S/\alpha_B \geq \rho_B/\rho_S = 4/3$. Convection alone produces no Mpemba effect, only raising the skin temperature slightly.
3) Highly nonadiabatic ambient conditions are necessary to ensure immediate energy dissipation at the source–drain interface. The Mpemba crossing temperature is sensitive to the volume of liquid source being cooled, the drain temperature and skin radiation.
4) The Mpemba effect takes place with a characteristic relaxation time that drops exponentially with increased initial temperature, or with initial energy storage in the O:H-O bond.
5) O:H-O bond memory may have implications for living cells, in which the O:H-O bond relaxation dominates the signaling, messaging, and damage recovering.

11   Prospects and perspectives

- *Negative thermal expansion at freezing may apply to other materials having short-range interactions and strong correlation.*
- *The Hofmeister series for protein dissociation, activation and deactivation of ion channeling could be controllable by mediating O--O repulsion.*
- *Melting and anti-icing could be adjustable by $E_H$ modulation through electro- and magneto-field.*
- *Dielectric relaxation of water ice may vary with changes in $\rho$, $E_H$, and polarization.*

11.1    Negative thermal expansion

The vast majority of materials have a positive coefficient of thermal expansion ($\alpha(\theta) > 0$) and their volume increases on heating. There is also another very large number of materials that display the opposite behavior: their volume contracts on heating, that is, they have a negative thermal expansion (NTE) coefficient [346-349]. A typical specimen is cubic $ZrW_2O_8$ that contracts over a temperature range exceeding 1000 K [350]. NTE also appears in diamond, silicon and germanium at very low temperatures (< 100 K) [351], and in glass in the titania-silicate family, Kevlar fiber, carbon fibers, anisotropic Invar Fe–Ni alloys, and certain kinds of molecular networks at room temperature. The NTE of graphite [352], graphene oxide paper [353], and $ZrWO_3$ [350] all share the NTE attribute of water at freezing, see Figure 53. NTE materials may be combined with other materials with a positive thermal expansion coefficient to fabricate composites having an overall zero thermal expansion (ZTE). ZTE materials are useful because they do not undergo thermal shock on rapid heating or cooling.

Typical of the models that explain the NTE effect suggests that NTE arises from the transverse thermal vibrations of the bridging oxygen in the M-O-M linkages inside $ZrW_2O_8$, $HfW_2O_8$, $SC_2W_3O_{12}$, $AlPO_{4-17}$, and faujasite-$SiO_2$ [354,355]. The phonon modes (centered around 30 meV or 3200 cm$^{-1}$) [356] can propagate without distorting the $WO_4$ tetrahedron or the $ZrO_6$ octahedron, termed the 'rigid-unit mode'. The rigid-unit mode also accounts for the structural phase transition of $ZrW_2O_8$ and $ZrV_2O_7$ [357].

Extending the NTE mechanism to water–ice transition from the perspective of bond relaxation may complement existing models in understanding NTE in general. Phonon spectroscopes have the capability to monitor the relaxation process easily and directly, as they do for water. Current understanding indicates that NTE results from the involvement of at least two kinds of coupled, short-range interactions and the associated specific-heat disparity. In the instance of graphite, the (0001) intralayer covalent bond and the interlayer vdW interactions may play certain roles, in much the same way as the O:H-O bond does in water. O, N and F all create lone pairs of electrons upon reaction, which create the weaker short-range nonbonding interaction.

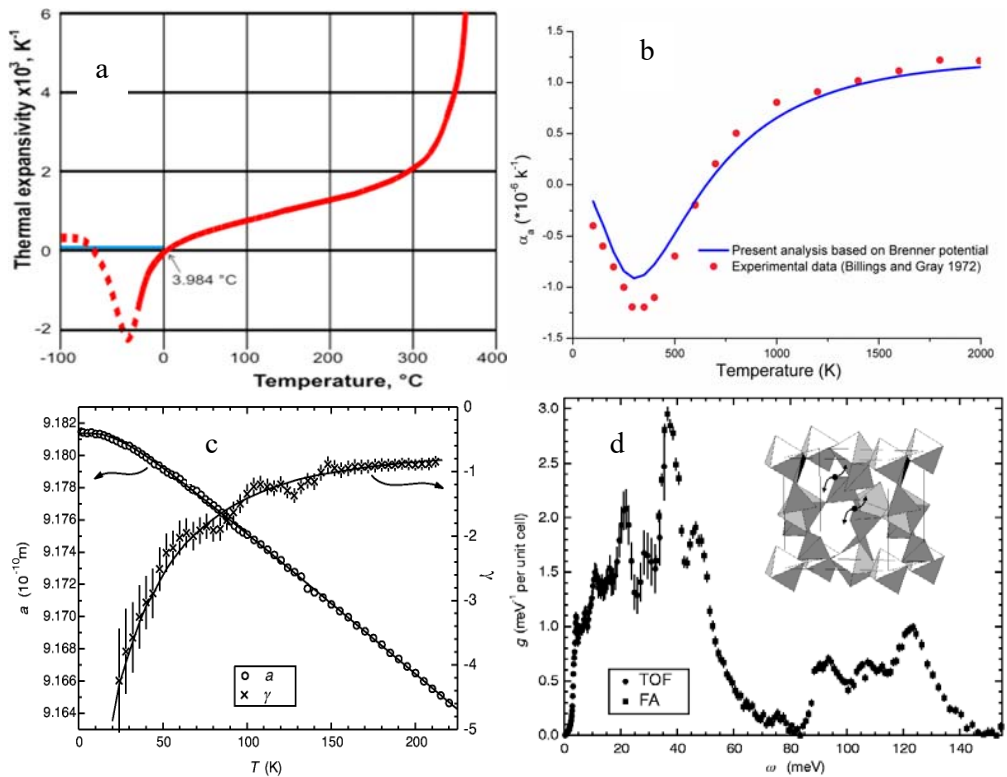

Figure 53. The NTE of (a) $H_2O$; (b) graphite; and (c) $ZrW_2O_8$ with thermal expansion coefficient $\alpha$ (open circles) and Grüneisen parameter $\gamma = 3\alpha B/C_v$ (crosses), where $B$ is the bulk modulus and $C_v$ is the specific heat at constant volume; (d) shows the associated phonon spectrum measured at $T = 300$ K. The inset illustrates the 'rigid rotation model' model (reprinted with permission from [82,352,356]). These NTEs share the same behavior as water freezing, but at different temperature ranges, which is evidence of the essentiality of two types of coupled short-range interactions with specific-heat disparity to these materials.

## 11.2 Electro-, magneto-, and mechanico-freezing

Electric fields affect water freezing [358,359]. For example, the rate of ice nucleation from the vapor phase substantially increases in electric fields above $10^4$ Vm$^{-1}$ with respect to the normal growth rate [360]. A substrate charged with unlike charges has opposite effects on water freezing. Positively charged LiTaO$_3$ and SrTiO$_3$ films promote ice nucleation by elevating $T_m$; if negatively charged, they demote nucleation by lowering the freezing temperature [361]. Water droplets cooled down on a negatively charged LiTaO$_3$ surface remain liquid at -11°C; if the surface is positively charged, the droplets freeze immediately at -8°C, beginning at the solid/water interface; conversely, nucleation begins at the air/water interface if the surface is negatively charged [361].

The freezing temperature of water is also altered by electrical fields in narrow cracks at the hydrophobic faces of a-amino acid single crystals [362]. Ice forms between the STM tip and the substrate [358]. MD calculations predict that a strong electrical field ($> 10^9$ V m$^{-1}$) align water dipoles and crystallizes the water into polar cubic ice [363,364]. According to MD calculations (see Figure 54), the $T_m$ of monolayer ice decreases from 325 K to 278 K when the external electric field rises to $10^9$ Vm$^{-1}$ [138].

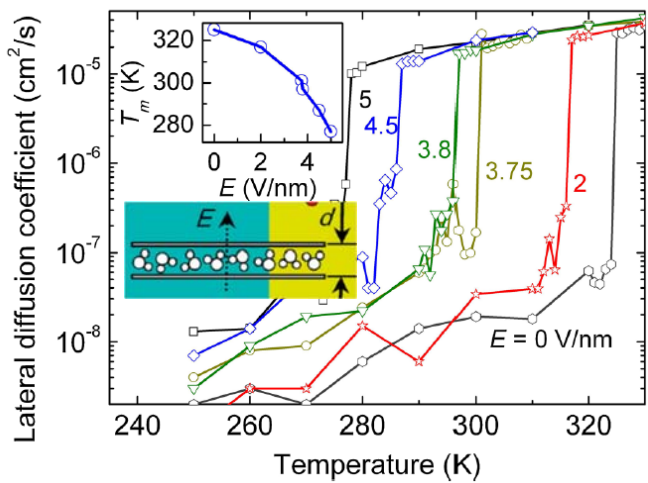

Figure 54. Temperature-dependence of the in-plane diffusion coefficient of the confined monolayer water of $d = 0.79$ nm thick (inset) under different external electric fields. The inset also shows the $T_m$ as a function of $E$. (Reprinted with permission from [138].)

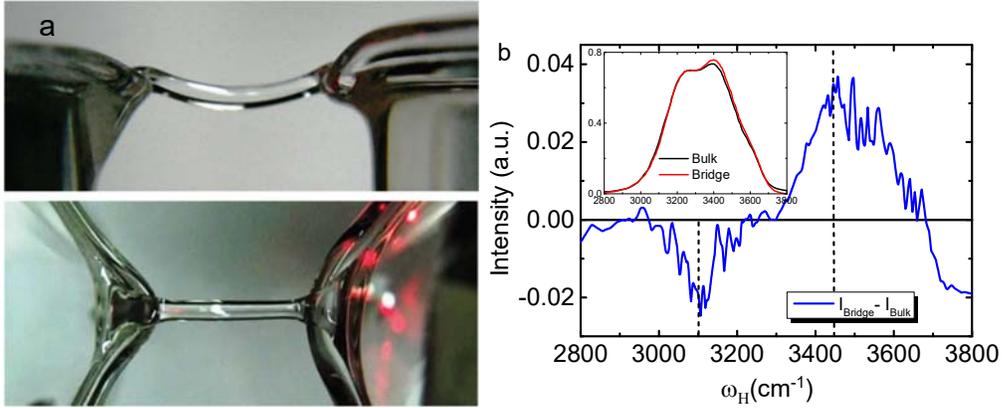

Figure 55. (a) Front and top views of the floating water bridge (14 mm long) formed between two beakers in a $10^6$ Vm$^{-1}$ field. (b) The residual phonon spectrum of the bridge (inset are the raw spectra) with respect to that of bulk water at 20°C revealed that the bridge exhibited skin supersolidity with $\omega_H$ from 3100/3150 cm$^{-1}$ to 3450 cm$^{-1}$ or slightly higher. (Reprinted with permission from [365,366].)

The floating water bridge is another interesting phenomenon, first reported by Armstrong [367] in 1893 and by Fuchs et al. [368] in 2007. Two beakers filled with deionized water form a bridge between them when they are subject to a d/c [365] or a/c (up to 1000 Hz) [366] voltage higher than 10 kV ($10^6$ Vm$^{-1}$ field) at room temperature. The bridge lasted for hours and was > 2 cm long (see Figure 55a) [365,369]. This experiment is stable and reproducible, which leads to the special condition that the water in the bridge can be assessed under high voltages and different atmospheric conditions [370].

There are two views on the forces equilibrating the bridge. One is tension along the bridge caused by the electric field within the dielectric material [371], and the other is surface tension [372]. The latter indicates that the electric field avoids the breakup of the bridge into small droplets and maintains its stability.

The tension due to the electric field in a dielectric medium is given by [371]:

$$T_{DE} = \varepsilon_0 (\varepsilon_r - 1) E^2 A,$$

(42)

where $A = \pi D^2/4$ is the cross-sectional area of the bridge, $\varepsilon_r$ is the relative permittivity of water, and $\varepsilon_0$ is the permittivity of the vacuum. If a tension $T_{DE}$ is acting on a curved bridge of curvature $\xi$, the vertical force produced per unit length of the bridge is $\xi T_{DE}$, while the gravitational force per unit length is $A\rho g$. Thus the ratio of the dielectric force and the gravitational force will be:

$$R_{DE} = \frac{\varepsilon_0 (\varepsilon_r - 1) E^2 \xi}{\rho g}.$$

(43)

The second mechanism suggests that the force holding the bridge is the surface tension only. According to Aerov [372], the electric tension along the bridge is zero. The electric field causes stability of the bridge and avoids it from breaking into droplets. The tension caused by surface tension is the sum of the tension on the sides ($\gamma P$) and the repulsion caused by the pressure jump at the surface ($-\gamma P/2$):

$$T_{ST} = \frac{\gamma P}{2},$$

(44)

where $P = \pi D$ is the perimeter of the cross-section of the bridge. According to this notation, the ratio between the upwards surface-tension force and the gravitational force is then:

$$R_{ST} = \frac{2\gamma\xi}{\rho g D}.$$

(45)

According to Namin et al. [365], surface tension and electrical tension contribute equally to holding the bridge.

However, the Raman residual phonon spectrum in Figure 55b reveals that the bridge formation raises the $\omega_H$ from 3100 cm$^{-1}$ to 3500 cm$^{-1}$, slightly higher than the characteristic peak of the supersolid skin at 3450 cm$^{-1}$. Therefore, polarization and viscoelasticity due to skin supersolidity may dominate in holding the bridge. The $\omega_H$ offset from 3450 to 3500 cm$^{-1}$ may indicate that the electric effect enhances skin supersolidity by promoting molecular polarization that shortens and stiffens the H-O bond further. Further investigation from the perspective of joint electrical and skin supersolidity would be interesting and revealing.

Magnetic fields also modulate water freezing in a much more irregular manner. MD calculations reveal that a 10 T magnetic field can raise the freezing temperature of a hydrophobically confined water nanodroplet up to 340 K [373]. Increasing the magnetic field up to 10 T, the surface tension of regular water at 298 K increases from 71.7 to 73.3 mN/m and that of D$_2$O increases to 74.0 mN/m [374] in a B$^2$ manner. $^1$H-NMR measurements revealed that a 0.01–1.0 T magnetic field reduces the surface tension but raises the viscosity of water [375]. However, a 60 × 10 T d/c magnetic field lowers the freezing point of the ambient water to -7°C [376]. Confirmation of the magnetic field effect on water freezing would be very interesting.

When subjected to about 1 GPa compression, liquid water turns to ice-VII at room temperature [82]. Figure 56 shows the pressure-dependent Raman spectra of water at 25°C. During the phase transition, the pressure suddenly drops from 1.35 to 0.86 GPa although the volume of the diamond compression cell containing the water shrinks continuously [377]. In the ice phase, the $\omega_x$ shifts cooperatively, following the trend of compressed ice at lower temperature. The sudden drop in pressure upon freezing may indicate a new mechanism for the O:H-O relaxation at the transition point.

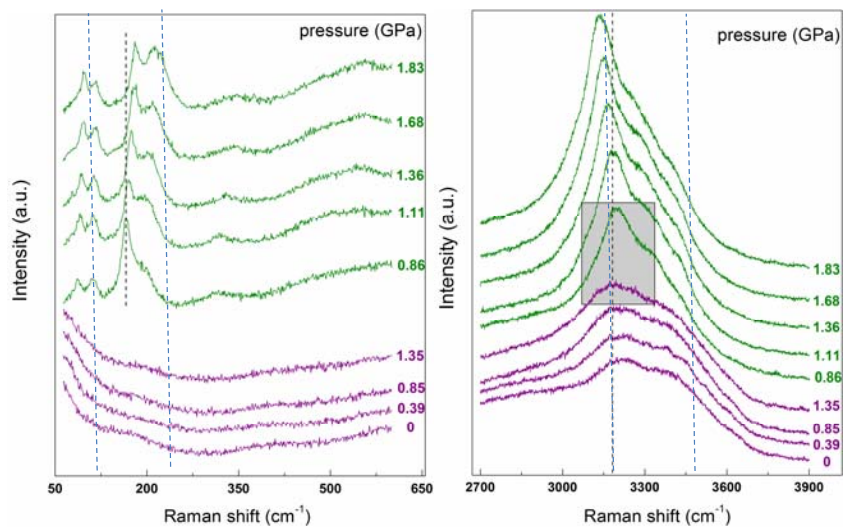

Figure 56. Raman spectra of the mechanico-freezing of ambient water reveals a sudden drop in pressure at freezing from 1.35 to 0.86 GPa. Hatched vertical lines denote the $\omega_x$ of the skin (75 and 3450 cm$^{-1}$), bulk (< 300 and 3200 cm$^{-1}$) and the $\omega_H$ for bulk ice (3150 cm$^{-1}$). The skin feature at 3450 cm$^{-1}$ persists throughout the range of pressures, which indicates that water contacts diamond hydrophobically. (Reprinted with permission from [377].)

How the electrical, magnetic and mechanical fields mediate $E_H$ could be the key to these issues, as $E_H$ determines the $T_m$ uniquely. The electrical and magnetic fields affect only moving electrons by polarization that may adjust the short-range potentials in their own manner. Further investigation on these multifield effects on O:H-O bond relaxation would be more viable.

11.3    Hofmeister series, de-icing, and anti-icing

The behavior of aqueous ions has a profound impact on biological molecules such as proteins and DNA, and thus have implications for health care and disease curing [378]. Ions added to protein solution in the form of salts, acids, sugars or buffer reagents are crucial to maintaining protein stability. Different ions may help to prevent aggregation and self-association of DNA by activating or deactivating ion channeling.

Hofmeister [379] discovered in late 1800s (see Figure 57) that certain aqueous ions follow a peculiar order in increasing or decreasing the ability of water to precipitate egg white in solution. Anions in particular, such as $SO_4^{2-}$, Cl$^-$ and SCN$^-$ follow a seemingly arbitrary sequence. In this order, they increasingly de-nature and dissolve proteins, and have either increasing or decreasing effects on many other solution properties, such as surface tension.

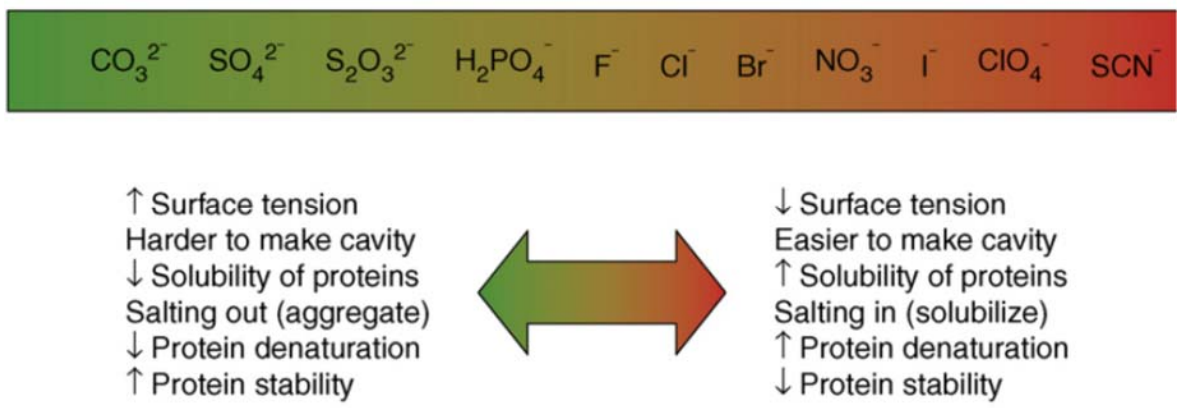

Figure 57. The Hofmeister series. (Reprinted with permission from [379].)

Randall and Failey [380-382] realized that the efficiency of common cations as salting-out agents follows these orders:

For anions:
$OH^- > SO_4^{2-}, CO_3^{2-} > ClO_4^- > BrO_3^- > Cl^- > CH_3COO^- > IO_3^-, IO_4^- > Br^-, I^- > NO_3^-$;

For cations:
$Na^+ > K^+ > Li^+ > Ba^{2+} > Rb^+ > Ca^{2+} > Ni^{2+} > Co^{2+} > Mg^{2+} > Fe^{2+} > Zn^{2+} > Cs^+ > Mn^{2+} > Al^{3+} > Fe^{3+}$, $Cr^{3+} > NH4^+ > H^+$

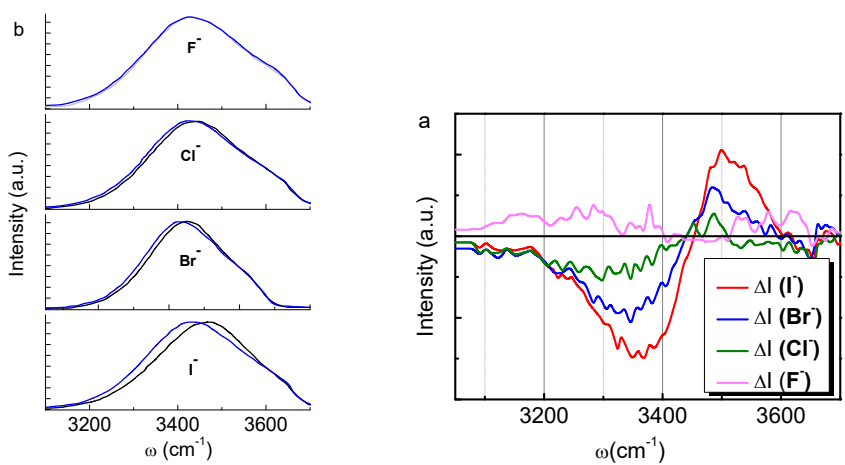

Figure 58. (a) Comparison of normalized $\omega_H$ for 1 M KX (X = $F^-$, $Cl^-$, $Br^-$, and $I^-$) in HOD/D$_2$O (black lines) with pure HOD/D$_2$O (blue lines); and (b) the residual phonon spectra with grid denoting the frequency offset due to ion replacement. (Reprinted with permission from [343].)

The effect of adding salts to solutions of nonelectrolytes is very complex, due to different types of intermolecular interactions that involve the ions, the solvent, and the solute molecules [383]. Figure 58 compares the H-O spectra of HOD/D$_2$O with and without presence of 1 M KX (X = $F^-$, $Cl^-$, $Br^-$, and $I^-$) [343]. Results show that larger ions with lower electronegativity shift the $\omega_H$ more than otherwise.

Using ultrafast 2DIR spectroscopy and MD simulations, Park and coworkers [345] found that 5% NaBr addition to HOD in $H_2O$ shifts the O-D stretching frequency from 2509 to 2539 cm$^{-1}$ depending on the relative number (8, 16, 32) of $H_2O$ molecules attached to a Br$^-$ ion. Adding varied concentrations of NaBr in the aqueous NaBr solutions disrupts the O:H-O bond network to some extent. HOD molecules are H-bonded to ions as HOD–Br$^-$, DOH–Br$^-$, and HDO–Na$^+$ in the hydration shells around the ions. Salt ions such as NaCl [344], NaBe [345], LiCl [384], $NaClO_4$ and $Mg(ClO_4)_2$ [385-387] also shift positively the vibration frequency of the hydroxyl group (-OH or -OD). Aqueous LiCl performs the same as adding $H_2O$/LiCl ratio from 100 to 6.7, by dropping the supercooling temperature from 248 K to 190 K [384]. The vibration frequency also changes with the pH value of aqueous solutions containing organic compounds such as $NH_4H_2PO_4$ [388]. These features share the same attribute as heating on the blueshift of the H-O stretching phonons [113,344]. Detailed examination of the type and concentration effect on the full-spectrum of phonon relaxation in combination the Lagrangian transformation would be an efficient method resolving the Hofmeister phenomena.

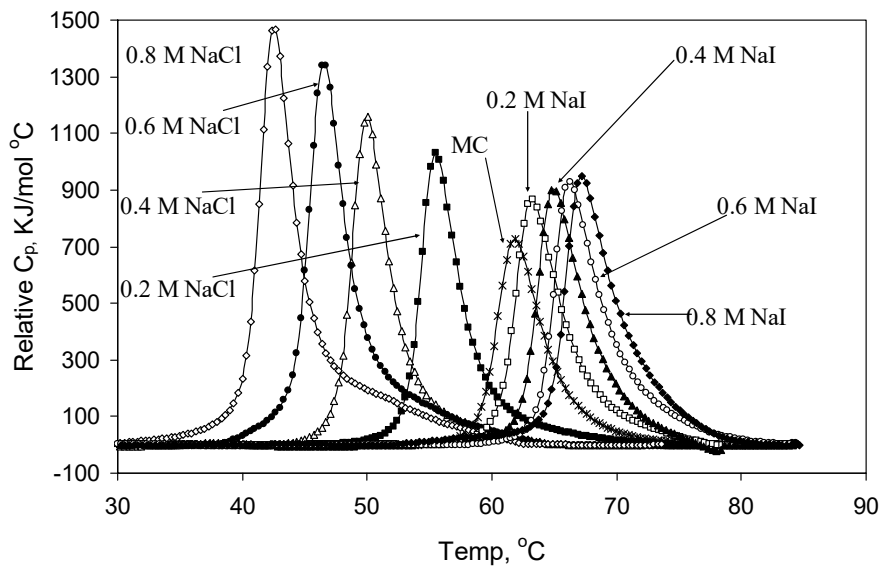

Figure 59. Salt concentration (mol/L) dependence of the heat capacity $C_p$ peaks for methylcellulose (MC) solutions shows the salt-out (NaCl) and salt-in (NaI) effect on sol–gel transition thermal dynamics. (Reprinted with permission from [389].)

Likewise, salt anions also affect the phase transition characterized by the $C_p$ peaks and enthalpies of methylcellulose (MC) aqueous solutions. Xu et al. [389,390] found in their differential scanning calorimetric measurement that NaCl exhibits the salt-out and NaI salt-in effect, as shown in Figure 59. Increasing NaCl concentration lowers the $C_p$ peak temperature and raises the peak intensity, but NaI has the opposite effect. The peak corresponds to the transition between the sol and the gel state of the material. They also found that NaScN has the same salt-in effect as NaI at a slight higher rate of the linear concentration dependence. Other salts show the salt-out effect, with the slopes in the order $NaNO_3 < NaBr < NaCl < NaSO_4 < Na_3PO_4$. The effect of salt-in and salt-out on the thermal behavior of the MC solution may share some common mechanism with the Hofmeister series. Interplay between the organic and the inorganic may be helpful in identifying the effect of anion substitution on O:H-O bond relaxation and its consequences for the performance of the relevant materials.

There are two possible mechanisms that might explain the Hofmeister series from the perspective of the length scale of interactions [378]. One is that ions produce long-range effects on the structure of water, leading to changes in the tendency for proteins to be precipitated from a water solution or remain dissolved. The general current view is that the Hofmeister effects stem largely from the varying capabilities of different salt ions to replace water at nonpolar molecular or macroscopic surfaces, but neither theoretical framework can yet predict these actions [383] nor explain them from the perspective of ion-replacement-induced O:H-O relaxation.

We have attempted to explore this problem from the perspective of Coulomb mediation of O:H-O cooperativity [121]. Figure 60 shows the residual FTIR $\omega_{B1}$ (≈ 550 cm$^{-1}$, corresponding to the ∠O:H-O bending mode) and $\omega_H$ (3200–3450 cm$^{-1}$) spectra of water upon being heated and salted with NaCl. The salted $\omega_H$ spectra are the same as those reported in [344,391]. Encouragingly, heating and salting have the same effect in relaxing $\omega_x$ phonons (although with a slight difference in the $\Delta\omega_L$ between heating and salting), which indicates that both heating and salting modulate the Coulomb repulsion between the electrons of ions by means of some common but as yet unknown mechanism.

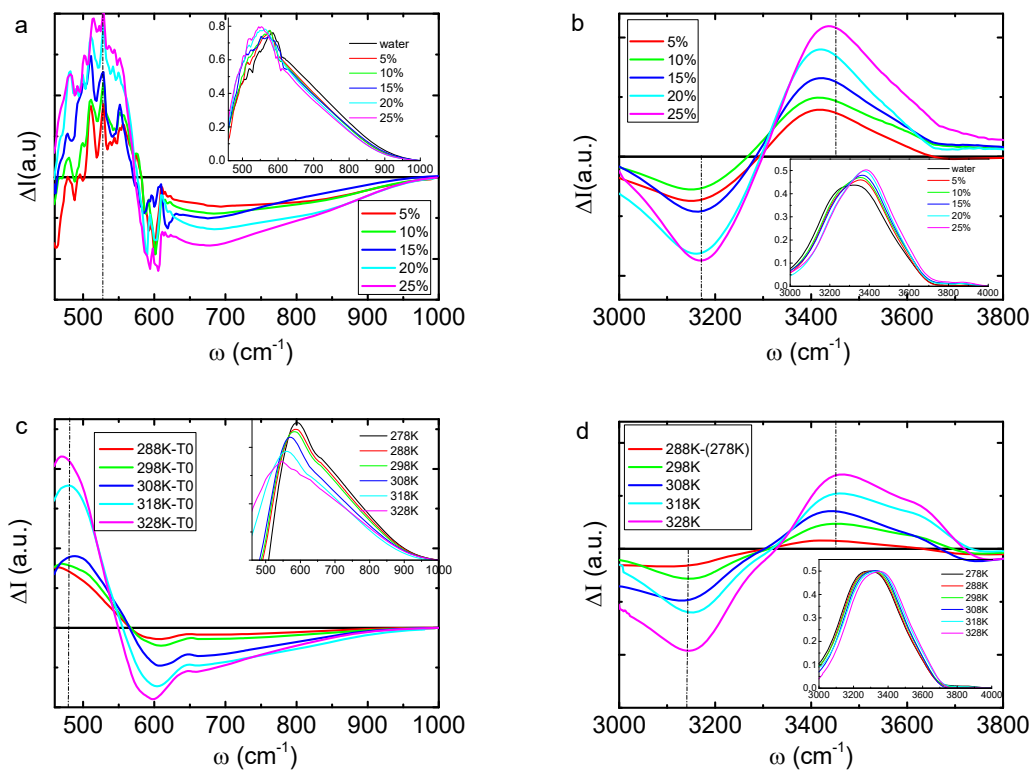

Figure 60. (a), (c) FTIR residual $\omega_{B1}$; and (b), (d) $\omega_H$ spectra of water as a function of (a), (b) salinity (mass% NaCl) and (c), (d) temperature with respect to the reference spectrum collected from unsalted water at 278 K. The insets show the raw spectra [392]. Heating shifts $\omega_{B1}$ from 600 to 480 cm$^{-1}$, while salting shifts $\omega_{B1}$ to 530 cm$^{-1}$. The $\omega_H$ shifts from 3150 to 3450 cm$^{-1}$, and is slightly sensitive to temperature and salting concentration.

However, dependence of the contact angle on temperature and NaCl concentration in Figure 61 reveals that heating lowers the surface tension but salting raises it, although both relax the $\omega_x$ phonon in the same manner. Replacement of the $O^{2-}$ ions by $Cl^-$ enhances the polarization within the O:H-O bond but heating weakens the O:H strength by depolarization [186]. The slight difference in the $\Delta\omega_L$ in Figure 61 may indicate heating depolarization.

The mechanism of $\omega_H$ heating stiffening may explain the salt effect on the $\omega_H$ stiffening in liquid water. As demonstrated in [5], heating lengthens the $d_{O-O}$, weakening the Coulomb repulsion by O:H nonbond thermal expansion, which shortens the H-O bond and stiffens the $\omega_H$ phonon. Larger ions with lower ionicity would weaken the Coulomb repulsion and soften the $\omega_H$ more. The extent of the shift in the order $F^- < Cl^- < Br^- < I^-$ (Figure 60) indeed follows this expectation. This realization may be the starting point for unlocking the Hofmeister series; further investigation is in progress.

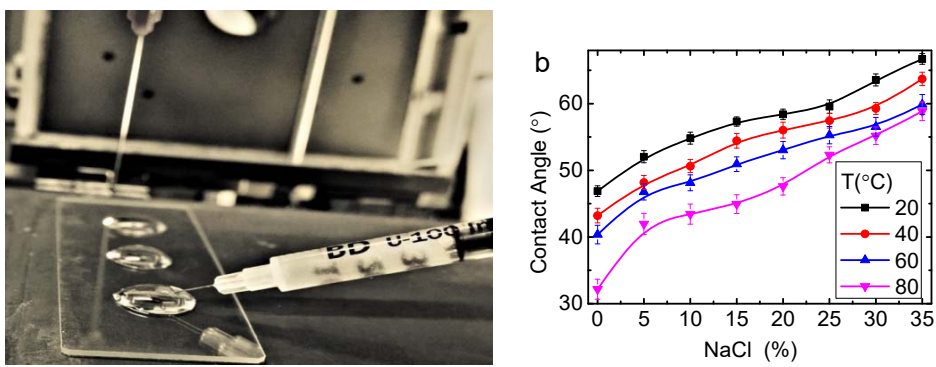

Figure 61. (a) Contact angle measurements of water droplets on a glass substrate as a function of (b) NaCl concentration and heating temperature. These results reveal that the evolution trends of the contact angle are opposite to IR vibration frequencies. Heating weakens the polarization that modulates the stiffness of the skin, whereas salting enhances it [186].

Conversely, sweet solutions are obtained upon heating mixtures of carbohydrates, urea and inorganic salts to moderate temperatures, to give new chiral media for organic reactions [393]. The solubility of sugar increases with temperature but drops with pressure [394]. Salt assists the dissociation of snow by softening the O:H nonbond and stiffening the H-O bond. Therefore, sugars and salts may share some common but unknown mechanism for the anti-icing and de-icing effect.

The preliminary understanding of the heating and salting effect on surface tension and $\omega_x$ shift may explain why hot water is a better cleaning agent than cold water. Soaps, sugars and detergents may functionalize in a similar way to heating and salting in washing and cleaning. Replacing $O^{2-}$ with ions of salts, acids or sugars, together with heating, perhaps weaken the Coulomb repulsion between charged ions. According to the present understanding, reduced repulsion weakens the O:H bond by reducing $E_L$. Thermal softening of the O:H bond decreases $E_L$ and surface tension, making hot water a better 'wetting agent' that can penetrate into pores and fissures rather than bridging them by surface tension. Soaps and detergents may further weaken the O:H nonbond and help the cleaning process. Lowering surface tension by shortening the H-O bond makes $H_2O$ smaller and more readily soak into pores and soils. Salted soils, being harmful to plant growth, may prevent plants from completely absorbing salted water.

11.4    Dielectric relaxation anomalies

The transition of electrons from the valence band to the conduction band determines the relaxation of the static dielectric constant of a substance [198]. Band-gap $E_G$ expansion, lattice relaxation, and electron–phonon coupling processes all contribute to the dielectric constant relaxation. Enhancement of interatomic interaction lowers the dielectric constant [198,395]. The dielectric permittivity ($\chi = \varepsilon_r - 1$) of a semiconductor is approximately proportional to the inverse square of its $E_G$ [395-397]. The refractive index, $n = \varepsilon_r^{1/2} = (\chi + 1)^{1/2}$, drops accordingly when the specimen is compressed or cooled, since $E_G$ increases under such stimuli [398-400]. The dielectric constants of semiconductors also drop in the skin region, and with solid size [198].

However, the refractive index of liquid water at room temperature increases with pressure in the same trend as density (see Figure 62). The dielectric constant also increases with cooling. Both compression and cooling soften the O:H nonbond and stiffen the H-O bond and increase the dielectric constant. This is contrary to the dielectric behavior of 'normal' materials; for example, Si exhibits dielectric loss when cooled or compressed.

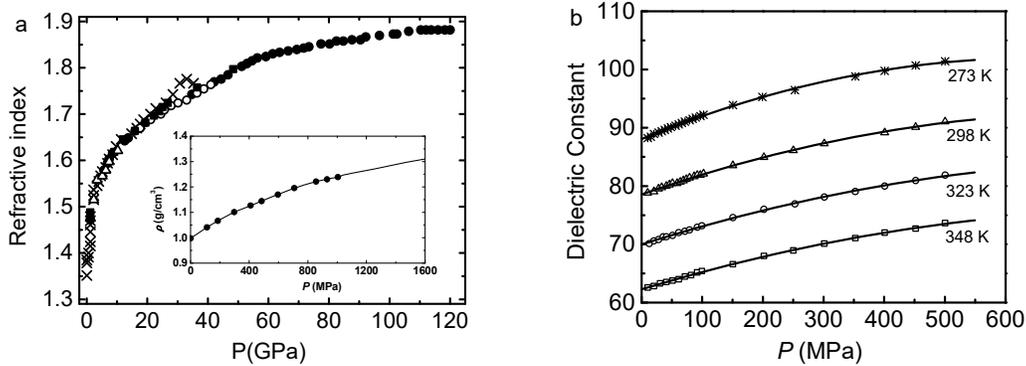

Figure 62. Enhancement of the refractive index in water by (a) compression, and (b) cooling. The inset in (a) shows the pressure-dependence of the mass density of water at 298 K. (Reprinted with permission from [21,401].)

The following possible mechanisms may be responsible for the dielectric enhancement of water when compressed or cooled. These stimuli shorten the O:H bond and lengthen the H-O bond, producing a gain in mass density yet a loss in $E_H$. Some or all of the following parameters may raise the dielectric constant of water:

$$\varepsilon_r \propto \begin{cases} \rho(d_{OO}) & (Mass\ density) \\ E_H^{-2} & (H-O\ energy) \\ P & (Polarization) \end{cases}$$

Generally, the dielectric constant is inversely proportional to the gradient of the band gap [395]. Band gap, electron–phonon interaction, polarization and bond length relaxation all contribute to the change of the dielectric constant. In the skin supersolid phase of water ice, how does the dielectric constant change with the molecular undercoordination? This question remains open for discussion. Formulation from the perspective of O:H-O bond relaxation is more meaningful and revealing than merely correlating the dielectric properties with mass density.

## 11.5 H$_2$O–cell and H$_2$O–DNA interaction

Understanding the interactions between water molecules and cells, membranes, proteins and so on is one of the most challenging areas in biology [8]. For instances, solvation water around proteins is denser than bulk water [27], implying that the H-O bond becomes longer and softer, and that the interaction is hydrophilic. Ice absorbs and entraps albumin protein in solution [10]. The geometry of the O:H-O network within the solvation layer differs from that in bulk water when interacting with a protein surface.

Figure 63 shows Raman $\omega_H$ spectra in normal (non-cancerous) and cancerous breast tissue (infiltrating ductal cancer) with bulk pure water [12]. These distinguish between the H-O features of cancerous tissue, healthy tissue and pure water, and reveal how the length and stiffness of the H-O bond change once water interacts with cancerous tissue.

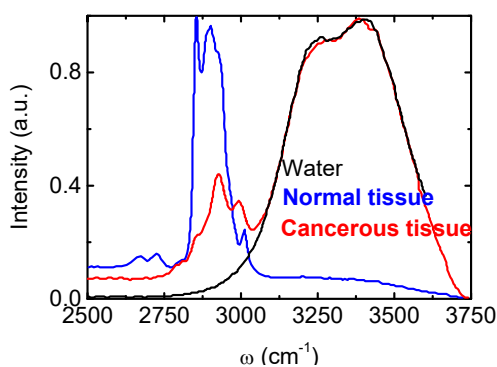

Figure 63. Raman $\omega_H$ spectra in normal (non-cancerous) and cancerous breast tissue compared with bulk pure water. (Reprinted with permission from [12].)

DNA is a most important bio-molecule because it stores all the information regarding the structure and function of every living cell. Interaction between DNA and water molecules largely determines its structure and function. However, little is known about the DNA hydration mechanism. Light-scattering from DNA and from proteins in general is very weak and it is in a different energy-transfer range. It is only possible to resolve vibrations that are largely due to water–water interactions with biomolecule perturbation [402]. The neutron diffraction spectrum (predominated by H motion) in the low-frequency region is mainly due to O:H-O bonding between water and DNA, or between water molecules. Inelastic incoherent neutron scattering provides information within the same energy ranges as IR and Raman spectra, but the intensity is sensitive to the phonon DOS.

Figure 64 shows the concentration-dependence of the neutron spectra of H$_2$O–DNA at 200 K. The H-O stretching modes of water around 400 meV shift to a higher frequency and the O:H stretching mode at < 5 meV approximately ($\omega_x$ = meV × 10$^5$ = meV × 80.7 cm$^{-1}$) shift to lower frequencies at lower hydration levels. Spectral peaks approach those of ice if the hydration level is sufficiently high. This effect is the same as for heating and salting, but in inverse order of concentration.

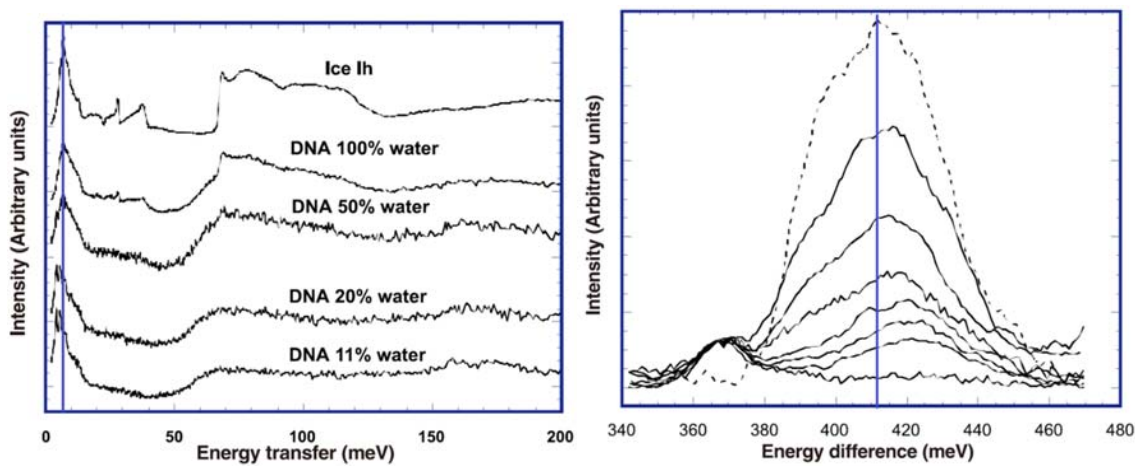

Figure 64. Neutron vibration spectra of $H_2O$–DNA at 200 K as a function of water concentration. (a) The percentage indicates grams of water per 100 g dry DNA. (b) Order of lyophilized DNA, upwards from the bottom curve, is 25, 50, 75, 100, 150, 200 g of water per 100 g DNA. Vertical lines denote the characteristic $\omega_x$ of net water (broken-line spectrum). (Reprinted with permission from [402].)

11.6    Wetting dynamics: Leidenfrost effect and droplet dancing

The Leidenfrost effect denotes the skittering of water when it hits a hotplate. When a liquid hits something that is very hot — about double the liquid's boiling point — it never comes directly in contact with its surface. The action is similar to that of liquid nitrogen on any flat surface. The liquid droplets each produce an insulating vapor layer that prevents the liquid from boiling rapidly, and acts as a barrier that appears to levitate the droplet. Water droplets will also 'climb' up a steep incline, the steepness increasing with the surface roughness, and the direction of the droplets' movement depends on the thermal field of the surface [403].

The dynamics of molecular evaporation and the momentum of the ejected molecule may help in understanding this phenomenon. A water molecule evaporates more readily at higher temperatures and less saturated vapour pressures. Both heating and unsaturated vapour pressure lengthen and soften the O:H bond with the memory effect. Therefore, the liquid‑vapor phase transition at the contact point ejects molecules with considerable momentum, applying a reaction impulse to the droplet. The direction of the impulse depends on the contact conditions: the component of the impulse parallel to an incline pushes the droplet up the incline; and the component normal to the horizontal hotplate surface separates the droplet from the hotplate. Theoretical formulation of the Leidenfrost effect, in particular the upward movement of the droplet, from the perspective of O:H-O bond memory and evaporation impulse could be of interest.

In dynamic electrical field experiments, Wang and Zhao [404,405] trapped tiny water droplets and caused them to 'dance', by combining the effects of surface tension, elastic force and electrical force to manipulate a flexible thin film and encapsulate or release a tiny droplet in a controlled and reversible manner. The film-supported droplet vibrated at twice the frequency of the input a/c signal. During this action, the droplet was observed to lie flat on the surface at the maximum applied voltage, and bent upward as the voltage was reduced.

Dynamic wetting of water droplets on either a hydrophobic or a hydrophilic surface of different roughnesses and at various temperatures is an interesting issue, as it is related to liquid–solid interactions. Yuan and Zhao [406,407] reported the development of a mechanism of multiple-scale dynamics of a moving contact line on lyophilic pillars with the scaling relation $R \approx t^x$ ($x = 1/3$ for a rough surface and 1/7 for a smooth surface), where $R$ is the spreading radius and $t$ is time. The spreading of a liquid drop on a hydrophilic, flexible pillared surface followed the same scaling relationship. The flexible pillars accelerated the liquid when the liquid approached, and trapping the liquid as it passed. The liquid deformed the pillars, resulting in energy dissipation at the moving contact line. The joint effect of the surface topology, intrinsic wettability and elasticity of a solid influenced the flow pattern and the flow field of the droplet on the pillar-arrayed surface.

11.7    O:H-O bond similarities in organic materials

Asymmetrical, short-range O:H-O bond potentials are intrinsic to specimens containing F, O, and N element. The short-range interactions and Coulomb coupling is applicable to the inter- and intramolecular interactions of these materials. For instance, Raman measurements have revealed coupled $\omega_L$ stiffening (110–290 cm$^{-1}$) and $\omega_H$ softening ($\approx$ 3000 cm$^{-1}$) in the O:H-N bonds in oxamide subjected to compression [408]. The pressure-trend of the Raman shifts of melamine-boric acid adduct ($C_3N_6H_{(6)} \cdot 2H_3BO_3$) super molecules [409], shown in Figure 65, exactly emulate the trend of compressed water ice [300].

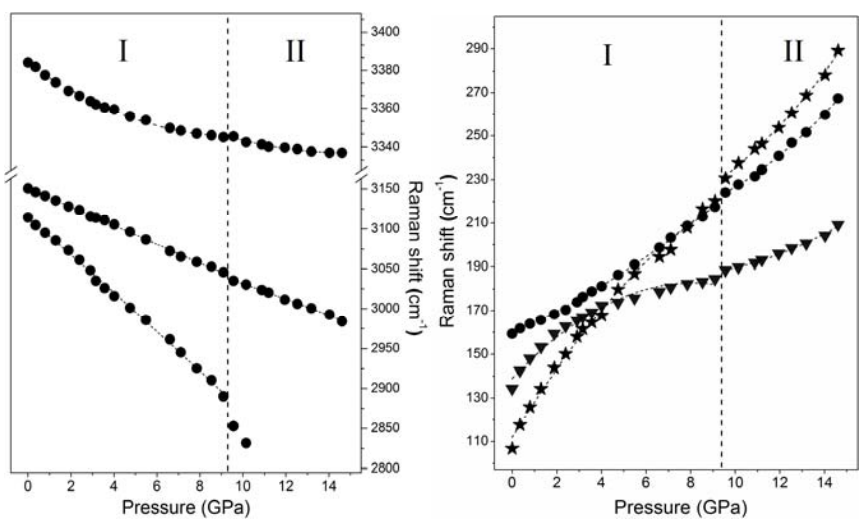

Figure 65. Compression stiffens the $\omega_L$ and softens the $\omega_H$ of the O:H-N bond in oxamide ($C_2H_4N_2O_2$) molecular crystal in addition to the abrupt phase transition at 9.5 GPa pressure. (Reprinted with permission from [408].)

Compression at pressures greater than 150 GPa also softens the phonons of hydrogen crystal ($\approx$ 4000 cm$^{-1}$) at various temperatures [410]. Computations reveal that compression symmetrizes the intra- and inter-$H_2$ molecular distance [411]. These findings may indicate that short-range inter- and intramolecular interactions and the Coulomb coupling exist in hydrogen crystals.

Therefore, O:H-O bond exists in a wide range of materials — $H_2O$, $NH_3$, HF, $H_2$, oxides, nitrides and fluorides — because of the presence of short-range interactions. N, O and F create nonbonding lone pairs upon reacting with atoms of other less-electronegative elements. Based on the current notation of O:H-O bond cooperativity, it is expected that asymmetrical relaxation in length and stiffness of the O:H-O bond dictates the functionality of species with O:H-O bond-like involvement, including biomolecules, organic materials, H crystals, among others.

## 12 Concluding remarks

The following tabulates the progress in quantitative and qualitative understanding of water ice in terms of O:H-O bond cooperative relaxation under the examined stimuli. Understanding these trends should be of use in controlling O:H-O bond relaxation dynamics and harnessing the performance of water ice and its solutions of salts and acids interaction with biomolecules.

| Quantities | | Bulk 4°C | Skin | Under-coordination | Cooling in | Solid/liquid cooling | Compressing |
|---|---|---|---|---|---|---|---|
| H-O length(Å) | $\Delta d_H$ | 1.000 | 0.840 | < 0 | | > 0 | |
| H-O energy (eV) | $\Delta E_H$ | 3.97 | 4.66 | | | | |
| O1s level shift (eV) | $\Delta E_{1s}$ | 536.6 | 538.1 | > 0 | | < 0 | |
| H-O frequency | $\Delta \omega_H$ | 3200 | 3450 | | | | |
| Critical $T_C$ (K) | $\Delta T_C/\Delta T_m$ | 273 | 315 | > 0 | – | – | < 0 |
| O:H length (Å) | $\Delta d_L$ | 1.695 | 2.180 | > 0 | | < 0 | |
| Mass density(g·cm$^{-3}$) | $\Delta \rho$ | 1 | 0.75 | | | | |
| O:H energy (eV) | $\Delta E_L$ | 0.1 | 0.08 | < 0 | | > 0 | |
| O:H frequency(cm$^{-1}$) | $\Delta \omega_L$ | 175 | 75 | | | | |
| Evaporation $T_C$ | $\Delta T_v$ | 373 | – | < 0 | – | – | > 0 |
| Debye T(K) | $\Delta \theta_D$ | 198 | – | | | | |
| Dipole moment | $\Delta P$ | | | > 0 dual-process polarization | | | > 0 |
| Elastic modulus | $\Delta Y$ | | – | | – | | |
| Viscosity | $\Delta \eta$ | | | | | | |
| Anomalies visited | Density unification of the length scale and geometry, potential paths, structure uniqueness of bulk mono-phase and nanodroplet bi-phase, H$^+$ centralization, $T_C$ depression, low compressibility, band gap expansion, skin supersolidity, slipperiness of ice, hydrophobicity and toughness of water skin, ice floating; 4-region $\rho$ and $\omega_x$ oscillation; H-bond memory, superheating, supercooling, Mpemba paradox, etc. | | | | | | |

The consistency between DFT-MD calculations, Lagrangian-Laplace transformation, Fourier fluid thermodynamic computation, and electron- and phonon spectroscopic observations has verified our hypotheses and expectations regarding the anomalies of water ice under mechanical compression, molecular undercoordination, thermal excitation, and their coupling. The following may update the understanding:

1) An extension of the Ice Rule produced the basic tetrahedron block that unifies the length scale, geometrical structure and mass density of molecular packing in water and ice, and the flexible, polarizable, and fluctuating O:H-O bond whose relaxation mediates the performances of water and ice.
2) The O:H-O bond is segmented into the stronger H-O polar-covalent bond and the weaker O:H nonbond with mechanical and thermal disparities. Such a master–slave-segmented O:H-O bond performs like a pair of asymmetrical, coupled, H-bridged oscillators with short-range interactions and memory, which represents all O-O interactions in water and ice, regardless of topological configuration or phase structure.
3) Inter-oxygen repulsion and O:H-O bond disparity discriminate water ice from other 'normal' materials in their response to external stimuli; if one becomes shorter and stronger, the other becomes longer and softer; the O:H always relaxes more than the H-O does; the O:H and H-O bond cooperative relaxations determine the physical anomalies of water ice.
4) H-O bond relaxation determines the $\omega_H$, $\Delta E_{1s}$, $E_H$ and $T_C$ for phase transitions except evaporation. H-O contraction not only raises the density and entraps bonding electrons but also polarizes the nonbonding electrons in a dual process. O:H relaxation determines the $E_L$ and $\omega_L$; and polarization increases elasticity, stiffness, viscosity and the dipole moment.
5) Compression shortens the softer O:H nonbond and simultaneously lengthens the stiffer H-O bond through Coulomb repulsion, leading to low-compressibility proton centralization. Reproduction of the $T_C$–$P$ profile confirms that H-O energy determines the $T_C$, $T_m$ and $E_H$ of 3.97 eV for bulk water and ice.
6) Molecular undercoordination has an opposite effect to compression, resulting in a supersolid phase pertaining to water molecules with fewer than four neighbors. The supersolid phase is elastic, polarized, hydrophobic, frictionless, and thermally more stable, and has ultra-low mass density.
7) Water and ice share the common supersolid skin ($\omega_H \approx 3450$ cm$^{-1}$) that causes the slipperiness of ice and toughens water skin with a strong repulsive force.
8) The shorter and stronger H-O bond between undercoordinated molecules dictates the $T_C$ for phase transition except for evaporation; Compression lengthens the H-O bond, which results in the drop of the freezing point of water to -22°C by 220 MPa compression.
9) Thermal excitation and the O:H-O bond-specific disparity result in oscillation of mass density and phonon stiffness across the full temperature range of water and ice. O:H relaxation dominates in both the liquid and solid phases, whereas H-O relaxation dominates at the solid–liquid phase transition. At temperatures below 80 K, segmental length and stiffness change insignificantly because of the extremely low specific heat of both segments.
10) Critical temperatures for extreme density vary with droplet size, because the stiffness of the H-O and the O:H bond change the respective Debye temperature of their specific-heat curves. Water nanodroplets undergo supercooling at freezing and superheating at melting.
11) Cooling stretching of the ∠O:H-O angle contributes positively to mass density at the freezing transition, but negatively at solid phase. Angle relaxation does not influence the macroscopic properties of water and ice.
12) The extraordinarily high heat capacity of liquid water arises from the stronger H-O bond (3.97 eV) than the O:H lone pair networks (≤ 0.1 eV).

13) Water prefers a tetrahedrally coordinated monophase structure with a supersolid skin of 0.75 g·cm$^{-3}$ density; however, at the nanometer scale, water droplets or bubbles prefer a two-phase 'core–shell' structure when the core and the skin volumes are compatible.
14) The cooperative O:H-O relaxation may be mediated by applying an electrical or magnetic field, or by replacing O$^{2-}$ with other species such as biomolecules or salts or acids, resulting in the Hofmeister series.
15) The Mpemba paradox arises intrinsically from O:H-O bond memory and water–skin supersolidity. The rate of heat emission depends on the initial energy storage, and the skin supersolidity promotes thermal diffusivity in the source liquid favoring an outward heat flow. Since it is sensitive to the source volume, the radiation rate and the drain temperature, the Mpemba effect occurs only in strictly non-adiabatic cycling systems.
16) Lagrangian solutions map the potential paths of O:H-O bond relaxation with $d_x$ and $\omega_x$ as input.

**Nomenclature**

| | |
|---|---|
| $\theta_{Dx}$ | O:H-O segmental Debye temperature (x = L for O:H nonbond; x = H for H-O bond) |
| $\Delta E_{1s}$ | O 1s core-level shift |
| $\omega_x$ | O:H-O segmental phonon frequency |
| $\eta_x$ | Specific heat of the x-th segment |
| BOLS | Bond order-length-strength correlation |
| NEP | Nonbonding electron polarization |
| CN(z) | Coordination number |
| DOS | Density of states |
| $d_x$ | O:H-O segmental length |
| $E_x$ | O:H-O segmental cohesive energy |
| FTIR | Fourier transformation infrared spectroscopy |
| H/LDL | High/low density liquid |
| $\rho$ | Mass density |
| NTE | Negative thermal expansion |
| RLS | Residual length spectroscopy |
| RPS | Residual phonon spectroscopy |
| SFG | Sum frequency generation |
| STM/S | Scanning tunneling microscopy/spectroscopy |
| $T_C$ | Critical temperature for phase transtion except for evaporation |
| $\theta$ | Temperature |
| $T_V/T_m$ | Evaporation/melting temperature |
| XPS | X-ray photoelectron spectroscopy |

Acknowledgments
Critical readings by Yi Sun and John Colligon, inspiring discussions with Jichen Li, Philip Ball, Even Robertson, Jer-Lai Kuo, Toshiaki Itaka, and financial support by NSF (Nos.: 21273191, 11274198, and 51102148) China are all gratefully acknowledged.

Affiliations
1. Key Laboratory of Low-Dimensional Materials and Application Technologies (Ministry of Education) and Faculty of Materials, Optoelectronics and Physics, Xiangtan University, Hunan 411105, China
2. NOVITAS, School of Electrical and Electronic Engineering, Nanyang Technological University,

Singapore 639798
3. Center for Coordination Bond and Electronic Engineering, College of Materials Science and Engineering, China Jiliang University, Hangzhou 310018, China
4. Institute of Nanosurface Science and Engineering, Shenzhen University, Shenzhen 518060, China
5. School of Materials Science, Jilin University, Changchun 130012, China
6. State Key Laboratory of New Ceramics and Fine Processing, Department of Materials Science and Engineering, Tsinghua University, Beijing 100084, China

Footnote

*These authors contribute equally

# Correspondence to: Ecqsun@ntu.edu.sg;

CQ is associated with honorary appointments at rest affiliations.

References
[1] Y. Huang, X. Zhang, Z. Ma, Y. Zhou, J. Zhou, W. Zheng, C.Q. Sun, Sci. Rep. 3 (2013) 3005.
[2] C.Q. Sun, X. Zhang, W.T. Zheng, Chem Sci 3 (2012) 1455-1460.
[3] C.Q. Sun, X. Zhang, J. Zhou, Y. Huang, Y. Zhou, W. Zheng, J. Phys. Chem.Lett. 4 (2013) 2565-2570.
[4] X. Zhang, Y. HUang, Z. Ma, J. Zhou, Y. Zhou, W. Zheng, C.Q. Sun, PCCP (2014) DOI: 10.1039/C1034CP02516D.
[5] C.Q. Sun, X. Zhang, X. Fu, W. Zheng, J.-l. Kuo, Y. Zhou, Z. Shen, J. Zhou, J. Phy. Chem. Lett. 4 (2013) 3238-3244.
[6] Y. Huang, X. Zhang, Z. Ma, Y. Zhou, G. Zhou, C.Q. Sun, J. Phys. Chem. B 117 (2013) 13639-13645.
[7] X. Zhang, Y. Huang, Z. Ma, Y. Zhou, W. Zheng, Q. Jiang, C.Q. Sun, PCCP (2014) DOI: 10.1039/C1034CP03669G.
[8] G.H. Zuo, J. Hu, H.P. Fang, Phys. Rev. E 79 (2009) 031925.
[9] J.L. Kulp, D.L. Pompliano, F. Guarnieri, J. Am. Chem. Soc. 133 (2011) 10740-10743.
[10] A. Twomey, R. Less, K. Kurata, H. Takamatsu, A. Aksan, J. Phys. Chem. B 117 (2013) 7889-7897.
[11] P. Ball, Nature 452 (2008) 291-292.
[12] I.V. Stiopkin, C. Weeraman, P.A. Pieniazek, F.Y. Shalhout, J.L. Skinner, A.V. Benderskii, Nature 474 (2011) 192-195.
[13] D. Marx, M.E. Tuckerman, J. Hutter, M. Parrinello, Nature 397 (1999) 601-604.
[14] Y. Yoshimura, S.T. Stewart, M. Somayazulu, H. Mao, R.J. Hemley, J. Chem. Phys. 124 (2006) 024502.
[15] D. Kang, J. Dai, Y. Hou, J. Yuan, J. Chem. Phys. 133 (2010) 014302.
[16] J.W.M. Frenken, T.H. Oosterkamp, Nature 464 (2010) 38-39.
[17] J.M. Headrick, E.G. Diken, R.S. Walters, N.I. Hammer, R.A. Christie, J. Cui, E.M. Myshakin, M.A. Duncan, M.A. Johnson, K.D. Jordan, Science 308 (2005) 1765-1769.
[18] J.K. Gregory, D.C. Clary, K. Liu, M.G. Brown, R.J. Saykally, Science 275 (1997) 814-817.
[19] N. Bjerrum, Science 115 (1952) 385-390.
[20] A.K. Soper, J. Teixeira, T. Head-Gordon, PNAS 107 (2010) E44-E44.
[21] C.S. Zha, R.J. Hemley, S.A. Gramsch, H.K. Mao, W.A. Bassett, J. Chem. Phys. 126 (2007) 074506.
[22] T. Bartels-Rausch, V. Bergeron, J.H.E. Cartwright, R. Escribano, J.L. Finney, H. Grothe, P.J. Gutiérrez, J. Haapala, W.F. Kuhs, J.B.C. Pettersson, S.D. Price, C.I. Sainz-Díaz, D.J. Stokes, G. Strazzulla, E.S. Thomson, H. Trinks, N. Uras-Aytemiz, Rev. Mod. Phys. 84 (2012) 885-944.
[23] R.J. Bakker, M. Baumgartner, Cent Eur J Geo 4 (2012) 225-237.
[24] R.J. Bakker, Can. Mineral. 42 (2004) 1283-1314.
[25] M. Smyth, J. Kohanoff, Phys. Rev. Lett. 106 (2011) 238108.

[26] P. Baaske, S. Duhr, D. Braun, Appl. Phys. Lett. 91 (2007) 133901.
[27] A. Kuffel, J. Zielkiewicz, J. Phys. Chem. B 116 (2012) 12113-12124.
[28] C. Castellano, J. Generosi, A. Congiu, R. Cantelli, Appl. Phys. Lett. 89 (2006) 233905.
[29] J.H. Park, N.R. Aluru, Appl. Phys. Lett. 96 (2010) 123703.
[30] F. Garczarek, K. Gerwert, Nature 439 (2006) 109-112.
[31] P. Ball, Chem. Rev. 108 (2008) 74-108.
[32] Y.B. Shan, E.T. Kim, M.P. Eastwood, R.O. Dror, M.A. Seeliger, D.E. Shaw, J. Am. Chem. Soc. 133 (2011) 9181-9183.
[33] J. Ostmeyer, S. Chakrapani, A.C. Pan, E. Perozo, B. Roux, Nature 501 (2013) 121-124.
[34] G. Malenkov, J Phys Condens Matter 21 (2009) 283101.
[35] H.M. Lee, S.B. Suh, J.Y. Lee, P. Tarakeshwar, K.S. Kim, J. Chem. Phys. 112 (2000) 9759.
[36] H.G. Lu, Y.K. Wang, Y.B. Wu, P. Yang, L.M. Li, S.D. Li, J. Chem. Phys. 129 (2008) 124512.
[37] C.K. Lin, C.C. Wu, Y.S. Wang, Y.T. Lee, H.C. Chang, J.L. Kuo, M.L. Klein, PCCP 7 (2005) 938-944.
[38] A. Lenz, L. Ojamae, J. Chem. Phys. 131 (2009) 134302.
[39] S.O.N. Lill, J. Phys. Chem. A 113 (2009) 10321-10326.
[40] S.N. Steinmann, C. Corminboeuf, Journal of Chemical Theory and Computation 7 (2011) 3567-3577.
[41] K. Kobayashi, M. Koshino, K. Suenaga, Phys. Rev. Lett. 106 (2011) 206101.
[42] A. Hermann, P. Schwerdtfeger, Phys. Rev. Lett. 106 (2011) 187403.
[43] W. Chen, X.F. Wu, R. Car, Phys. Rev. Lett. 105 (2010) 017802.
[44] Y. Wang, H. Liu, J. Lv, L. Zhu, H. Wang, Y. Ma, Nat Commun 2 (2011) 563.
[45] M. Abu-Samha, K.J. Borve, J. Chem. Phys. 128 (2008) 154710.
[46] O. Bjorneholm, F. Federmann, S. Kakar, T. Moller, J. Chem. Phys. 111 (1999) 546-550.
[47] G. Ohrwall, R.F. Fink, M. Tchaplyguine, L. Ojamae, M. Lundwall, R.R.T. Marinho, A.N. de Brito, S.L. Sorensen, M. Gisselbrecht, R. Feifel, T. Rander, A. Lindblad, J. Schulz, L.J. Saethre, N. Martensson, S. Svensson, O. Bjorneholm, J. Chem. Phys. 123 (2005) 054310.
[48] S. Hirabayashi, K.M.T. Yamada, J. Mol. Struct. 795 (2006) 78-83.
[49] P. Andersson, C. Steinbach, U. Buck, European Physical Journal D 24 (2003) 53-56.
[50] S. Maheshwary, N. Patel, N. Sathyamurthy, A.D. Kulkarni, S.R. Gadre, J. Phys. Chem. A 105 (2001) 10525-10537.
[51] A. Nilsson, L.G.M. Pettersson, Chem. Phys. 389 (2011) 1-34.
[52] G.N.I. Clark, C.D. Cappa, J.D. Smith, R.J. Saykally, T. Head-Gordon, Mol. Phys. 108 (2010) 1415-1433.
[53] C. Vega, J.L.F. Abascal, M.M. Conde, J.L. Aragones, Faraday Discuss. 141 (2009) 251-276.
[54] R. Ludwig, Chemphyschem 8 (2007) 938-943.
[55] B. Santra, A. Michaelides, M. Fuchs, A. Tkatchenko, C. Filippi, M. Scheffler, J. Chem. Phys. 129 (2008) 194111.
[56] K. Liu, J.D. Cruzan, R.J. Saykally, Science 271 (1996) 929-933.
[57] V. Buch, S. Bauerecker, J.P. Devlin, U. Buck, J.K. Kazimirski, Int. Rev. Phys. Chem. 23 (2004) 375-433.
[58] J.H.E. Cartwright, B. Escribano, C.I. Sainz-Diaz, Astrophys. J. 687 (2008) 1406-1414.
[59] Y. Li, G.A. Somorjai, J. Phys. Chem. C 111 (2007) 9631-9637.
[60] A.-M. Kietzig, S.G. Hatzikiriakos, P. Englezos, J. Appl. Phys. 107 (2010) 081101-081115.
[61] S.K. Sikka, S.M. Sharma, Phase Transitions 81 (2008) 907-934.
[62] P. Pruzan, J.C. Chervin, E. Wolanin, B. Canny, M. Gauthier, M. Hanfland, J. Raman Spectrosco. 34 (2003) 591-610.
[63] K. Davitt, E. Rolley, F. Caupin, A. Arvengas, S. Balibar, J. Chem. Phys. 133 (2010) 174507.

[64] M. Chaplin, Water 1 (2009).
[65] Y.R. Shen, V. Ostroverkhov, Chem. Rev. 106 (2006) 1140-1154.
[66] M. Faubel, K.R. Siefermann, Y. Liu, B. Abel, Acc. Chem. Res. 45 (2011) 120-130.
[67] A. Morita, T. Ishiyama, PCCP 10 (2008) 5801-5816.
[68] J.L. Skinner, P.A. Pieniazek, S.M. Gruenbaum, Acc. Chem. Res. 45 (2012) 93-100.
[69] C.H. Sun, L.M. Liu, A. Selloni, G.Q. Lu, S.C. Smith, J. Mater. Chem. 20 (2010) 10319-10334.
[70] M.A. Henderson, Surf. Sci. Rep. 46 (2002) 5-308.
[71] A. Hodgson, S. Haq, Surf. Sci. Rep. 64 (2009) 381-451.
[72] A. Verdaguer, G.M. Sacha, H. Bluhm, M. Salmeron, Chem. Rev. 106 (2006) 1478-1510.
[73] J. Carrasco, A. Hodgson, A. Michaelides, Nature Materials 11 (2012) 667-674.
[74] Y. Marcus, Chem. Rev. 109 (2009) 1346-1370.
[75] Y.R. Shen, J. Phys. Chem. C 116 (2012) 15505–15509.
[76] K.R. Wilson, B.S. Rude, T. Catalano, R.D. Schaller, J.G. Tobin, D.T. Co, R.J. Saykally, J. Phys. Chem. B 105 (2001) 3346-3349.
[77] S.N. Wren, D.J. Donaldson, PCCP 12 (2010) 2648-2654.
[78] T.F. Kahan, J.P. Reid, D.J. Donaldson, J. Phys. Chem. A 111 (2007) 11006-11012.
[79] F. Zaera, Chem. Rev. (2012).
[80] H.J. Bakker, J.L. Skinner, Chem. Rev. 110 (2010) 1498-1517.
[81] C.M. Johnson, S. Baldelli, Chem. Rev. DOI: 10.1021/cr4004902 (2014).
[82] M. Chaplin, Water structure and science.
[83] E.s. commentary, Science 309 (2005) 78-102.
[84] T. Head-Gordon, G. Hura, Chem. Rev. 102 (2002) 2651-2669.
[85] V. Molinero, E.B. Moore, J. Phys. Chem. B 113 (2009) 4008-4016.
[86] P.T. Kiss, A. Baranyai, J. Chem. Phys. 137 (2012) 084506-084508.
[87] J. Alejandre, G.A. Chapela, H. Saint-Martin, N. Mendoza, PCCP 13 (2011) 19728-19740.
[88] L. Pauling, J. Am. Chem. Soc. 57 (1935) 2680-2684.
[89] J. Teixeira, Nature 392 (1998) 232-233.
[90] P. Wernet, D. Nordlund, U. Bergmann, M. Cavalleri, M. Odelius, H. Ogasawara, L.A. Naslund, T.K. Hirsch, L. Ojamae, P. Glatzel, L.G.M. Pettersson, A. Nilsson, Science 304 (2004) 995-999.
[91] A.K. Soper, J Phys Condens Matter 17 (2005) S3273-S3282.
[92] K.T. Wikfeldt, M. Leetmaa, M.P. Ljungberg, A. Nilsson, L.G.M. Pettersson, J. Phys. Chem. B 113 (2009) 6246-6255.
[93] M. Leetmaa, K.T. Wikfeldt, M.P. Ljungberg, M. Odelius, J. Swenson, A. Nilsson, L.G.M. Pettersson, J. Chem. Phys. 129 (2008) 084502.
[94] R. Aswani, J.C. Li, J. Mol. Liq. 134 (2007) 120-128.
[95] T. Yokono, S. Shimokawa, M. Yokono, H. Hattori, Water 1 (2009) 29-34.
[96] J. Guo, X. Meng, J. Chen, J. Peng, J. Sheng, X.-Z. Li, L. Xu, J.-R. Shi, E. Wang, Y. Jiang, Nat Mater 13 (2014) 184-189.
[97] C.Q. Sun, Prog. Mater Sci. 48 (2003) 521-685.
[98] C.Q. Sun, Relaxation of the Chemical Bond, Springer press; ISBN: 978-981-4585-20-0, Berlin, 2014
[99] Water Thermal Properties - The Engineering Toolbox [Online]. Available: http://www.engineeringtoolbox.com/water-thermal-properties-d_162.html.
[100] J.R. Cooper, IAPWS Release (1994).
[101] M. Zhao, W.T. Zheng, J.C. Li, Z. Wen, M.X. Gu, C.Q. Sun, Phys. Rev. B 75 (2007) 085427.
[102] M.T. Suter, P.U. Andersson, J.B. Pettersson, J. Chem. Phys. 125 (2006) 174704.
[103] C. Vega, J.L.F. Abascal, P.G. Debenedetti, PCCP 13 (2011) 19660-19662.
[104] C. Vega, J.L.F. Abascal, PCCP 13 (2011) 19663-19688.

[105]    P.W. Atkins, Physical Chemistry, Oxford University Press 1990.
[106]    S.A. Harich, D.W.H. Hwang, X. Yang, J.J. Lin, X. Yang, R.N. Dixon, J Chem Phys 113 (2000) 10073-10090.
[107]    T.D. Kuhne, R.Z. Khaliullin, Nat commun 4 (2013) 1450.
[108]    V. Petkov, Y. Ren, M. Suchomel, J. Phys.: Condens. Matter 24 (2012) 155102.
[109]    A. Nilsson, C. Huang, L.G.M. Pettersson, J. Mol. Liq. 176 (2012) 2-16.
[110]    J.D. Bernal, R.H. Fowler, J. Chem. Phys. 1 (1933) 515-548.
[111]    W.T. Zheng, C.Q. Sun, Prog. Solid State Chem. 34 (2006) 1-20.
[112]    M. Hus, T. Urbic, J. Chem. Phys. 136 (2012) 144305.
[113]    P.C. Cross, J. Burnham, P.A. Leighton, J. Am. Chem. Soc. 59 (1937) 1134-1147.
[114]    C.Q. Sun, Vacuum 48 (1997) 525-530.
[115]    R. Hoffmann, American Scientist 102 (2014) 94.
[116]    C.Q. Sun, Nanoscale 2 (2010) 1930-1961.
[117]    R.F. McGuire, F.A. Momany, H.A. Scheraga, J. Phys. Chem. 76 (1972) 375-393.
[118]    N. Kumagai, K. Kawamura, T. Yokokawa, Mol. Simulat. 12 (1994) 177-186.
[119]    Y. Liu, J. Wu, J. Chem. Phys. 139 (2013) 041103.
[120]    X.Z. Li, B. Walker, A. Michaelides, PNAS 108 (2011) 6369-6373.
[121]    X. Zhang, C.Q. Sun, PCCP (2014) Communicated.
[122]    C.Q. Sun, Prog. Solid State Chem. 35 (2007) 1-159.
[123]    E. Roduner, Chem. Soc. Rev. 35 (2006) 583-592.
[124]    J.W. Li, S.Z. Ma, X.J. Liu, Z.F. Zhou, C.Q. Sun, Chem. Rev. 112 (2012) 2833-2852.
[125]    W.T. Zheng, C.Q. Sun, Energy & Environ Sci 4 (2011) 627-655.
[126]    V.M. Goldschmidt, Ber. Der Deut. Chem. Ges. 60 (1927) 1263-1296.
[127]    P.J. Feibelman, Phys. Rev. B 53 (1996) 13740-13746.
[128]    W.J. Huang, R. Sun, J. Tao, L.D. Menard, R.G. Nuzzo, J.M. Zuo, Nature Materials 7 (2008) 308-313.
[129]    C.Q. Sun, Prog. Mater Sci. 54 (2009) 179-307.
[130]    M.X. Gu, Y.C. Zhou, C.Q. Sun, J. Phys. Chem. B 112 (2008) 7992-7995.
[131]    M.A. Omar, Elementary Solid State Physics: Principles and Applications, Addison-Wesley, New York, 1993.
[132]    F.A. Lindemann, Phys. Z. 11 (1910) 609-612.
[133]    F. Mallamace, C. Branca, M. Broccio, C. Corsaro, C.Y. Mou, S.H. Chen, PNAS 104 (2007) 18387-18391.
[134]    M. Erko, D. Wallacher, A. Hoell, T. Hauss, I. Zizak, O. Paris, PCCP 14 (2012) 3852-3858.
[135]    K. Rottger, A. Endriss, J. Ihringer, S. Doyle, W.F. Kuhs, Acta Crystallographica B 50 (1994) 644-648.
[136]    E.B. Moore, V. Molinero, Nature 479 (2011) 506-508.
[137]    S.V. Lishchuk, N.P. Malomuzh, P.V. Makhlaichuk, Phys. Lett. A 375 (2011) 2656-2660.
[138]    H. Qiu, W. Guo, Phys. Rev. Lett. 110 (2013) 195701.
[139]    C. Wang, H. Lu, Z. Wang, P. Xiu, B. Zhou, G. Zuo, R. Wan, J. Hu, H. Fang, Phys. Rev. Lett. 103 (2009) 137801-137804.
[140]    H. Sun, J. Phys. Chem. B 102 (1998) 7338-7364.
[141]    V.F. Petrenko, R.W. Whitworth, Physics of ice, Clarendon Press, 1999.
[142]    P. Pruzan, J.C. Chervin, B. Canny, J. Chem. Phys. 97 (1992) 718-721.
[143]    P. Pruzan, J.C. Chervin, B. Canny, J. Chem. Phys. 99 (1993) 9842-9846.
[144]    J. Wang, Q.H. Qin, Y.L. Kang, X.Q. Li, Q.Q. Rong, Mech. Mater. 42 (2010) 537-547.
[145]    J.A. Hayward, J.R. Reimers, J. Chem. Phys. 106 (1997) 1518-1529.
[146]    H.C. Andersen, J. Chem. Phys. 72 (1980) 2384-2393.

[147] J.P. Perdew, Y. Wang, Phys. Rev. B 45 (1992) 13244-13249.
[148] F. Ortmann, F. Bechstedt, W.G. Schmidt, Phys. Rev. B 73 (2006) 205101.
[149] E.B. Wilson, J.C. Decius, P.C. Cross, Molecular Vibrations, Dover, New York, 1980.
[150] S.J. Clark, M.D. Segall, C.J. Pickard, P.J. Hasnip, M.J. Probert, K. Refson, M.C. Payne, Zeitschrift Fur Kristallographie 220 (2005) 567-570.
[151] J.P. Perdew, K. Burke, M. Ernzerhof, Phys. Rev. Lett. 78 (1997) 1396-1396.
[152] X. Su, Z.J. Zhang, M.M. Zhu, Applied Physics Letters 88 (2006) 061913-061913.
[153] C.Q. Sun, Atomic scale purification of electron spectroscopic information (USA patent: publication: Dec, 22 2011: WO 2011/159252), USA, 2011.
[154] L.X. Dang, T.-M. Chang, J. Chem. Phys. 106 (1997) 8149-8159.
[155] M.J.P. Nijmeijer, A.F. Bakker, C. Bruin, J.H. Sikkenk, J. Chem. Phys. 89 (1988) 3789-3792.
[156] M.S. Green, J. Chem. Phys. 20 (1952) 1281.
[157] R. Kubo, J. Phys. Soc. Jpn. 12 (1957) 570.
[158] L.N. Hand, J.D. Finch, Analytical Mechanics, Cambridge University Press, 2008.
[159] J. Fourier, The Analytical Theory of Heat Dover Publications, New York, 1955.
[160] C.Q. Sun, Y. Sun, Y.G. Nie, Y. Wang, J.S. Pan, G. Ouyang, L.K. Pan, Z. Sun, J. Phys. Chem. C 113 (2009) 16464-16467.
[161] C.Q. Sun, Phys. Rev. B 69 (2004) 045105.
[162] C.Q. Sun, Y. Nie, J. Pan, X. Zhang, S.Z. Ma, Y. Wang, W. Zheng, RSC Advances 2 (2012) 2377-2383.
[163] C.Q. Sun, Y. Wang, Y.G. Nie, Y. Sun, J.S. Pan, L.K. Pan, Z. Sun, J. Phys. Chem. C 113 (2009) 21889-21894.
[164] G. Ouyang, G.W. Yang, C.Q. Sun, W.G. Zhu, Small 4 (2008) 1359-1362.
[165] M. Faraday, Proc. R. Soc. London 10 (1859) 440-450.
[166] J. Thomson, Proc. R. Soc. London 11 (1860) 198-204.
[167] J.D. Goddard, AIChE J. 60 (2014) 1488-1498.
[168] D.T.F. Moehlmann, Cryobiology 58 (2009) 256-261.
[169] D. Errandonea, B. Schwager, R. Ditz, C. Gessmann, R. Boehler, M. Ross, Phys. Rev. B 63 (2001) 132104.
[170] Z.W. Chen, C.Q. Sun, Y.C. Zhou, O.Y. Gang, J. Phys. Chem. C 112 (2008) 2423-2427.
[171] K. Aoki, H. Yamawaki, M. Sakashita, Phys. Rev. Lett. 76 (1996) 784-786.
[172] M. Song, H. Yamawaki, H. Fujihisa, M. Sakashita, K. Aoki, Phys. Rev. B 60 (1999) 12644.
[173] T. Hynninen, V. Heinonen, C.L. Dias, M. Karttunen, A.S. Foster, T. Ala-Nissila, Phys. Rev. Lett. 105 (2010).
[174] D. Petely, http://ihrrblog.org/2013/11/08/our-strange-desire-to-find-a-landslide-trigger/ (2013).
[175] C.Q. Sun, Communicated.
[176] G.M. Marion, S.D. Jakubowski, Cold Regions Science and Technology 38 (2004) 211-218.
[177] W. Holzapfel, J. Chem. Phys. 56 (1972) 712.
[178] A.F. Goncharov, V.V. Struzhkin, H.-k. Mao, R.J. Hemley, Phys. Rev. Lett. 83 (1999) 1998-2001.
[179] M. Benoit, D. Marx, M. Parrinello, Nature 392 (1998) 258-261.
[180] P. Loubeyre, R. LeToullec, E. Wolanin, M. Hanfland, D. Husermann, Nature 397 (1999) 503-506.
[181] A.F. Goncharov, V.V. Struzhkin, M.S. Somayazulu, R.J. Hemley, H.K. Mao, Science 273 (1996) 218-220.
[182] D.D. Kang, J. Dai, H. Sun, Y. Hou, J. Yuan, Sci. Rep. 3 (2013) 3272.
[183] M. Song, H. Yamawaki, H. Fujihisa, M. Sakashita, K. Aoki, Phys. Rev. B 68 (2003) 014106.
[184] K. Aoki, H. Yamawaki, M. Sakashita, H. Fujihisa, Phys. Rev. B 54 (1996) 15673-15677.

[185] V.V. Struzhkin, A.F. Goncharov, R.J. Hemley, H.K. Mao, Phys. Rev. Lett. 78 (1997) 4446-4449.
[186] D.D. Kang, J.Y. Dai, J.M. Yuan, J. Chem. Phys. 135 (2011) 024505.
[187] K. Umemoto, R.M. Wentzcovitch, S. de Gironcoli, S. Baroni, Chem. Phys. Lett. 499 (2010) 236-240.
[188] I.A. Ryzhkin, J. Exp. Ther. Phys. 88 (1999) 1208-1211.
[189] F.H. Stillinger, K.S. Schweizer, J. Phys. Chem. 87 (1983) 4281-4288.
[190] L.N. Tian, A.I. Kolesnikov, J.C. Li, J. Chem. Phys. 137 (2012) 204507.
[191] M.X. Gu, Y.C. Zhou, L.K. Pan, Z. Sun, S.Z. Wang, C.Q. Sun, J. Appl. Phys. 102 (2007) 083524.
[192] M.X. Gu, L.K. Pan, T.C.A. Yeung, B.K. Tay, C.Q. Sun, J. Phys. Chem. C 111 (2007) 13606-13610.
[193] C. Yang, Z.F. Zhou, J.W. Li, X.X. Yang, W. Qin, R. Jiang, N.G. Guo, Y. Wang, C.Q. Sun, Nanoscale 4 (2012) 1304-1307.
[194] Y. Yoshimura, S.T. Stewart, M. Somayazulu, H.K. Mao, R.J. Hemley, J. Phys. Chem. B 115 (2011) 3756-3760.
[195] T. Okada, K. Komatsu, T. Kawamoto, T. Yamanaka, H. Kagi, Spectrochimica Acta A 61 (2005) 2423-2427.
[196] B. Santra, J. Klimeš, D. Alfè, A. Tkatchenko, B. Slater, A. Michaelides, R. Car, M. Scheffler, Phys. Rev. Lett. 107 (2011) 185701.
[197] R. Ludwig, Angew. Chem. Int. Ed. 40 (2001) 1808-1827.
[198] L.K. Pan, S.Q. Xu, W. Qin, X.J. Liu, Z. Sun, W.T. Zheng, C.Q. Sun, Surf. Sci. Rep. 68 (2013) 418-455.
[199] C.Q. Sun, H.L. Bai, B.K. Tay, S. Li, E.Y. Jiang, J. Phys. Chem. B 107 (2003) 7544-7546.
[200] F.N. Keutsch, R.J. Saykally, PNAS 98 (2001) 10533-10540.
[201] F. Li, Y. Liu, L. Wang, J. Zhao, Z. Chen, Theor. Chem. Acc. 131 (2012).
[202] A. Michaelides, K. Morgenstern, Nature Materials 6 (2007) 597-601.
[203] L. Turi, W.S. Sheu, P.J. Rossky, Science 309 (2005) 914-917.
[204] J.R.R. Verlet, A.E. Bragg, A. Kammrath, O. Cheshnovsky, D.M. Neumark, Science 307 (2005) 93-96.
[205] N.I. Hammer, J.W. Shin, J.M. Headrick, E.G. Diken, J.R. Roscioli, G.H. Weddle, M.A. Johnson, Science 306 (2004) 675-679.
[206] C. Perez, M.T. Muckle, D.P. Zaleski, N.A. Seifert, B. Temelso, G.C. Shields, Z. Kisiel, B.H. Pate, Science 336 (2012) 897-901.
[207] T. Ishiyama, H. Takahashi, A. Morita, J. Phys. Chem.Lett. 3 (2012) 3001-3006.
[208] F. Li, L. Wang, J. Zhao, J.R.-H. Xie, K.E. Riley, Z. Chen, Theor. Chem. Acc. 130 (2011) 341-352.
[209] M.L. Lakhanpal, B.R. Puri, Nature 172 (1953) 917-917.
[210] L. Li, Y. Kazoe, K. Mawatari, Y. Sugii, T. Kitamori, J. Phys. Chem.Lett. (2012) 2447-2452.
[211] K. Xu, P.G. Cao, J.R. Heath, Science 329 (2010) 1188-1191.
[212] D. Xu, K.M. Liechti, K. Ravi-Chandar, Langmuir 25 (2009) 12870-12873.
[213] P.B. Miranda, L. Xu, Y.R. Shen, M. Salmeron, Phys. Rev. Lett. 81 (1998) 5876-5879.
[214] F. McBride, G.R. Darling, K. Pussi, A. Hodgson, Phys. Rev. Lett. 106 (2011) 226101.
[215] S. Meng, E.G. Wang, S.W. Gao, Phys. Rev. B 69 (2004) 195404.
[216] J.C. Johnston, N. Kastelowitz, V. Molinero, J. Chem. Phys. 133 (2010) 154516.
[217] S. Strazdaite, J. Versluis, E.H. Backus, H.J. Bakker, J Chem Phys 140 (2014) 054711.
[218] A. Uysal, M. Chu, B. Stripe, A. Timalsina, S. Chattopadhyay, C.M. Schlepütz, T.J. Marks, P. Dutta, Phys. Rev. B 88 (2013) 035431.
[219] F.G. Alabarse, J. Haines, O. Cambon, C. Levelut, D. Bourgogne, A. Haidoux, D. Granier, B.

Coasne, Phys. Rev. Lett. 109 (2012) 035701.
[220]	E.B. Moore, E. de la Llave, K. Welke, D.A. Scherlis, V. Molinero, PCCP 12 (2010) 4124-4134.
[221]	Q. Yuan, Y.P. Zhao, Proc. R. Soc. A 468 (2011) 310-322.
[222]	M. Kasuya, M. Hino, H. Yamada, M. Mizukami, H. Mori, S. Kajita, T. Ohmori, A. Suzuki, K. Kurihara, J. Phys. Chem. C 117 (2013) 13540-13546.
[223]	M. James, T.A. Darwish, S. Ciampi, S.O. Sylvester, Z.M. Zhang, A. Ng, J.J. Gooding, T.L. Hanley, Soft Matter 7 (2011) 5309-5318.
[224]	G. Free, Surface Tension Droplets at 2500fps; http://www.youtube.com/watch?v=ynk4vJa-VaQ, 2013.
[225]	C. Antonini, I. Bernagozzi, S. Jung, D. Poulikakos, M. Marengo, Phys. Rev. Lett. 111 (2013) 014501.
[226]	E. Pennisi, Science 343 (2014) 1194-1197.
[227]	O. Marsalek, F. Uhlig, T. Frigato, B. Schmidt, P. Jungwirth, Phys. Rev. Lett. 105 (2010) 043002.
[228]	S. Liu, J. Luo, G. Xie, D. Guo, J. Appl. Phys. 105 (2009) 124301-124304.
[229]	K.R. Siefermann, Y. Liu, E. Lugovoy, O. Link, M. Faubel, U. Buck, B. Winter, B. Abel, Nature Chemistry 2 (2010) 274-279.
[230]	D.H. Paik, I.R. Lee, D.S. Yang, J.S. Baskin, A.H. Zewail, Science 306 (2004) 672-675.
[231]	R. Vacha, O. Marsalek, A.P. Willard, D.J. Bonthuis, R.R. Netz, P. Jungwirth, J. Phy. Chem. Lett. 3 (2012) 107-111.
[232]	F. Baletto, C. Cavazzoni, S. Scandolo, Phys. Rev. Lett. 95 (2005) 176801.
[233]	M. Abu-Samha, K.J. Borve, M. Winkler, J. Harnes, L.J. Saethre, A. Lindblad, H. Bergersen, G. Ohrwall, O. Bjorneholm, S. Svensson, Journal of Physics B-Atomic Molecular and Optical Physics 42 (2009) 055201.
[234]	K. Nishizawa, N. Kurahashi, K. Sekiguchi, T. Mizuno, Y. Ogi, T. Horio, M. Oura, N. Kosugi, T. Suzuki, PCCP 13 (2011) 413-417.
[235]	B. Winter, E.F. Aziz, U. Hergenhahn, M. Faubel, I.V. Hertel, J. Chem. Phys. 126 (2007) 124504.
[236]	L. Belau, K.R. Wilson, S.R. Leone, M. Ahmed, J. Phys. Chem. A 111 (2007) 10075-10083.
[237]	J. Ceponkus, P. Uvdal, B. Nelander, J. Chem. Phys. 129 (2008) 194306.
[238]	J. Ceponkus, P. Uvdal, B. Nelander, J. Chem. Phys. 134 (2011) 064309.
[239]	J. Ceponkus, P. Uvdal, B. Nelander, J. Phys. Chem. A 116 (2012) 4842-4850.
[240]	X.X. Yang, J.W. Li, Z.F. Zhou, Y. Wang, L.W. Yang, W.T. Zheng, C.Q. Sun, Nanoscale 4 (2012) 502-510.
[241]	X.J. Liu, L.K. Pan, Z. Sun, Y.M. Chen, X.X. Yang, L.W. Yang, Z.F. Zhou, C.Q. Sun, J. Appl. Phys. 110 (2011) 044322.
[242]	U. Buck, F. Huisken, Chem. Rev. 100 (2000) 3863-3890.
[243]	K.E. Otto, Z. Xue, P. Zielke, M.A. Suhm, PCCP 16 (2014) 9849-9858.
[244]	Q. Sun, Vib. Spectrosc 51 (2009) 213-217.
[245]	C.C. Pradzynski, R.M. Forck, T. Zeuch, P. Slavicek, U. Buck, Science 337 (2012) 1529-1532.
[246]	X. Wei, P. Miranda, Y. Shen, Phys. Rev. Lett. 86 (2001) 1554-1557.
[247]	D.D.D. Ma, C.S. Lee, F.C.K. Au, S.Y. Tong, S.T. Lee, Science 299 (2003) 1874-1877.
[248]	K.R. Wilson, R.D. Schaller, D.T. Co, R.J. Saykally, B.S. Rude, T. Catalano, J.D. Bozek, J. Chem. Phys. 117 (2002) 7738-7744.
[249]	U. Bergmann, A. Di Cicco, P. Wernet, E. Principi, P. Glatzel, A. Nilsson, J. Chem. Phys. 127 (2007) 174504.
[250]	E.G. Solveyra, E. de la Llave, V. Molinero, G. Soler-Illia, D.A. Scherlis, J. Phys. Chem. C 117 (2013) 3330-3342.

[251] R.M. Townsend, S.A. Rice, J. Chem. Phys. 94 (1991) 2207-2218.
[252] Y.I. Tarasevich, Colloid J. 73 (2011) 257-266.
[253] B.H. Chai, H. Yoo, G.H. Pollack, J. Phys. Chem. B 113 (2009) 13953-13958.
[254] M. Faraday, Experimental researches in chemical and physics, Tayler and Francis London, 1859.
[255] I.A. Ryzhkin, V.F. Petrenko, J. Exp. Ther. Phys. 108 (2009) 68-71.
[256] Y. Furukawa, M. Yamamoto, T. Kuroda, Journal of Crystal Growth 82 (1987) 665-677.
[257] J. Thomson, Camb. Dubl. Math. J. 11 (1850) 248-255.
[258] F.P. Bowden, T.P. Hughes, Proc. R. Soc. London A172 (1939) 280.
[259] R. Rosenberg, Phys. Today (2005) 50-55.
[260] S.C. Colbeck, L. Najarian, H.B. Smith, Am. J. Phys. 65 (1997) 488.
[261] Y.B. Fan, X. Chen, L.J. Yang, P.S. Cremer, Y.Q. Gao, J. Phys. Chem. B 113 (2009) 11672-11679.
[262] C. Huang, K.T. Wikfeldt, D. Nordlund, U. Bergmann, T. McQueen, J. Sellberg, L.G.M. Pettersson, A. Nilsson, PCCP 13 (2011) 19997-20007.
[263] S.A. Deshmukh, S.K. Sankaranarayanan, D.C. Mancini, J. Phys. Chem. B 116 (2012) 5501-5515.
[264] M. Sulpizi, M. Salanne, M. Sprik, M.-P. Gaigeot, J. Phys. Chem. Lett. 4 (2012) 83-87.
[265] C.Q. Sun, Y. Sun, Y.G. Ni, X. Zhang, J.S. Pan, X.H. Wang, J. Zhou, L.T. Li, W.T. Zheng, S.S. Yu, L.K. Pan, Z. Sun, J. Phys. Chem. C 113 (2009) 20009-20019.
[266] F. Yang, X. Wang, M. Yang, A. Krishtal, C. van Alsenoy, P. Delarue, P. Senet, PCCP 12 (2010) 9239-9248.
[267] L. Vrbka, P. Jungwirth, J. Phys. Chem. B 110 (2006) 18126-18129.
[268] D. Donadio, P. Raiteri, M. Parrinello, J. Phys. Chem. B 109 (2005) 5421-5424.
[269] S.R. Friedman, M. Khalil, P. Taborek, Phys. Rev. Lett. 111 (2013) 226101.
[270] D.P. Singh, J.P. Singh, Appl. Phys. Lett. 102 (2013) 243112.
[271] J. Li, Y.X. Li, X. Yu, W.J. Ye, C.Q. Sun, J. Phys. D: Appl. Phys. 42 (2009) 045406.
[272] M.J. Holmes, N.G. Parker, M.J.W. Povey, J. Phys.: Conf. Ser. 269 (2011) 012011.
[273] K.B. Jinesh, J.W.M. Frenken, Phys. Rev. Lett. 101 (2008) 036101.
[274] M. Whitby, L. Cagnon, M. Thanou, N. Quirke, Nano Lett. 8 (2008) 2632-2637.
[275] J.A. Thomas, A.J.H. McGaughey, Nano Lett. 8 (2008) 2788-2793.
[276] X.C. Qin, Q.Z. Yuan, Y.P. Zhao, S.B. Xie, Z.F. Liu, Nano Lett. 11 (2011) 2173-2177.
[277] C.Q. Sun, B.K. Tay, S.P. Lau, X.W. Sun, X.T. Zeng, S. Li, H.L. Bai, H. Liu, Z.H. Liu, E.Y. Jiang, J. Appl. Phys. 90 (2001) 2615-2617.
[278] C. Lu, Y.W. Mai, P.L. Tam, Y.G. Shen, Philos. Mag. Lett. 87 (2007) 409-415.
[279] T. Head-Gordon, M.E. Johnson, PNAS 103 (2006) 7973-7977.
[280] G.N. Clark, G.L. Hura, J. Teixeira, A.K. Soper, T. Head-Gordon, PNAS 107 (2010) 14003-14007.
[281] A.J. Stone, Science 315 (2007) 1228-1229.
[282] K. Stokely, M.G. Mazza, H.E. Stanley, G. Franzese, PNAS 107 (2010) 1301-1306.
[283] C. Huang, K.T. Wikfeldt, T. Tokushima, D. Nordlund, Y. Harada, U. Bergmann, M. Niebuhr, T.M. Weiss, Y. Horikawa, M. Leetmaa, M.P. Ljungberg, O. Takahashi, A. Lenz, L. Ojamäe, A.P. Lyubartsev, S. Shin, L.G.M. Pettersson, A. Nilsson, PNAS 106 (2009) 15214-15218.
[284] N.J. English, J.S. Tse, Phys. Rev. Lett. 106 (2011) 037801.
[285] F. Mallamace, M. Broccio, C. Corsaro, A. Faraone, D. Majolino, V. Venuti, L. Liu, C.Y. Mou, S.H. Chen, PNAS 104 (2007) 424-428.
[286] O. Mishima, H.E. Stanley, Nature 396 (1998) 329-335.
[287] M. Matsumoto, Phys. Rev. Lett. 103 (2009) 017801.

[288] J.C. Li, A.I. Kolesnikov, J. Mol. Liq. 100 (2002) 1-39.
[289] Y.S. Tu, H.P. Fang, Phys. Rev. E 79 (2009) 016707.
[290] I. Calizo, A.A. Balandin, W. Bao, F. Miao, C.N. Lau, Nano Lett. 7 (2007) 2645-2649.
[291] H.Q. Zhou, C.Y. Qiu, H.C. Yang, F. Yu, M.J. Chen, L.J. Hu, Y.J. Guo, L.F. Sun, Chem. Phys. Lett. 501 (2011) 475-479.
[292] X.X. Yang, J.W. Li, Z.F. Zhou, Y. Wang, W.T. Zheng, C.Q. Sun, Appl. Phys. Lett. 99 (2011) 133108.
[293] J.W. Li, L.W. Yang, Z.F. Zhou, X.J. Liu, G.F. Xie, Y. Pan, C.Q. Sun, J. Phys. Chem. B 114 (2010) 1648-1651.
[294] M.X. Gu, L.K. Pan, B.K. Tay, C.Q. Sun, J. Raman Spectrosco. 38 (2007) 780-788.
[295] J.D. Smith, C.D. Cappa, K.R. Wilson, R.C. Cohen, P.L. Geissler, R.J. Saykally, PNAS 102 (2005) 14171-14174.
[296] F. Paesani, J. Phys. Chem. A 115 (2011) 6861-6871.
[297] M. Paolantoni, N.F. Lago, M. Albertí, A. Laganà, J. Phys. Chem. A 113 (2009) 15100-15105.
[298] Y. Marechal, J. Chem. Phys. 95 (1991) 5565-5573.
[299] G.E. Walrafen, J. Chem. Phys. 47 (1967) 114-126.
[300] Y. Yoshimura, S.T. Stewart, H.K. Mao, R.J. Hemley, J. Chem. Phys. 126 (2007) 174505.
[301] I. Durickovic, R. Claverie, P. Bourson, M. Marchetti, J.M. Chassot, M.D. Fontana, J. Raman Spectrosco. 42 (2011) 1408-1412.
[302] K. Furic, V. Volovsek, J. Mol. Struct. 976 (2010) 174-180.
[303] H. Suzuki, Y. Matsuzaki, A. Muraoka, M. Tachikawa, J. Chem. Phys. 136 (2012) 234508.
[304] Y. Marechal, J. Mol. Struct. 1004 (2011) 146-155.
[305] X. Xue, Z.-Z. He, J. Liu, J. Raman Spectrosco. 44 (2013) 1045-1048.
[306] J.J. Shephard, J.S.O. Evans, C.G. Salzmann, J. Phys. Chem.Lett. (2013) 3672-3676.
[307] C. Medcraft, D. McNaughton, C.D. Thompson, D.R.T. Appadoo, S. Bauerecker, E.G. Robertson, PCCP 15 (2013) 3630-3639.
[308] C. Medcraft, D. McNaughton, C.D. Thompson, D. Appadoo, S. Bauerecker, E.G. Robertson, The Astrophysical Journal 758 (2012) 17.
[309] G.P. Johari, H.A.M. Chew, T.C. Sivakumar, J. Chem. Phys. 80 (1984) 5163.
[310] T. Tokushima, Y. Harada, O. Takahashi, Y. Senba, H. Ohashi, L.G.M. Pettersson, A. Nilsson, S. Shin, Chem. Phys. Lett. 460 (2008) 387-400.
[311] J.H. Guo, Y. Luo, A. Augustsson, J.E. Rubensson, C. Såthe, H. Ågren, H. Siegbahn, J. Nordgren, Phys. Rev. Lett. 89 (2002) 137402.
[312] X. Liu, Y. Wang, C.Q. Sun, Chem. Rev. (2014) Accepted on 30/09.
[313] L.B. Skinner, C. Huang, D. Schlesinger, L.G. Pettersson, A. Nilsson, C.J. Benmore, J. Chem. Phys. 138 (2013) 074506.
[314] K.T. Wikfeldt, M. Leetmaa, A. Mace, A. Nilsson, L.G.M. Pettersson, J. Chem. Phys. 132 (2010) 104513.
[315] J. Morgan, B.E. Warren, J. Chem. Phys. 6 (1938) 666-673.
[316] L.A. Naslund, D.C. Edwards, P. Wernet, U. Bergmann, H. Ogasawara, L.G.M. Pettersson, S. Myneni, A. Nilsson, J. Phys. Chem. A 109 (2005) 5995-6002.
[317] L. Orgel, Rev. Mod. Phys. 31 (1959) 100-102.
[318] A.H. Narten, W.E. Thiessen, L. Blum, Science 217 (1982) 1033-1034.
[319] L. Fu, A. Bienenstock, S. Brennan, J Chem Phys 131 (2009) 234702.
[320] J.L. Kuo, M.L. Klein, W.F. Kuhs, J Chem Phys 123 (2005) 134505.
[321] A.K. Soper, J Phys Condens Matter 19 (2007) 335206.
[322] M. Hakala, K. Nygård, S. Manninen, L.G.M. Pettersson, K. Hämäläinen, Phys. Rev. B 73 (2006) 035432.

[323] A.K. Soper, Pure Appl. Chem. 82 (2010) 1855-1867.
[324] J.W.G. Tyrrell, P. Attard, Phys. Rev. Lett. 87 (2001) 176104.
[325] Aristotle, Meteorology http://classics.mit.edu/Aristotle/meteorology.1.i.html, 350 B.C.E.
[326] E.B. Mpemba, D.G. Osborne, Phys. Educ. 14 (1979) 410-413.
[327] D. Auerbach, Am. J. Phys. 63 (1995) 882-885.
[328] M. Jeng, Am. J. Phys. 74 (2006) 514.
[329] C.A. Knight, Am. J. Phys. 64 (1996) 524-524.
[330] M. Vynnycky, S.L. Mitchell, Heat Mass Transfer. 46 (2010) 881-890.
[331] H. Heffner, http://www.mtaonline.net/~hheffner/Mpemba.pdf (2001).
[332] M. Vynnycky, N. Maeno, Int. J. Heat Mass Transfer 55 (2012) 7297-7311.
[333] J.I. Katz, Am. J. Phys. 77 (2009) 27-29.
[334] J.D. Brownridge, Am. J. Phys. 79 (2011) 78.
[335] P. Ball, Physics world 19 (2006) 19-21.
[336] N. Bregović, http://www.rsc.org/images/nikola-bregovic-entry_tcm18-225169.pdf (2012).
[337] L.B. Kier, C.K. Cheng, Chemistry & Biodiversity 10 (2013) 138-143.
[338] J.R. Welty, C.E. Wicks, R.E. Wilson, G.L. Rorrer, Fundamentals of Momentum, Heat and Mass transfer, John Wiley and Sons, 2007.
[339] G.R. Desiraju, Angew. Chem. Int. Ed. 50 (2011) 52-59.
[340] L.R. Falvello, Angew. Chem. Int. Ed. 49 (2010) 10045-10047.
[341] M. Freeman, Phys. Educ. 14 (1979) 417-421.
[342] B. Wojciechowski, Cryst. Res. Technol 23 (1988) 843-848.
[343] J.D. Smith, R.J. Saykally, P.L. Geissler, J Am Chem Soc 129 (2007) 13847-13856.
[344] Q. Sun, Vib. Spectrosc 62 (2012) 110-114.
[345] S. Park, M.D. Fayer, PNAS 104 (2007) 16731-16738.
[346] S. Iikubo, K. Kodama, K. Takenaka, H. Takagi, M. Takigawa, S. Shamoto, Phys. Rev. Lett. 101 (2008) 205901.
[347] A.L. Goodwin, M. Calleja, M.J. Conterio, M.T. Dove, J.S.O. Evans, D.A. Keen, L. Peters, M.G. Tucker, Science 319 (2008) 794-797.
[348] A.C. McLaughlin, F. Sher, J.P. Attfield, Nature 436 (2005) 829-832.
[349] J.S.O. Evans, J. Chem. Soc.-Dalton Trans. (1999) 3317-3326.
[350] T.A. Mary, J.S.O. Evans, T. Vogt, A.W. Sleight, Science 272 (1996) 90-92.
[351] S. Stoupin, Y.V. Shvyd'ko, Phys. Rev. Lett. 104 (2010) 085901.
[352] Q.H. Tang, T.C. Wang, B.S. Shang, F. Liu, Sci China-Phys Mech Astron G 55 (2012) 933.
[353] Y.J. Su, H. Wei, R.G. Gao, Z. Yang, J. Zhang, Z.H. Zhong, Y.F. Zhang, Carbon 50 (2012) 2804-2809.
[354] A.W. Sleight, Inorg. Chem. 37 (1998) 2854-2860.
[355] J.S.O. Evans, T.A. Mary, T. Vogt, M.A. Subramanian, A.W. Sleight, Chem. Mater. 8 (1996) 2809-2823.
[356] G. Ernst, C. Broholm, G.R. Kowach, A.P. Ramirez, Nature 396 (1998) 147-149.
[357] A.K.A. Pryde, K.D. Hammonds, M.T. Dove, V. Heine, J.D. Gale, M.C. Warren, J. Phys.: Condens. Matter 8 (1996) 10973-10982.
[358] E.-M. Choi, Y.-H. Yoon, S. Lee, H. Kang, Phys. Rev. Lett. 95 (2005) 085701.
[359] D. Rzesanke, J. Nadolny, D. Duft, R. Muller, A. Kiselev, T. Leisner, PCCP 14 (2012) 9359-9363.
[360] J.T. Bartlett, A.P. Vandenheuvel, B.J. Mason, Zeitschrift Fur Angewandte Mathematik Und Physik 14 (1963) 599-&.
[361] D. Ehre, E. Lavert, M. Lahav, I. Lubomirsky, Science 327 (2010) 672-675.
[362] M. Gavish, J.L. Wang, M. Eisenstein, M. Lahav, L. Leiserowitz, Science 256 (1992) 815-818.

[363]	I.M. Svishchev, P.G. Kusalik, J. Am. Chem. Soc. 118 (1996) 649-654.
[364]	R. Zangi, A.E. Mark, J. Chem. Phys. 120 (2004) 7123-7130.
[365]	R.M. Namin, S.A. Lindi, A. Amjadi, N. Jafari, P. Irajizad, Phys. Rev. E 88 (2013) 033019.
[366]	R.C. Ponterio, M. Pochylski, F. Aliotta, C. Vasi, M.E. Fontanella, F. Saija, Journal of Physics D-Applied Physics 43 (2010) 175405.
[367]	W. Armstrong, The Electrical Engineer 10 (1893) 153.
[368]	E.C. Fuchs, J. Woisetschlager, K. Gatterer, E. Maier, R. Pecnik, G. Holler, H. Eisenkolbl, Journal of Physics D-Applied Physics 40 (2007) 6112-6114.
[369]	J. Woisetschlager, K. Gatterer, E.C. Fuchs, Exp. Fluids 48 (2010) 121-131.
[370]	E.C. Fuchs, Water 2 (2010) 381-410.
[371]	A. Widom, J. Swain, J. Silverberg, S. Sivasubramanian, Y.N. Srivastava, Phys. Rev. E 80 (2009) 016301.
[372]	A.A. Aerov, Phys. Rev. E 84 (2011) 036314.
[373]	G. Zhang, W. Zhang, H. Dong, J. Chem. Phys. 133 (2010) 134703.
[374]	Y. Fujimura, M. Iino, J. Appl. Phys. 103 (2008) 2940128.
[375]	R. Cai, H. Yang, J. He, W. Zhu, J. Mol. Struct. 938 (2009) 15-19.
[376]	Z. Zhou, H. Zhao, J. Han, CIESC Journal 63 (2012) 1408-1410
[377]	X. Zhang, T. Yan, B. Zou, C.Q. Sun, http://arxiv.org/abs/1310.1441.
[378]	W.J. Xie, Y.Q. Gao, J. Phys. Chem.Lett. (2013) 4247-4252.
[379]	E.K. WILSON, Chemical & Engineering News Archive 90 (2012) 42-43.
[380]	M. Randall, C.F. Failey, Chem. Rev. 4 (1927) 271-284.
[381]	M. Randall, C.F. Failey, Chem. Rev. 4 (1927) 285-290.
[382]	M. Randall, C.F. Failey, Chem. Rev. 4 (1927) 291-318.
[383]	P. Lo Nostro, B.W. Ninham, Chem. Rev. 112 (2012) 2286-2322.
[384]	F. Aliotta, M. Pochylski, R. Ponterio, F. Saija, G. Salvato, C. Vasi, Phys. Rev. B 86 (2012) 134301.
[385]	S. Park, M.B. Ji, K.J. Gaffney, J. Phys. Chem. B 114 (2010) 6693-6702.
[386]	S. Park, M. Odelius, K.J. Gaffney, J. Phys. Chem. B 113 (2009) 7825-7835.
[387]	K.J. Gaffney, M. Ji, M. Odelius, S. Park, Z. Sun, Chem. Phys. Lett. 504 (2011) 1-6.
[388]	C. Sun, D. Xu, D. Xue, CrystEngComm 15 (2013) 7783-7791.
[389]	Y.R. Xu, L. Li, P.J. Zheng, Y.C. Lam, X. Hu, Langmuir 20 (2004) 6134-6138.
[390]	Y. Xu, C. Wang, K.C. Tam, L. Li, Langmuir 20 (2004) 646-652.
[391]	M. Baumgartner, R.J. Bakker, Mineralogy and Petrology 95 (2008) 1-15.
[392]	X. Zhang, C.Q. Sun, http://arxiv.org/abs/1310.0893 (2013).
[393]	G. Imperato, E. Eibler, J. Niedermaier, B. Konig, Chem. Commun. (2005) 1170-1172.
[394]	M.D.A. Saldaña, V.H. Alvarez, A. Haldar, J. Chem. Thermoldynamics 55 (2012) 115-123.
[395]	L.K. Pan, C.Q. Sun, T.P. Chen, S. Li, C.M. Li, B.K. Tay, Nanotechnology 15 (2004) 1802-1806.
[396]	L.K. Pan, H.T. Huang, C.Q. Sun, J. Appl. Phys. 94 (2003) 2695-2700.
[397]	R. Tsu, D. Babic, Appl. Phys. Lett. 64 (1994) 1806-1808.
[398]	J.W. Li, L.W. Yang, Z.F. Zhou, P.K. Chu, X.H. Wang, J. Zhou, L.T. Li, C.Q. Sun, J. Phys. Chem. C 114 (2010) 13370-13374.
[399]	L.K. Pan, Y.K. Ee, C.Q. Sun, G.Q. Yu, Q.Y. Zhang, B.K. Tay, J. Vac. Sci. Technol. B 22 (2004) 583-587.
[400]	G. Ouyang, C.Q. Sun, W.G. Zhu, J. Phys. Chem. C 113 (2009) 9516-9519.
[401]	W.B. Floriano, M.A.C. Nascimento, Brazilian Journal of Physics 34 (2004) 38-41.
[402]	I. Michalarias, I. Beta, R. Ford, S. Ruffle, J.C. Li, Appl. Phys. A 74 (2002) s1242-s1244.
[403]	E. Zolfagharifard, Mail online (2014).
[404]	Z. Wang, F.C. Wang, Y.P. Zhao, Proc. Roy. Soc. A 468 (2012) 2485-2495.

[405] S. Ceurstemont, Newscientist: http://www.newscientist.com/blogs/nstv/one-minute-physics/ (2013).
[406] Q. Yuan, Y.-P. Zhao, J. Fluid Mech 716 (2013) 171-188.
[407] Q.Z. Yuan, Y.P. Zhao, Sci. Rep. 3 (2013) 1944.
[408] T. Yan, S. Li, K. Wang, X. Tan, Z. Jiang, K. Yang, B. Liu, G. Zou, B. Zou, J. Phys. Chem. B 116 (2012) 9796-9802.
[409] K. Wang, D. Duan, R. Wang, A. Lin, Q. Cui, B. Liu, T. Cui, B. Zou, X. Zhang, J. Hu, G. Zou, H.K. Mao, Langmuir 25 (2009) 4787-4791.
[410] C.-S. Zha, Z. Liu, R. Hemley, Phys. Rev. Lett. 108 (2012) 146402.
[411] H. Liu, H. Wang, Y. Ma, J. Phys. Chem. C 116 (2012) 9221-9226.